%% file: paper.tex
\documentclass[a4paper,fleqn]{cas-dc}
\usepackage{enumitem}

\usepackage[authoryear]{natbib}
\usepackage{graphicx}
\usepackage{booktabs}
\usepackage{footnote}
\usepackage{multirow}
\usepackage{listings}
\usepackage{caption}
\usepackage{subcaption}
\usepackage{hyperref}
\usepackage{import}
\usepackage{xspace}
\usepackage{comment}
\usepackage{amsmath, nccmath}
\usepackage{fancyhdr}
\usepackage{listings}
\usepackage[frozencache,cachedir=.]{minted}

\DeclareUnicodeCharacter{2212}{-}
\DeclareUnicodeCharacter{223C}{~}
\DeclareUnicodeCharacter{2217}{*}

\def\tsc#1{\csdef{#1}{\textsc{\lowercase{#1}}\xspace}}
\tsc{WGM}
\tsc{QE}
\tsc{EP}
\tsc{PMS}
\tsc{BEC}
\tsc{DE}
\begin{document}
\let\WriteBookmarks\relax
\def\floatpagepagefraction{1}
\def\textpagefraction{.001}

\shorttitle{Empirical study on bug severity estimation}

%\shortauthors{CV Radhakrishnan et~al.}

\title [mode = title]{An Empirical Study on Bug Severity Estimation using Source Code Metrics and Static Analysis}                      
\tnotemark[1,2]

%\pagestyle{fancy}
%\fancyfoot{}

\ExplSyntaxOn
\cs_gset:Npn \__first_footerline:
  { \group_begin: \small \sffamily \group_end: }
\ExplSyntaxOff 

%\tnotetext[1]{This document is the results of the research
%   project funded by the National Science Foundation.}

%\tnotetext[2]{The second title footnote which is a longer text matter
%   to fill through the whole text width and overflow into
%   another line in the footnotes area of the first page.}

\author[1]{{Ehsan Mashhadi}}

%[type=editor, auid=000,bioid=1, prefix=Sir, role=Researcher, orcid=0000-0001-7511-2910]

%\cormark[1]


%\ead{cvr_1@tug.org.in}

%\ead[url]{www.cvr.cc, cvr@sayahna.org}

\credit{Conceptualization of this study, Methodology, Software}

\affiliation[1]{organization={University of Calgary},
    %addressline={Radarweg 29}, 
    city={Calgary},
    % citysep={}, % Uncomment if no comma needed between city and postcode
    %postcode={1043 NX}, 
    state={Alberta},
    country={Canada}}

\affiliation[2]{organization={University of Manitoba},
    %addressline={Radarweg 29}, 
    city={Winnipeg},
    % citysep={}, % Uncomment if no comma needed between city and postcode
    %postcode={1043 NX}, 
    state={Manitoba},
    country={Canada}}    

\author[2]{{Shaiful Chowdhury}}%[style=chinese]

\author[1]{{Somayeh Modaberi}}
%[role=Co-ordinator,suffix=Jr,]
%\ead{cvr3@sayahna.org}
%\ead[URL]{www.sayahna.org}

\credit{Data curation, Writing - Original draft preparation}

\affiliation[3]{organization={ York University},
    city={Toronto},
    % citysep={}, % Uncomment if no comma needed between city and postcode
    %postcode={695014}, 
    state={Ontario},
    country={Canada}}

%\cormark[2]
%\fnmark[1,3]
% \cormark[1]
% \ead{hossein.ahmadvand@ucalgary.ca}
%\ead[URL]{www.stmdocs.in}

\author[1,3]{{Hadi Hemmati}}%[style=chinese]
\cormark[1]
\ead{hemmati@yorku.ca}

\author[1]{{Gias Uddin}}%[style=chinese]

\cortext[cor1]{Corresponding author}
%\cortext[cor2]{Principal corresponding author}

%\fntext[fn1]{This is the first author footnote. but is common to third author as well.}
%\fntext[fn2]{Another author footnote, this is a very long footnote and it should be a really long footnote. But this footnote is not yet sufficiently long enough to make two lines of footnote text.}

%\nonumnote{This note has no numbers. In this work we demonstrate $a_b$ the formation Y\_1 of a new type of polariton on the interface between a cuprous oxide slab and a polystyrene micro-sphere placed on the slab.}
\begin{abstract}
In the past couple of decades, significant research efforts have been devoted to the prediction of software bugs (i.e., defects). In general, these works leverage a diverse set of metrics, tools, and techniques to predict which classes, methods, lines, or commits are buggy. However, most existing work in this domain treats all bugs the same, which is not the case in practice. The more severe the bugs the higher their consequences. Therefore, it is important for a defect prediction method to estimate the severity of the identified bugs, so that the higher severity ones get immediate attention. 
In this paper, we provide a quantitative and qualitative study on two popular datasets (Defects4J and Bugs.jar), using 10 common source code metrics, and two popular static analysis tools (SpotBugs and Infer) for analyzing their capability to predict defects and their severity. We studied 3,358 buggy methods with different severity labels from 19 Java open-source projects. Results show that although code metrics are useful in predicting buggy code (Lines of the Code, Maintainable Index, FanOut, and Effort metrics are the best), they cannot estimate the severity level of the bugs. In addition, we observed that static analysis tools have weak performance in both predicting bugs (F1 score range of 3.1\%-7.1\%) and their severity label (F1 score under 2\%). 
We also manually studied the characteristics of the severe bugs to identify possible reasons behind the weak performance of code metrics and static analysis tools in estimating their severity. Also, our categorization shows that Security bugs have high severity in most cases while Edge/Boundary faults have low severity. Finally, we discuss the practical implications of the results and propose new directions for future research. 
\end{abstract}

% \includegraphics{figs/grabs.pdf}
\begin{comment}
\begin{highlights}
\item Research highlights item 1
\item Research highlights item 2
\item Research highlights item 3
\end{highlights}
\end{comment}

% Keywords

\maketitle

\input{introduction}
\input{experiments}
\input{discussion}
\input{limitations}
\input{related_work}

\input{conclusion}
\section{Acknowledgement}
This work was partially supported by the NSERC Discovery Grant (RGPIN/04552-2020) and the NSERC and Alberta Innovates Alliance Grant (ALLRP/568643-2021).


\bibliographystyle{cas-model2-names}

\bibliography{paper}


\end{document}

%% file: introduction.tex
\section{Introduction}
\label{intro}

Software maintenance is one of the most challenging and expensive phases in the software development life cycle \cite{kafura1987use}. Handling bugs (including detecting, localizing, fixing, etc.) is the most typical challenge associated with the software maintenance step \cite{borstler2016role,bennett2000software}. Consequently, both practitioners and researchers are trying to make this tedious task as automated as possible from different aspects such as defect prediction, test generation, fault-localization, and program repair   \cite{kondo2020impact,shin2010evaluating,tosun2010practical,zhou2010ability,mashhadi2021applying}.

While there has been much research in handling bugs using different techniques like search-based \cite{le2011genprog}, pattern-based \cite{long2016automatic} and ML-based techniques \cite{mashhadi2021applying}, there is little research that focuses explicitly on the severe bugs. In other words, most of the research implicitly assumes that all bugs have the same importance \cite{shamshiri2015automatically,mashhadi2021applying,wong2016survey,pearson2017evaluating}. However, the bug severity indicates the intensity of the impact the bug has on system operation \cite{neysiani2020efficient}. Critical bugs may cause a system to crash completely or cause non-recoverable conditions such as data loss. High-severity bugs affect major system components that prevent users from working with some parts of the system. Presumably, fixing severe bugs is typically more challenging compared to the medium or low severity bugs, where few components are affected and there is an easy workaround \cite{vucevic2012testing}. Therefore, in practice, bugs with higher severity tend to be fixed sooner than other less severe bugs ~\cite{uddin2017survey,kanwal2012bug,saha2014empirical}.

In general, software practitioners have several means to detect bugs. These can range from QA practices such as code review \cite{kononenko2016code,mantyla2008types,bacchelli2013expectations}, and inspection to different testing approaches, and even development methodologies such as pair programming \cite{williams2000strengthening,nawrocki2001experimental,sun2015effectiveness}, and Test Driven Development (TDD) \cite{bhat2006evaluating,martin2007professionalism,aniche2013asserts,rafique2012effects}. 
When a bug is detected, it is common for the development teams to consider the most severe/important bugs first which helps them to prevent extreme consequences. Some issue tracking systems, such as Jira \cite{Jira2023}, have a specific field named Severity/Priority which is assigned during the bug reporting and helps the development teams to consider bug importance during debugging.

It is important to note that bug severity does not have a standard formal definition. The definition depends on the context of the software, the nature of the bug, the current state of the project, the ratio of affected users, potential harm to users, and many other factors. The current state of the practice to identify and record the severity of bugs is through a manual process in issue tracking systems, where the severity of a bug has its own field (with options such as \textit{Blocker}, \textit{Critical}, \textit{Major}, \textit{Minor}, and \textit{Trivial} or sometimes with numbers ranging from zero to 20). The field is manually populated by the person (e.g., a developer, tester, or user) who documents the bug, which, however, may change by the technical team during the bug reporting review process. Thus, the whole process of bug labeling is quite expensive. To alleviate this difficulty, significant research has focused on automatically labeling bug severity to a bug report (e.g.,~\cite{tian2012information, tian2013drone, ramay2019deep, lamkanfi2010predicting}).

Predicting the severity of bug reports, however, does not help in detecting bugs and their severity from the beginning of the software development life cycle (SDLC). Detecting bugs at the later stages of SDLC is much more expensive than detecting and fixing them early~\cite{ibm}. In this paper, we aim to study the feasibility of predicting software bugs and their severity using source code metrics and different static analysis tools. Code metrics have shown a great deal of success in predicting code smell \cite{tufano2015and}, maintenance effort \cite{polo2001using} and defects \cite{giger2012method,ferenc2020deep}. Static analysis tools such as SpotBugs~\cite{spotbugs}, the successor to the popular FindBugs \cite{findbugs} tool, Facebook Infer \cite{infer}, and Google Error Prone \cite{error_prone} have also been successfully used in related research \cite{habib2018many,tomassi2018bugs}. These tools use different techniques such as AST-based patterns or data-flow analysis to find a bug's existence and to predict the bug's type and severity.   

Our approach focuses on method-level granularity in contrast to class/module-level, since many studies have shown that developers find the class/module level granularity impractical (too coarse-grained to be useful for remedial actions) \cite{shihab2012industrial,pascarella2020performance,grund2021codeshovel,hata2012bug}. 
We use two popular datasets: Defects4J~\cite{just2014defects4j}, and Bugs.jar~\cite{saha2018bugs} which contain real bugs from different open-source Java projects. We studied 19 projects containing 1,668 bugs (3,358 buggy methods) for our quantitative study. Furthermore, we studied 140 randomly sampled bugs from both datasets for our qualitative study to find out when and why different code metrics and the existing static analysis tools fail to distinguish bug severity. 

To guide our study, we aim to answer the following research questions (RQ):

\begin{itemize}
    \item \textbf{RQ1: Are source code metrics good indicators of buggyness and bug severity?} 
    The results show that most of the code metrics (e.g, Lines of Code, McCabe, McClure, Nested Block Depth, Proxy Indentation, FanOut, Readability, Difficulty, and Effort) are good indicators of buggyness, but they perform quite poorly for predicting bug severity. For example, the Line of Code metric shows excellent performance in finding buggyness, but very poor performance in finding bug severity. The Halstead Difficulty and Effort metrics show good performance in identifying bug severity when compared to other metrics.
    \item \textbf{RQ2: What is the capability of static analysis tools in finding bugs and their severity?} Results show that the studied static analysis tools (SpotBugs, and Infer) are not yet powerful enough to find many bug types, and in many cases, they mislabel the bug severity. Based on our experiments, these tools often fail to identify numerous bugs due to insufficient built-in patterns. Additionally, they struggle to accurately assess the severity of bugs, as they assign fixed severity values to each bug type without considering the contextual factors and potential consequences that could classify them as severe or non-severe bugs.
    \item \textbf{RQ3: What are the characteristics of bugs with different severity values?} Results reveal that the severity of bugs is mostly related to the software's specification, which is not predictable solely based on code metrics or static analysis. Also, we found no direct relationship between method complexity and its severity value. Many low-severity bugs exist in the quite complex methods according to the code metrics, but these functions handle trivial functionalities, such as GUI, or they do not lead to a crash or unauthorized access. 
\end{itemize}

The findings of this paper can help researchers and practitioners to better understand the characteristics of severe bugs and guide future research/tools on how to advance the field to better predict bug severity. We provide concrete future directions in this regard in Section \ref{sec:discussion}.
All the data and source code of this study are also publicly available \cite{Ehsan-Empirical2023} for replication studies.

The remainder of the paper is organized as follows: We provide our dataset, and experiments setup in Section \ref{sec:Methodology}. Experiment motivation, design, and results are discussed in Section \ref{sec:Desing_Result}. Discussions about the results are provided in Section \ref{sec:discussion}. Threats to the validity of our work are described in Section \ref{sec:Threats to Validity}. Related works to this paper are described in Section \ref{sec:Related Work}. We conclude this paper with possible future work in Section \ref{sec:Conclusion}.

%% file: experiments.tex
\section{Study Setup}
\label{sec:Methodology}

\begin{figure*}
\centering
\includegraphics[width=1\linewidth]{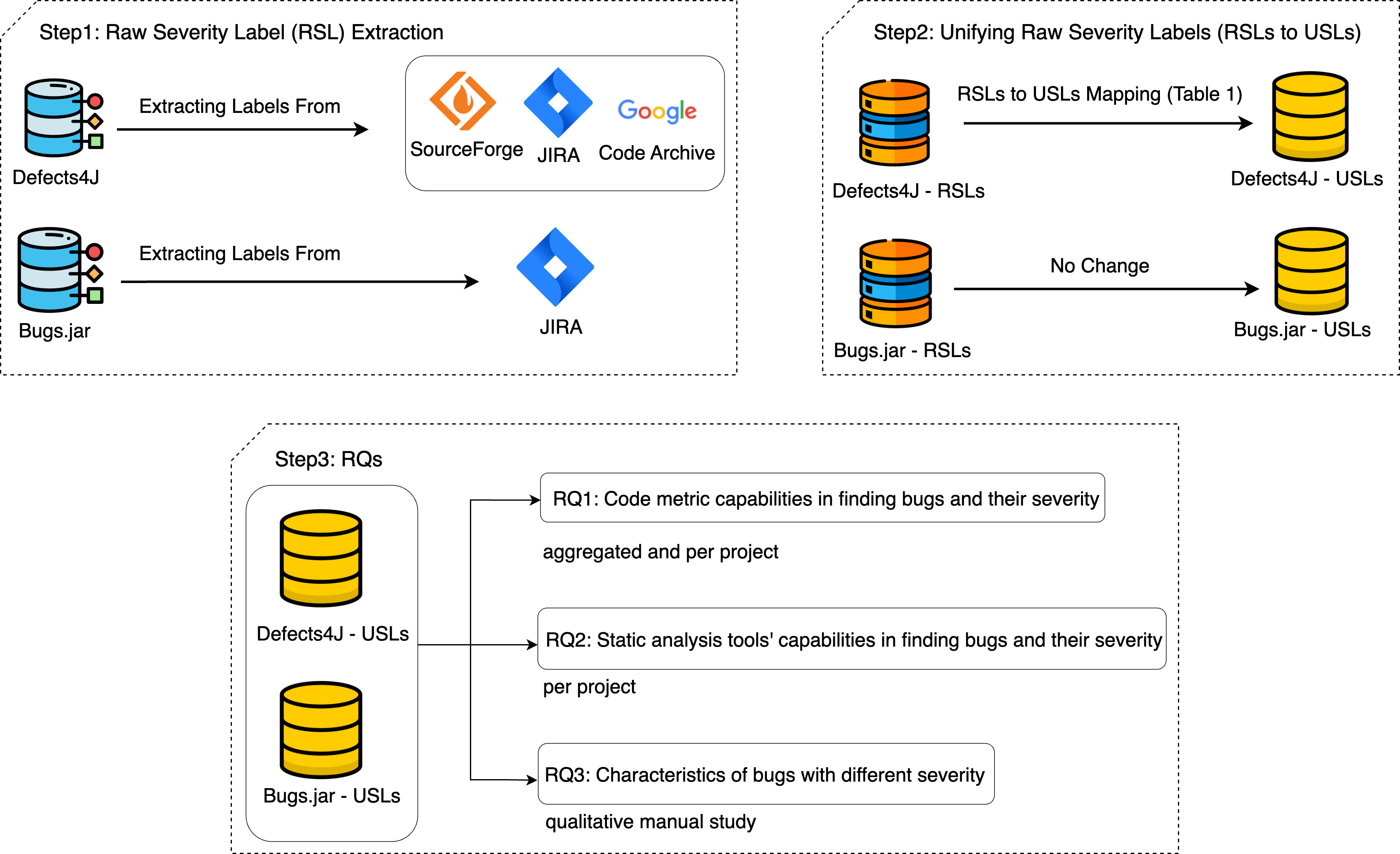}
   \caption{Overview of the setup of our empirical study.}
   \label{fig:overview} 
\end{figure*}

Fig \ref{fig:overview} summarizes our approach of collecting bug severity labels from two existing datasets (step 1), merging the raw severity labels to unified severity labels (step 2), and then formulating the research questions (step 3). We also discuss how we identify the set of buggy and not-buggy methods. Then we discuss the method-level source code metrics and the static analysis tools that we study in this paper.

\subsection{(Step1) Bug Datasets and Severity Extraction}
% \subsection{Dataset and Preprocessing}
\label{sec:Dataset}
We use Defect4J~\cite{just2014defects4j} and Bugs.jar~\cite{saha2018bugs} datasets for our study which contain bugs from different popular Java projects. Defect4J is used because it has been widely used in different automated software engineering research domains such as Test Generation \cite{shamshiri2015automatically}, Program Repair \cite{martinez2017automatic}, and Fault Localization \cite{pearson2017evaluating}. Also, this is a generic dataset containing real bugs from different projects with different domains. Bugs.jar has also been widely used in Program Repair~\cite{saha2017elixir} and contains many real bugs and there is a corresponding bug report for each sample in Jira \cite{Jira2023} containing the bug severity.

\begin{table}[h!]
    \caption{
\label{tab:preprocessing_d4j_severity} Unifying the raw severity labels of Defects4J dataset. For example, the third row means that bugs having raw severity labels of Medium or 5 are considered as Medium bugs.}
    \begin{center}
      \begin{tabular}{c|c}
        \toprule
        \textbf{Unified Severity Label (USL)} & \textbf{Raw Severity Label (RSL)}\\
        \midrule
        Critical & Critical \\
        High & High, Major, 3\\
        Medium & Medium, 5\\
        Low & Low, Trivial, Minor, 7, 8, 9 \\
        \bottomrule
      \end{tabular}
    \end{center}
  \end{table}
  
% \subsubsection{Defects4J} 
The latest version (2.0.0) of Defects4J contains 835 bugs from 17 Java projects. 
The dataset does not have any information regarding the bug severity itself, so we extract bug severity values from different issue-tracking systems. There are three software hosting services used for these projects: Jira, SourceForge, and Google Code Archive. We implemented scripts (available in our provided repository \cite{Ehsan-Empirical2023}) that either use open-source SDKs or scrap the web pages to extract the severity labels.
After extracting the severity values from these systems, we collected 510 bugs out of 835 bugs containing severity labels.

The Bugs.jar dataset contains 1,158 bugs from eight large, popular, and diverse open-source Java projects. 
All of these projects use Jira as an issue management system, and each of the bugs has an assigned severity level extracted in a similar way to the Defects4J dataset). The severity labels from this dataset are ``Blocker'', ``Critical'', ``Major'', ``Minor'', and ``Trivial''. Using the same technique mentioned above,

\subsection{(Step2) Unifying Raw Severity Labels (RSL to USL)}\label{sec:usls}
After extracting the raw severity labels (RSL) of the Defects4J dataset, we found that there are many inconsistencies between these values. Some projects use numerical values as an indicator for severity labels (smaller values indicate higher severity while larger values indicate lower severity) and others use categorical values due to the nature of their different issue-tracking systems. 
Therefore, we unified similar raw labels (RSL) to produce four meaningful unified severity labels (USL): ``Critical'', ``High'', ``Medium'', and ``Low'' severity labels.
Table \ref{tab:preprocessing_d4j_severity} shows the mapping between all RSL and the USL labels of the Defects4J dataset.\\
For the Bugs.jar dataset, we found that the extracted RSL values are consistent (since all of the bugs in the dataset are in the Jira issue-tracking system), so there is no need for a unifying process (RLS to USL) like what we did for the Defects4J dataset, and we considered all the RSLs as USLs.

It is important to note that, for both datasets, we have relied on human-labeled bug severity. In any bug-reporting system, some guidelines and examples exist to help practitioners and users decide what should be the label of a reported bug. For example, in Jira\footnote{\url{https://www.atlassian.com/incident-management/kpis/severity-levels}}, a severity level 1 (i.e., critical) is defined as, \emph{A critical incident with very high impact}. The presented examples are, \emph{A customer-facing service, like Jira Cloud, is down for all customers}; \emph{Confidentiality or privacy is breached}; and \emph{Customer data loss}. In contrast, a minor bug is defined as, \emph{A minor incident with low impact}, and one of the examples is, \emph{A minor inconvenience to customers, workaround available}. The relevant threats to validity in using human-labeled severity and their mitigation are discussed in Section~\ref{construct}.

\subsection{(Step3) Research Questions (RQs)}\label{sec:rqs}
After the mentioned preprocessing step we concluded with two datasets containing the USLs (RSLs are discarded and not used anymore). The Defect4J dataset contains ``Critical'', ``High'', ``Medium'',  and ``Low''. Bugs.jar dataset contains ``Blocker'', ``Critical'', ``Major'', ``Minor'', and ``Trivial''. This is our extracted and preprocessed dataset, but we will group some of these USLs in different RQs according to their experiment design requirements, which will be explained in their related sections accordingly. In general, we answer RQ1 and RQ2 with quantitative analysis, and RQ3 is answered with qualitative manual analysis.

\subsection{Buggy vs. Not Buggy}
Since we focus on the method-level granularity, we consider a method as a buggy method if it is modified/removed by a bug-fixing patch. If a method, however, is introduced with a bug-fix patch, we do not label this method as buggy. We discard the static initialization blocks and constructors since those are special types of methods that are mostly used for initializing the enclosed class, mostly.

A bug-fixing patch, however, can impact multiple methods, in that case, we consider all of them as buggy. This has been a common practice in earlier studies (e.g.,~\cite{chowdhury:2022,pascarella2020performance,Mo:2022}). We discuss the relevant threats in Section~\ref{sec:Threats to Validity}. Finally, from the 510 bugs of the Defects4J dataset, we found 742 buggy methods and from the 1,158 bugs of the Bugs.jar dataset, we found 2616 buggy methods.

Similar to earlier studies (e.g.,~\cite{chowdhury:2022,pascarella2020performance,Mo:2022}), we considered a method as non-buggy if it was not modified in the current bug fixing patch inside the buggy class. In this way, we are extracting methods that are contributing to the same functionality as the buggy methods since methods of a class have high cohesion.

We extracted 20,179 and 57,197 non-buggy methods from the Defects4J and Bugs.jar, respectively. The list of the buggy and non-buggy methods containing their project name, class name, start line, end line, and bug severity (corresponding bug severity for non-buggy methods) is available in our publicly shared GitHub repository \cite{Ehsan-Empirical2023}.

\begin{figure}
\centering
\begin{subfigure}[b]{1.0\columnwidth}
   \includegraphics[width=1\linewidth]{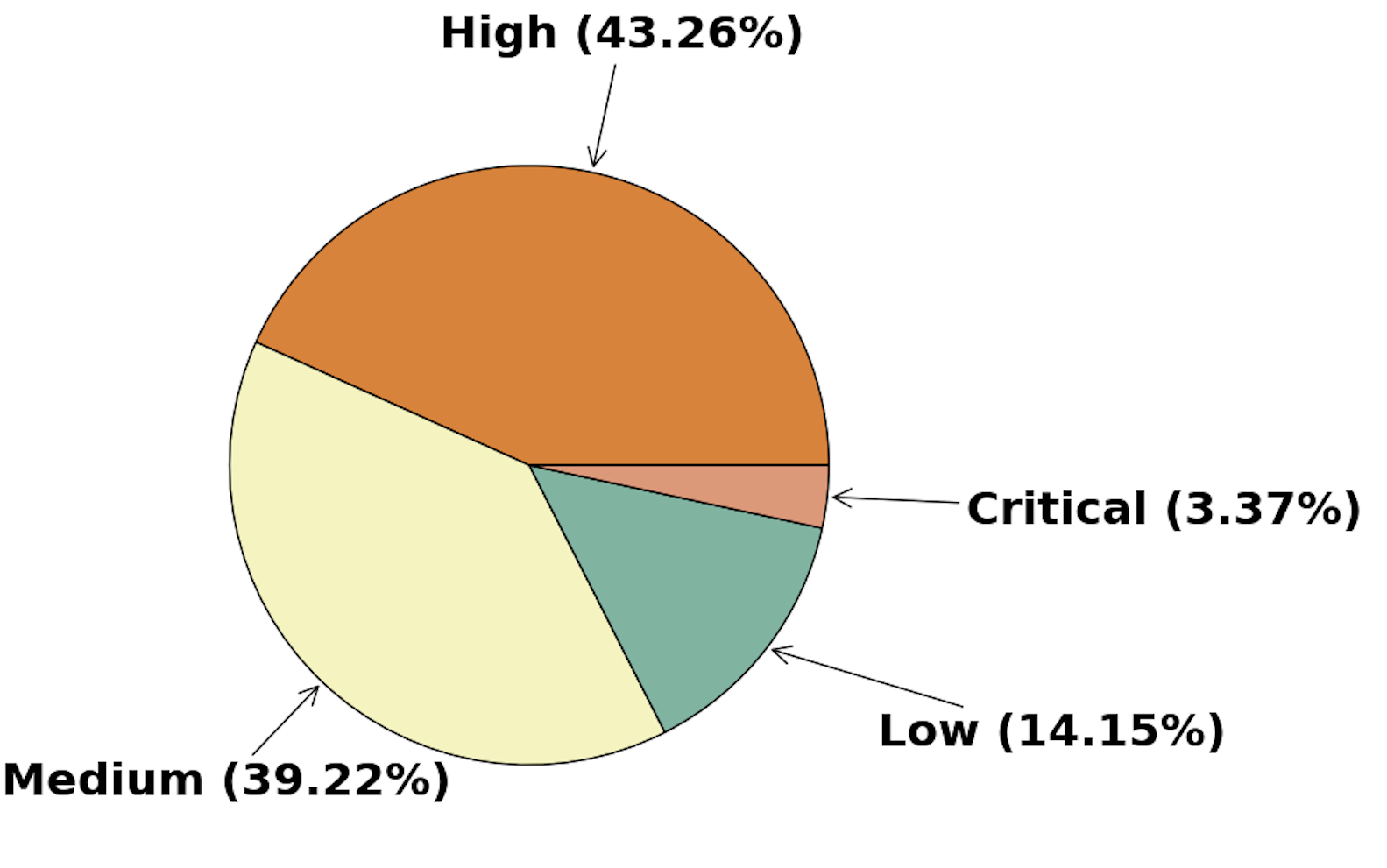}
   \caption{Defects4J}
\end{subfigure}

\begin{subfigure}[b]{1.0\columnwidth}
   \includegraphics[width=1\linewidth]{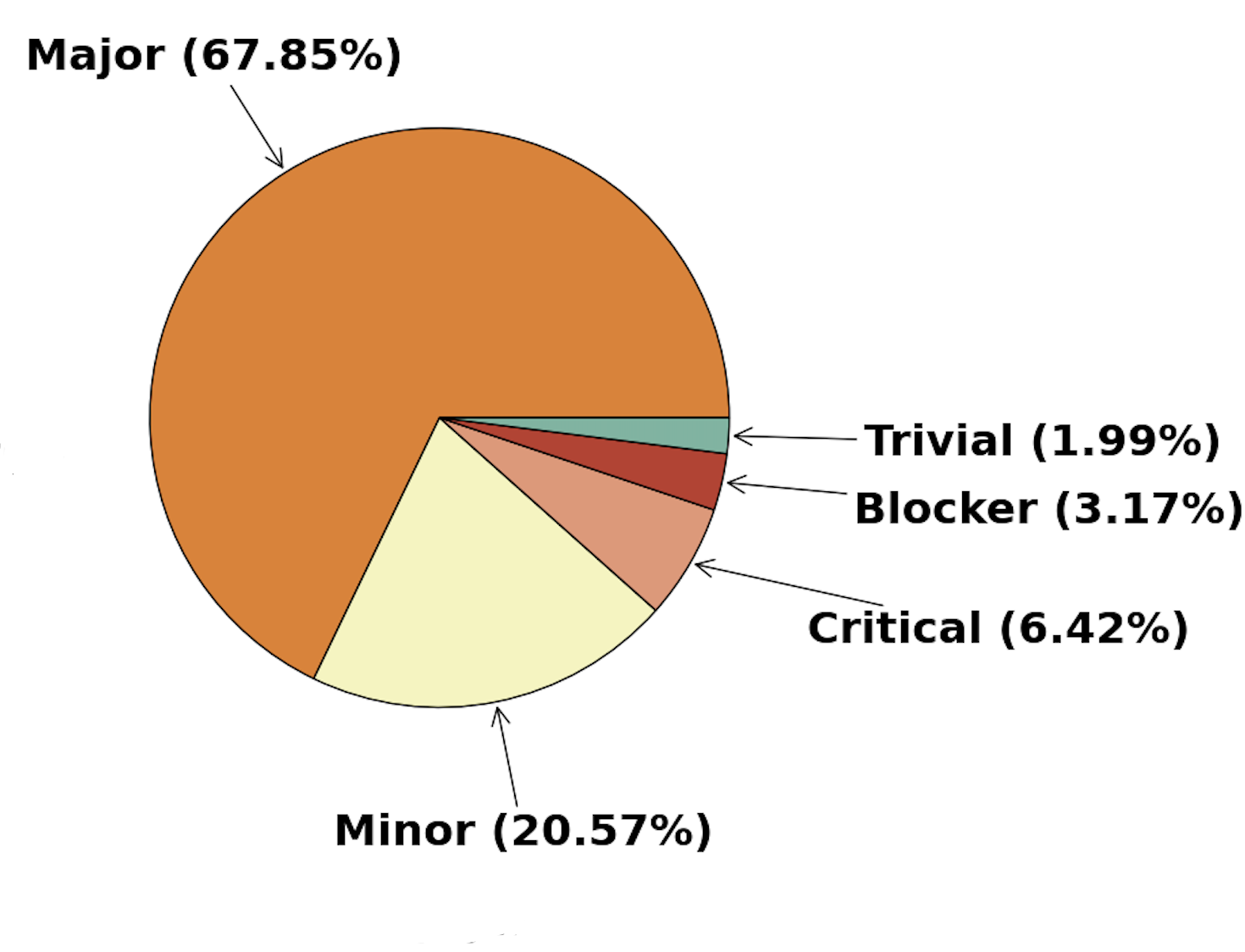}
   \caption{Bugs.jar}
\end{subfigure}

\caption[]{Buggy methods severity distributions of Defects4J and Bugs.jar datasets with their USL values.}
\label{fig:severity_distribution_buggymethods}
\end{figure}

\begin{figure*}
\centering
\includegraphics[width=1\linewidth]{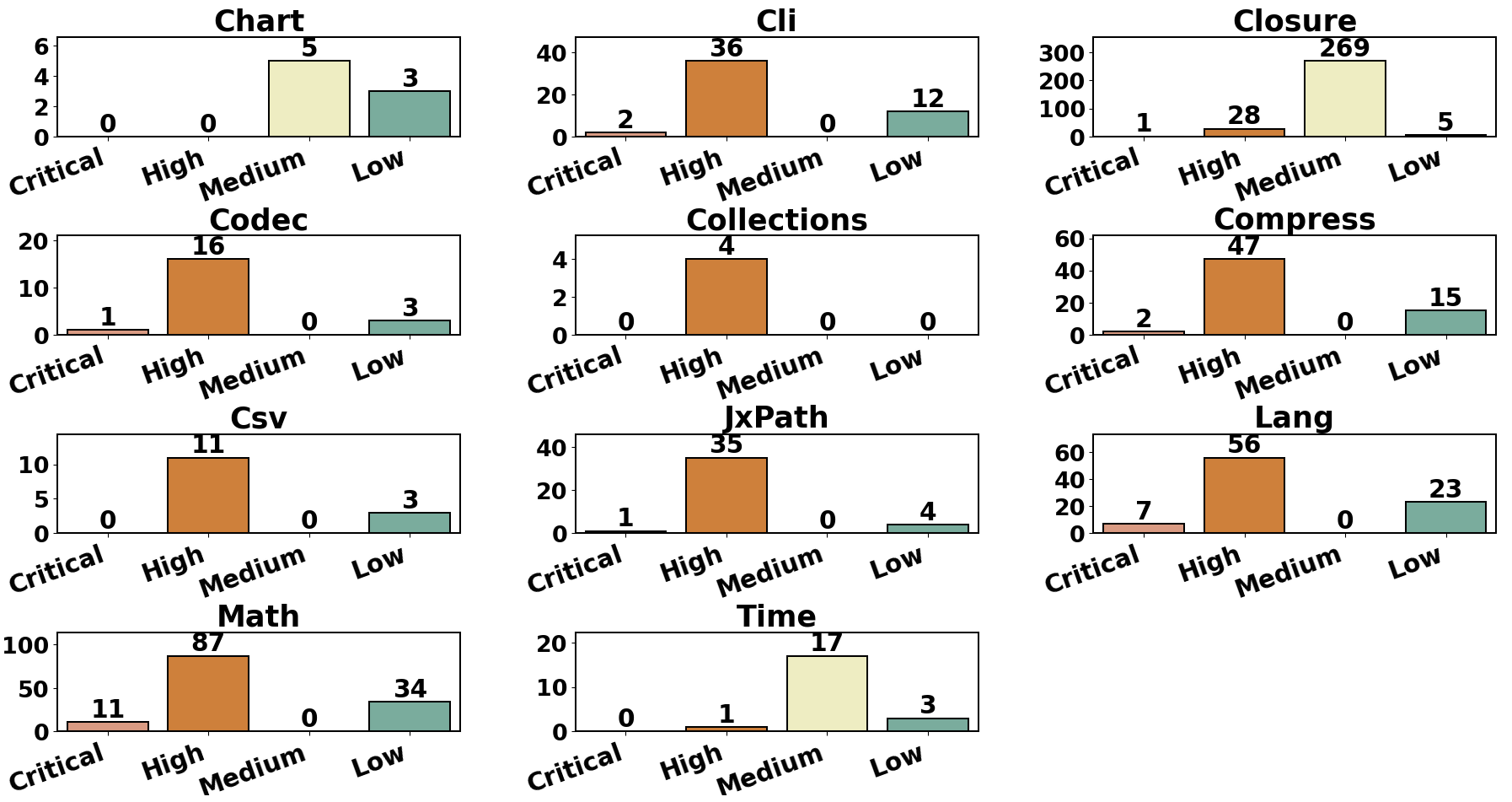}
   \caption{Buggy methods severity distributions in Defects4J dataset with the USL values.}
   \label{fig:d4j_project_severity_distribution} 
\end{figure*}

\begin{figure*}
\centering
\includegraphics[width=1\linewidth]{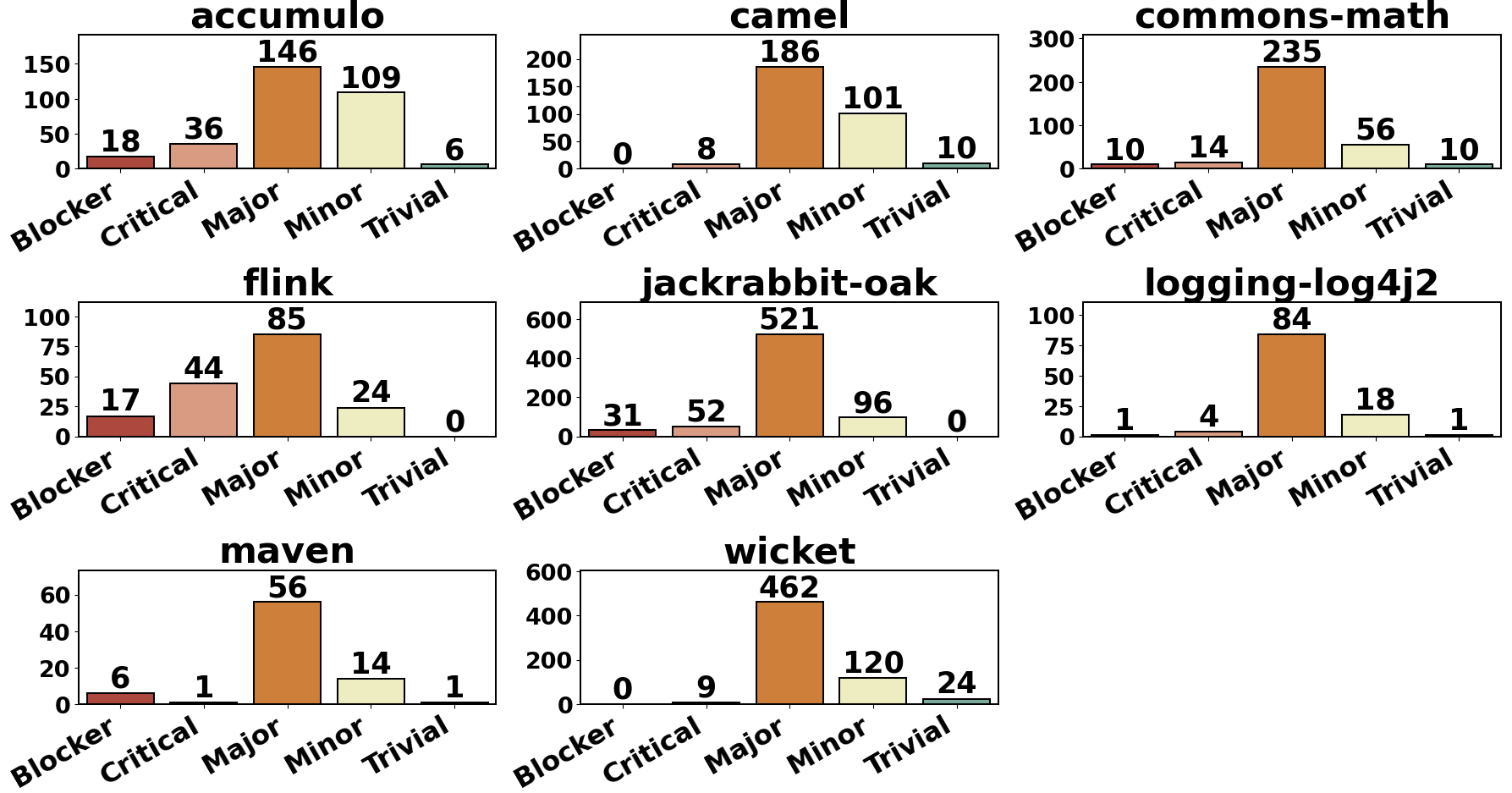}
   \caption{Buggy methods severity distributions in Bugs.jar dataset with the USL values.}
   \label{fig:bugs_jar_severity_distribution} 
\end{figure*}

The severity distributions of the buggy methods for Defects4J and Bugs.jar datasets are shown in Fig \ref{fig:severity_distribution_buggymethods}. We study these two datasets separately because of the difference in their labels. 
The severity distribution per project is shown in Fig \ref{fig:d4j_project_severity_distribution} and Fig \ref{fig:bugs_jar_severity_distribution}. The figures show that some projects do not have enough samples in all severity categories such as \verb|Chart| project, containing only a few bugs in ``Medium'' and ``Low'' severity groups, or the \verb|Collections| project, which only contains High severity bugs. Because of not having enough samples in each project/severity group, we may combine data in different RQs, which we discuss later. 

\subsection{Source Code Metrics}
Intuitively, if a code component implements a complicated functionality, it is more complex and may contain source code-induced code smells resulting in bugs. Consequently, different source code metrics with module-level, class-level, and method-level granularities have been used to measure software quality and predict bugs in previous research~\cite{giger2012method,pecorelli2019comparing,pascarella2020performance, gil2017correlation, chowdhury-method-2024, chowdhury2022revisiting}. These metrics generally focus on how large, complex, readable, and testable a code component is. The success of these code metrics in estimating maintenance efforts, such as predicting bugs, has been historically debated~\cite{chowdhury2022revisiting}. In recent research, however, it was discovered that although code metrics are not useful in understanding software maintenance at the class/file level~\cite{gil2017correlation}, they are very helpful in understanding maintenance in the method-level source code granularity~\cite{landman2014empirical, chowdhury2022revisiting}. Then again, this observation contradicts another recent study by ~\cite{pascarella2020performance}, who observed negative results while building code metrics-based method-level bug prediction models. None of these previous studies, however, investigate if code metrics are useful to understand bug severity. \emph{Perhaps, a bug inside a more complex method is more severe than a bug contained in a simpler method.} In this paper, we, therefore, not only investigate code metrics' effectiveness in detecting/predicting bugs at the method level but also their usefulness in understanding bug severity.   

We leverage most of the common method-level source code metrics that are used in the previous research (mentioned in the \hyperref[sec:Related Work]{Related Work} section) to see their capability in predicting buggy codes and their severity. Although the list of selected metrics is not exhaustive (given the numerous metrics explored in this field and the limitations of experiment size in one article), we made sure we have most of the metrics that have been shown effective in predicting the method-level buggyness, in the past. These metrics are defined as follows:

\textbf{Lines of Code (LC)}: Size, also known as lines of code (LC), is the most popular, easy to measure, and the most effective code metric for estimating software maintenance~\cite{gil2017correlation,Emam:2001,chowdhury:2022}.  
The use of LC as a proxy maintenance indicator is so prevalent that there are dedicated studies that completely focus on LC and its correlation with other quality metrics (e.g., ~\cite{gil2017correlation,landman2014empirical,chowdhury:2022}). LC has been extensively studied for bug prediction, fault localization, and for finding vulnerabilities (e.g.,~\cite{chowdhury-method-2024,pascarella2020performance,antinyan2014identifying,shin2010evaluating,chowdhury:2022}). In this paper, we calculate LC as the source lines of code without comments and blank lines, similar to~\cite{landman2014empirical,ralph2018construct,chowdhury:2022} to prevent the code formatting and comments effects.

\textbf{McCabe (MA)}: McCabe \cite{mccabe1976complexity,landman2014empirical}, also known as cyclomatic complexity, is another very popular metric that indicates the number of independent paths, and thus the logical complexity of a program. Intuitively, components with high McCabe values are more bug-prone. McCabe has been studied extensively to find bugs and locate suspicious code~\cite{antinyan2014identifying}, to understand its correlation with code quality~\cite{pantiuchina2018improving}, and to leverage its value for test generation methods, such as structured testing (path testing)~\cite{watson1996structured}. McCabe can be calculated as $1 + \#predicates$~\cite{mccabe1976complexity}.

\textbf{McClure (ML)}: McClure \cite{mcclure1978model,kafura1987use} was proposed as an improvement over McCabe. Unlike McCabe, McClure considers the number of control variables, and the number of comparisons in a predicate, which is not supported by McCabe. Intuitively, a predicate with multiple comparisons and multiple control variables would be more complex, and thus more bug-prone, than a predicate with only one comparison or only a single control variable.

\textbf{Nested Block Depth (NBD)}: McCabe and McClure do not consider nested depth. According to these metrics, there is no difference between two code snippets containing two identical \textit{for} loops if they are arranged serially or nested. NBD~\cite{kasto2013measuring,alenezi2019impact,zaw2020software} has been studied alongside McCabe and McClure to alleviate this issue. 

\textbf{Proxy Indentation (PI)}: Since McCabe-like complexity measures require a language-specific parser (for finding the predicates), Hindle et al. \cite{hindle2008reading} proposed Proxy Indentation metric as a proxy for McCabe-like complexity metrics. It was shown that, for measuring 
complexity, indentation measurement can perform very similar to more complex measurements such as McCabe, without requiring a language-specific
parser. Indentation measurement is done for each line, and then an aggregated value is calculated for the whole program component (e.g., a method).
Hindle et al. showed that the standard deviation as an aggregated value outperforms the mean, median, or max. Therefore, we only use the standard deviation form, as was also done by Chowdhury et al.~\cite{chowdhury2022revisiting}.

\textbf{FanOut (FO)}: This metric calculates the total number of methods called by a given method. This provides an estimate of the coupling---i.e., dependency of a particular method on other methods. It is observed that code components that are highly coupled are less maintainable and bug-prone~\cite{mo2016decoupling,alomar2019impact}.

\textbf{Readability (R)}: This metric combines different code features to calculate a single value for estimating code readability. We used the readability metric proposed by Buse et al.~\cite{buse2009learning} which generates a readability score for a given method. The readability scores range from 0 to 1 for specifying least readable code to most readable code, respectively. The authors concluded that this metric has a significant level of correlation with defects, code churn, and self-reported stability.

\textbf{Halstead Metrics}: The Halstead code metrics contain seven measures based on the number of operators and operands in a component \cite{halstead1977elements}. These metrics have been used in different research such as measuring code complexity perceived by developers~\cite{antinyan2017evaluating}, calculating the complexity of software maintenance tasks~\cite{curtis1979measuring,kafura1987use}, finding their correlation with indentation measures~\cite{hindle2008reading}, and estimating software readability ~\cite{posnett2011simpler}. Since all the Halstead metrics are highly correlated to each other, we consider only two of them: \textbf{Difficulty (D)} and \textbf{Effort (E)} which use other Halstead metrics in their formulas.
The Halstead Difficulty is calculated as shown in Eq\ref{Eq1}:

%\begin{displaymath}
\begin{equation} \label{Eq1}
D = \frac{n1}{2} * \frac{N2}{n2}
\end{equation}
%\end{displaymath}

Where ~$n1$ is the number of distinct operators, ~$n2$ is the number of distinct operands, and ~$N2$ is the total number of operands.\\

The Halstead Effort is calculated as shown in Eq\ref{Eq1}and \ref{Eq2}
%\begin{fleqn}
\begin{equation}\label{Eq2}
\begin{split}
%\begin{aligned}
&E = D * V\\
&V = N * \log_2 (n)\\
&N = N1 + N2\\
&n = n1 + n2
\end{split}
%\end{aligned}
\end{equation}
%\end{fleqn}

Where ~$N1$ is the total number of operators.\\

\textbf{Maintainability Index (MI)}: 
The Maintainability Index has been introduced by Omran and Hagemeister~\cite{oman1992metrics} where the authors defined metrics for measuring 
the maintainability of a software system and combine those metrics into a single value. This metric has evolved over time and was adopted by popular tools like Visual Studio~\cite{visual_studio}. 
MI calculated as shown in Eq\ref{Eq3}. 

\begin{equation}\label{Eq3}
\begin{split} 
%\begin{multline*}
&MI= 171- 5.2 ∗  ln(HalsteadVolume)- \\
&0.23 ∗ (McCabe)− 16.2 * ln(LC)
\end{split}
%\end{multline*}
\end{equation}
\\
Table \ref{tab:metrics_overview} provides a brief description of each of our 10 studied metrics with their abbreviation.

\begin{table*}[tbp]
    \caption{
\label{tab:metrics_overview} List of studied metrics and their brief description.}
    \begin{center}
      \begin{tabular}{c| p{12cm}}
        \toprule
        \textbf{Metric} & \textbf{Description}\\
        \midrule
        LC & Counts the number of source lines of code without comments and blank lines \\
        MA & Measures McCabe or the number of independent paths (cyclomatic complexity) \\
        ML & Similar to MA, except it also considers the number of control variables and comparisons in a predicate\\
        NBD & Counts the depth of the most nested block\\
        PI & Counts the indentation of source code lines\\
        FO & Counts the total number of methods called by a given method\\
        R& Measures the readability of a method in the range of 0-1 (least to most readable)\\
        D & Measures the Halstead difficulty of a method\\
        E & Measures the required Halstead effort of a method\\
        MI & Measures the maintainability of a method\\
        
        \bottomrule
      \end{tabular}
    \end{center}
  \end{table*}

\subsection{Static Analysis Tools}
Static analysis tools have been introduced for finding bugs by using different techniques, such as pattern matching. \textbf{FindBugs}~\cite{findbugs} is an open-source tool that analyzes Java byte code for finding bugs. \textbf{SpotBugs}~\cite{spotbugs} is a successor of FindBugs which is actively under maintenance. SpotBugs has more than 400 bug patterns and their description is available on the tool's website \cite{spotbugs}. \textbf{Infer} \cite{infer} is a static analysis tool for different programming languages such as Java, C++, Objective-C, and C, which is also actively maintained. Infer has more than 100 predefined issue types which are described on its website \cite{infer}.
\textbf{Error prone}~\cite{error_prone} is another static analysis tool from Google that is used for Java code to catch common programming faults at compile-time, and it is actively under maintenance. We have selected SpotBugs and Infer tools for our study because of their extensive usage in previous studies and in practice, and we exclude ErrorProne because its latest versions do not support older versions of Java projects that exist in our datasets. A more detailed explanation of our selected configuration for these tools is explained in Section~\ref{static-tools}.

\section{Experiment Design and Results}

In this section, we provide the motivation, experiment designs, and results for each RQ. Since each RQ has a few sub-RQs, they may have different experiment designs which are explained in their relevant sections.

\label{sec:Desing_Result}

\subsection{Code Metrics \& Bug Severity (RQ1)}
We aim to understand the capabilities of source code metrics in finding bugs and predicting their severity labels.

\subsubsection{Motivation}
One of the most important attributes in every issue tracking system is bug severity which is usually determined manually by the developers/QA team. This process is time-consuming and error-prone since the technical team should investigate the effects of the reported bug (e.g., number of affected users, number of crashes, probable consequence on the whole system) based on the bug description or other available data from analytical systems (e.g., crash report logging systems). 

There has been significant research that uses source code metrics for defect-related issues (more in Section~\ref{sec:Related Work}), but the severity prediction studies mostly use bug description which works in cases where there exists a well-written bug description. Therefore, in this RQ we study source code metrics' ability to distinguish among different types of bug severities.

\subsubsection{Approach}
To understand the effectiveness of source code metrics, we answer RQ1, which is divided into two sub-RQs.\\\\
\textbf{RQ1}: Are source code metrics good indicators of buggyness and bug severity?

\begin{itemize}
\itemsep-.2em
    \item \textbf{RQ1-1}: Do source code metrics distinguish between buggy and not-buggy code?
    \item \textbf{RQ1-2}: Do source code metrics distinguish between different bug severity?
\end{itemize}
\textbf{Design of RQ1:} In this RQ, we first assess code metrics' capabilities in finding bugs and then evaluate them further in terms of their potential in estimating the bugs' severity. To find the answer to this RQ, we apply a statistical test to see if distributions of methods (e.g., buggy vs non-buggy or critical vs low) are statistically different according to our selected code metrics.

After applying the Shapiro-Wilk test \cite{shaphiro1965analysis} using 5\% level of significance ($\alpha$ = 0.05), we found out that none of our distributions are normal, so we use the non-parametric Wilcoxon Rank-sum test for answering RQ1 with the 5\% level of significance ($\alpha$ = 0.05) with the following two null-hypotheses for RQ1-1 and RQ1-2:

\begin{itemize}
  \item RQ1-1: Source code distributions of buggy methods and non-buggy methods are not statistically different.
  \item RQ1-2: Source code distributions of buggy methods with different severity values are not statistically different.
\end{itemize}

To find the significance of the difference, we calculated Cliff’s Delta effect size. We followed the values provided by Hess and Kromrey \cite{hess2004robust} for interpreting the result of this value. We considered values smaller than 0.147 to be negligible, values in the range [0.147, 0.33) to be small, values in the range [0.33, 0.474) to be medium, and values greater than 0.474 to be large.

Since we are using the statistical analysis test for this RQ, it is important to make sure that each distribution has enough samples, and that distributions are diverse enough (having samples from different projects to reduce the bias problem) to make the results robust. Also, we discard the methods with $LC < 4$ in this RQ since they are mostly boilerplate code such as setters, getters, and constructors which are generated automatically.

We found that our studied datasets contain imbalanced distributions for RQ1-2 since some USLs contain only a few samples (e.g., the Critical USL in the Defects4J contains only 25 samples) or some USLs do not contain enough bugs from various projects to make that category diverse (e.g., 92\% of samples in the Medium USL of the Defect4J dataset are from \verb|Closure| project). Because of the mentioned problem, we merge some USLs (defined in Section~\ref{sec:Dataset}) one level further for RQ1-2 which is shown in Table \ref{tab:merging_rq1}.

\begin{table}[h!]
    \caption{
\label{tab:merging_rq1} Merging different USLs to create categories for Defect4J and Bugs.jar datasets. For example, the first row means that bugs having Critical or High USLs are considered in the High category}.
    \begin{center}
      \begin{tabular}{c|c|c}
        \toprule
        \textbf{Dataset} &
        \textbf{Unified Severity Label (USL)} & \textbf{Merged Category}\\
        \midrule
        \multirow{2}{*}{D4J} & Critical, High & High \\
        & Low, Medium & Low\\
        \midrule
        \multirow{3}{*}{Bugs.jar} & Blocker, Critical & Critical\\
         & Major & Major \\
         & Minor, Trivial & Minor\\
        \bottomrule
      \end{tabular}
    \end{center}
  \end{table}

As shown in Table \ref{tab:merging_rq1}, for the Defects4J dataset, the Critical USL is merged into the High, and also the Medium and Low USLs are merged, so we ended up with two categories of High and Low severity bugs. 

In the Bugs.jar dataset, the Blocker and Critical USLs are merged, and also the Trivial and Minor USLs are merged, so we concluded with three severity categories: Critical, Major, and Minor. The Blocker and Critical USLs are merged since the Critical USL does not contain enough samples, and the Minor and Trivial USLs are merged because of not enough samples in Trivial. We keep the Major USL as an independent category since it has enough samples.

\label{sec:Answer to RQ1}
\subsubsection{Results}

\begin{figure*}
\centering
\begin{subfigure}[b]{1.0\linewidth}
   \includegraphics[width=1\linewidth]{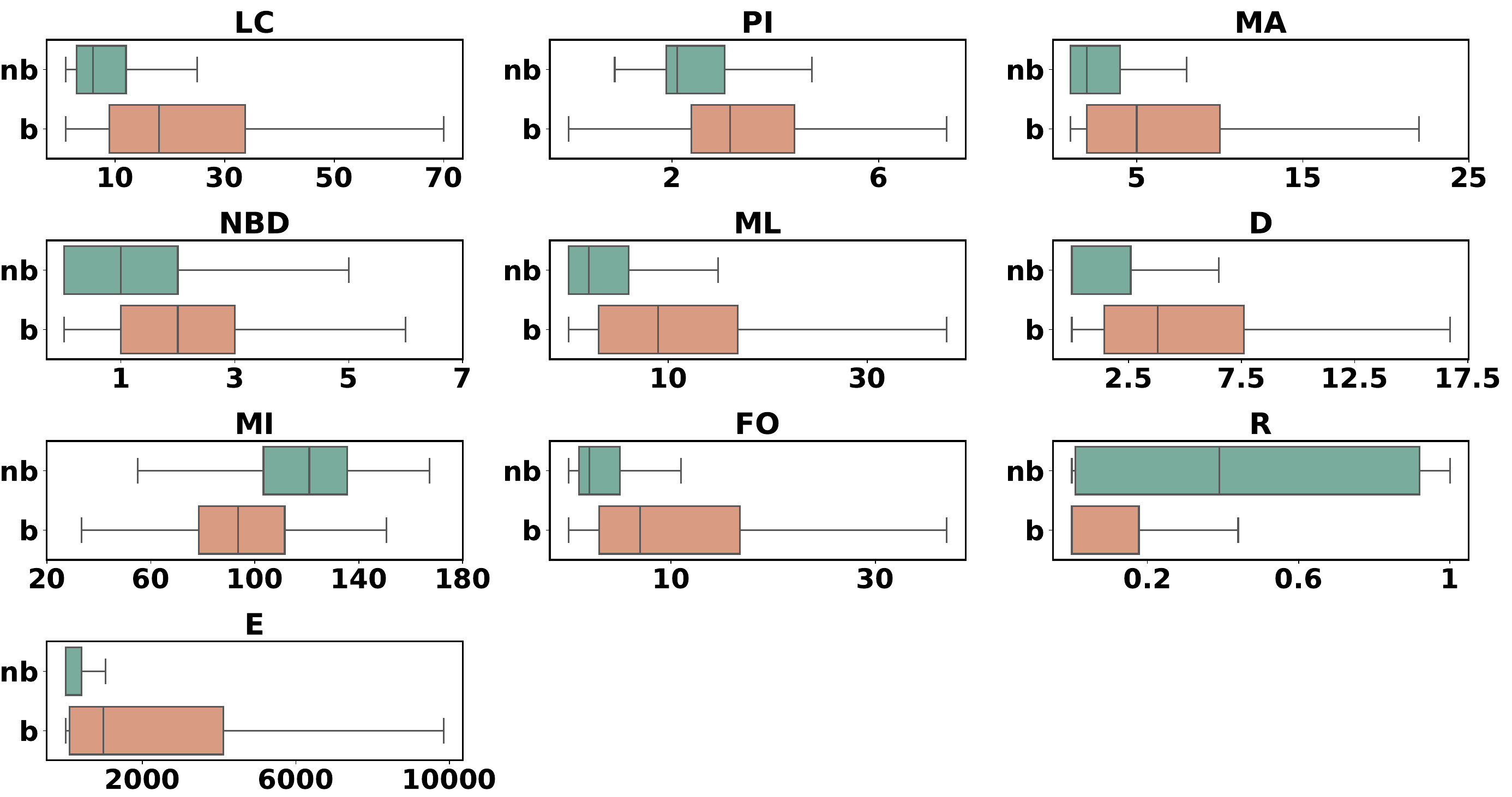}
   \caption{Defects4J}
   \label{fig:jxpath_14_a} 
\end{subfigure}

\begin{subfigure}[b]{1.0\linewidth}
   \includegraphics[width=1\linewidth]{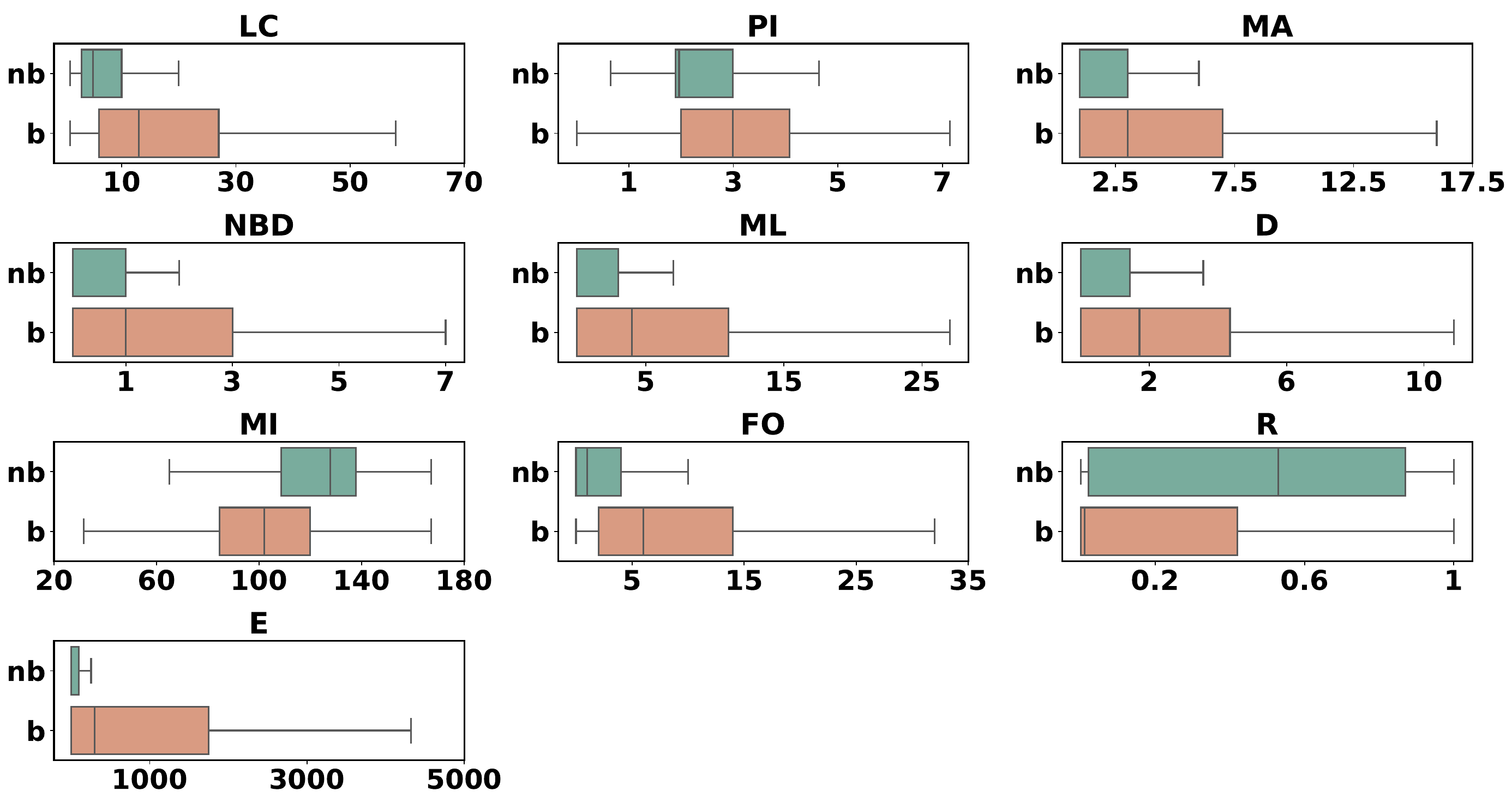}
   \caption{Bugs.jar}
   \label{fig:jxpath_14_b}
\end{subfigure}

\caption[]{Comparing source code metrics between buggy methods (b axis) and non-buggy methods (nb axis) using aggregated dataset.}
\label{fig:buggy_nonbuggy_metric_boxplot}
\end{figure*}

\textbf{RQ1-1 Answer}: Figure~\ref{fig:buggy_nonbuggy_metric_boxplot} compares the distribution of different code metrics between buggy and non-buggy methods (Figure~\ref{fig:jxpath_14_a} for the Defect4J, and Figure~\ref{fig:jxpath_14_b} for the Bugs.jar datasets). These are aggregated results, as we have combined data for all the projects in a specific dataset.
Results suggest that the median values of all code metrics, except for the Readability and Maintainable Index, in the buggy methods are larger than the median or Q3 (75$^{th}$ percentile) values in the non-buggy methods. Not surprisingly, for the Maintainability Index and Readability, the median of buggy methods is smaller than the median or Q3 (75$^{th}$ percentile) values in the non-buggy methods: less readable and maintainable code is more bug-prone.
Evidently, the buggy methods are generally larger (LC), more complex (e.g., McCabe, McClure), and less readable than the non-buggy methods. These figures imply that buggy and non-buggy distributions are different regarding all code metrics.

After applying the Wilcoxon Rank Sum Test to find if there is a statistical difference between these two distributions, we found that all of our comparisons for code metrics between buggy and non-buggy methods are statistically significant ($P \le 0.5$).  Therefore, the null hypothesis is rejected which means that these metrics are able to distinguish between buggy code and non-buggy code. We provided the Cliff's delta size values in Table \ref{tab:cliff_buggy_nonbuggy} which shows that effect size values are small or medium in all cases. In particular, LC, FanOut, Maintainable Index and, Halstead Effort exhibit better performance than other metrics, because for them the effect size is medium in both datasets.

\begin{table*}[tbp]
\centering
\caption{
\label{tab:cliff_buggy_nonbuggy} Cliff’s Delta Effect sizes for comparing code metrics between buggy and non-buggy methods. S refers to Small and M refers to Medium effect size. 
}
    \begin{center}
      \begin{tabular}{c|c|c|c|c|c|c|c|c|c|c}
        \toprule
        \textbf{Dataset} & \textbf{LC} & \textbf{PI}&\textbf{MA} & \textbf{NBD} & \textbf{ML} & \textbf{D} & \textbf{MI} & \textbf{FO} & \textbf{R} & \textbf{E}
        \\
        \midrule
        Defects4J & M & S & M & S& M & S & M & M & S & M \\              
        Bugs.jar & M & S & S & S & S & S & M & M & S & M               
        \\
        \bottomrule
      \end{tabular}
    \end{center}
  \end{table*}

With aggregated analysis, however, we can not observe if our results are true for all of the projects. The aggregated data can be highly influenced by a few large projects. Also, different external factors, such as code review policy~\cite{WANG:2019}, developers' commit patterns~\cite{Herzig:2013}, and expertise~\cite{Matter:2009}, can impact the distribution of code metrics and bug-proneness of a particular project.
Therefore, we now reproduce the results for each project individually.  

\begin{table*}[tbp]
\centering
\caption{
\label{tab:wicoxon_buggy_nonbuggy_per_project} Wilcoxon Rank-sum test results to show the percentage of projects where the distribution of different code metrics are statistically different between buggy and non-buggy methods.
}
    \begin{center}
      \begin{tabular}{c|c|c|c|c|c|c|c|c|c|c}
        \toprule
        \textbf{Dataset} & \textbf{LC} & \textbf{PI}&\textbf{MA} & \textbf{NBD} & \textbf{ML} & \textbf{D} & \textbf{MI} & \textbf{FO} & \textbf{R} & \textbf{E}
        \\
        \midrule
        Defects4J & 100\% & 50\% & 70\% & 80\% & 80\% & 70\% & 100\% & 100\% & 90\% & 80\% \\
        Bugs.jar & 100\% & 87.5\% & 100\% & 100\% & 100\% & 100\% & 100\% & 100\% & 100\% & 100\% \\
        \bottomrule
      \end{tabular}
    \end{center}
\end{table*}

\begin{table}[htbp]
\centering
\caption{
\label{tab:cliff_buggy_nonbuggy_per_project} Cliff’s Delta Effect sizes after comparing buggy vs non-buggy methods. N refers to Negligible, S refers to Small, M refers to Medium, and L refers to Large effect size. For example, the first row compares the LC distributions between buggy and non-buggy methods and shows the percent of differences with negligible, small, medium, and large effect sizes. Noticeably, the values for negligible (N) are zero for both datasets, implying that the difference in LC distributions between buggy and non-buggy methods is never negligible. This observation is also true for MI, FO, and R. 
}
    
\begin{center}
\begin{tabular}{c|cccc|cccc}
        \toprule
          & \multicolumn{4}{c}{\textbf{Defects4J}} & \multicolumn{4}{c}{\textbf{Bugs.jar}} \\
        \multirow{1}{*}{\textbf{Metrics}} & N & S & M & L  & N& S & M  & L\\
          \midrule
          \textbf{LC} & 0 & 40 & 30 & 30 & 0 & 25 & 62.5 & 12.5\\
          \textbf{PI} & 30 & 50 & 10 & 10 & 37.5 & 62.5 & 0 & 0\\
          \textbf{MA} & 20 & 40 & 20 & 20 & 12.5 & 37.5 & 50 & 0\\
          \textbf{NBD} & 20 & 50 & 10 & 20 & 12.5 & 75 & 12.5 & 0\\
          \textbf{ML} & 10 & 30 & 40 & 20 & 0 & 62.5 & 37.5 & 0\\
          \textbf{D} & 20 & 40 & 30 & 10 & 12.5 & 37.5 & 50 & 0\\
          \textbf{MI} & 0 & 30 & 40 & 30 & 0 & 25 & 62.5 & 12.5\\
          \textbf{FO} & 0 & 40 & 40 & 20 & 0 & 37.5 & 50 & 12.5\\
          \textbf{R} & 0 & 50 & 40 & 10 & 0 & 87.5 & 12.5 & 0\\
          \textbf{E} & 10 & 30 & 30 & 30 & 0 & 37.5 & 62.5 & 0\\
          \bottomrule
          
\end{tabular}
\end{center}
\end{table}

We calculated the percent of projects where the distributions of a particular metric are statistically different between buggy and non-buggy methods. 
Table \ref{tab:wicoxon_buggy_nonbuggy_per_project} shows the results for all the code metrics. The distribution of LC, Maintainable Index, and FanOut are statistically different between buggy and non-buggy methods in all the studied projects from both datasets (100\%). In the Defect4J dataset, the proxy indentation (PI) metric has the weakest performance since in 50\% of the projects, the result is not significantly different. Most of the metrics, however, show good performance in distinguishing the buggy method from the non-buggy method. We also calculated Cliff's delta effect sizes of the differences, as presented
and described in Table \ref{tab:cliff_buggy_nonbuggy_per_project}. The results show that Lines of Code, Maintainable Index, Fan Out, and Readability metrics do not exhibit negligible effect size for both datasets, but Proxy Indentation has the largest percentage of negligible effect size in both datasets.\\

\noindent\fbox{%
    \parbox{\columnwidth}{%
    \textbf{Summary of RQ1.1 Results:}
Our selected code metrics show high performance in identifying the buggyness of methods in both datasets (with both aggregated and individual project analysis). Line of Code, Maintainable Index, Fan-out, readability, and Effort metrics have the best performance while Proxy Indentation and Nested Block Depth exhibit very poor performance.
    }%
}\\\\

\textbf{RQ1.2 Answer:} After observing the somewhat known power of code metrics in distinguishing buggy and non-buggy methods, in this sub-RQ, we show their power in estimating the bug's severity. Although we wanted to perform the analysis for each project separately, there is not enough data in each severity category, except for \verb|accumulo| and \verb|jackrabbit-oak| projects. The \verb|accumulo| project has 44, 100, and 87, and the \verb|jackrabbit-oak| project has 71, 439, and 80 samples in Critical, Major, and High categories (we described in Section \ref{sec:Dataset}) respectively. The number of samples for each category in other projects was much lower, so we did not report their statistical test results separately.

\begin{figure*}
\centering
\begin{subfigure}[htbp]{1.0\linewidth}
   \includegraphics[width=1\linewidth]{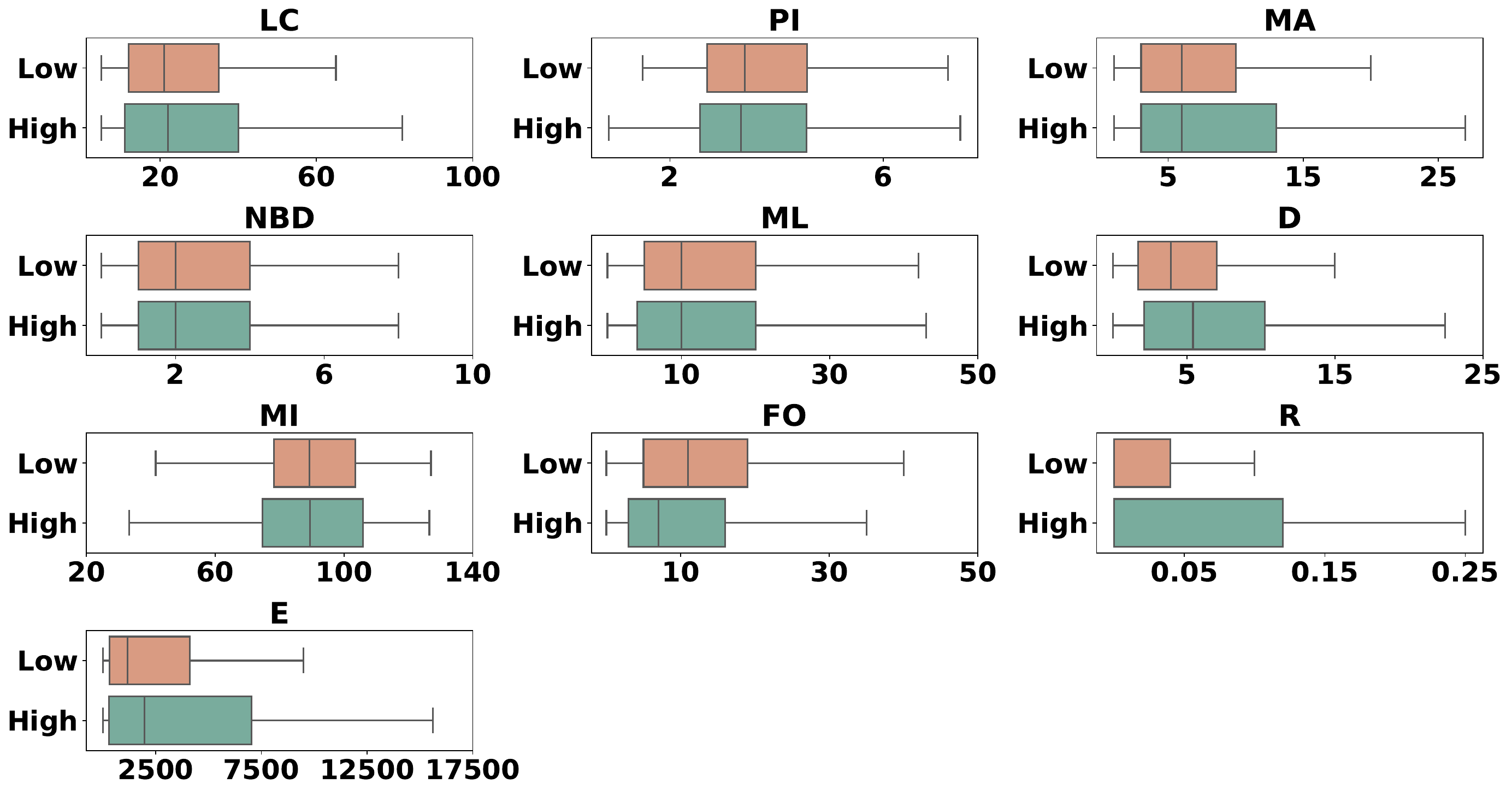}
   \caption{Defects4J}
   \label{fig:severity_d4j} 
\end{subfigure}

\begin{subfigure}[b]{1.0\linewidth}
   \includegraphics[width=1\linewidth]{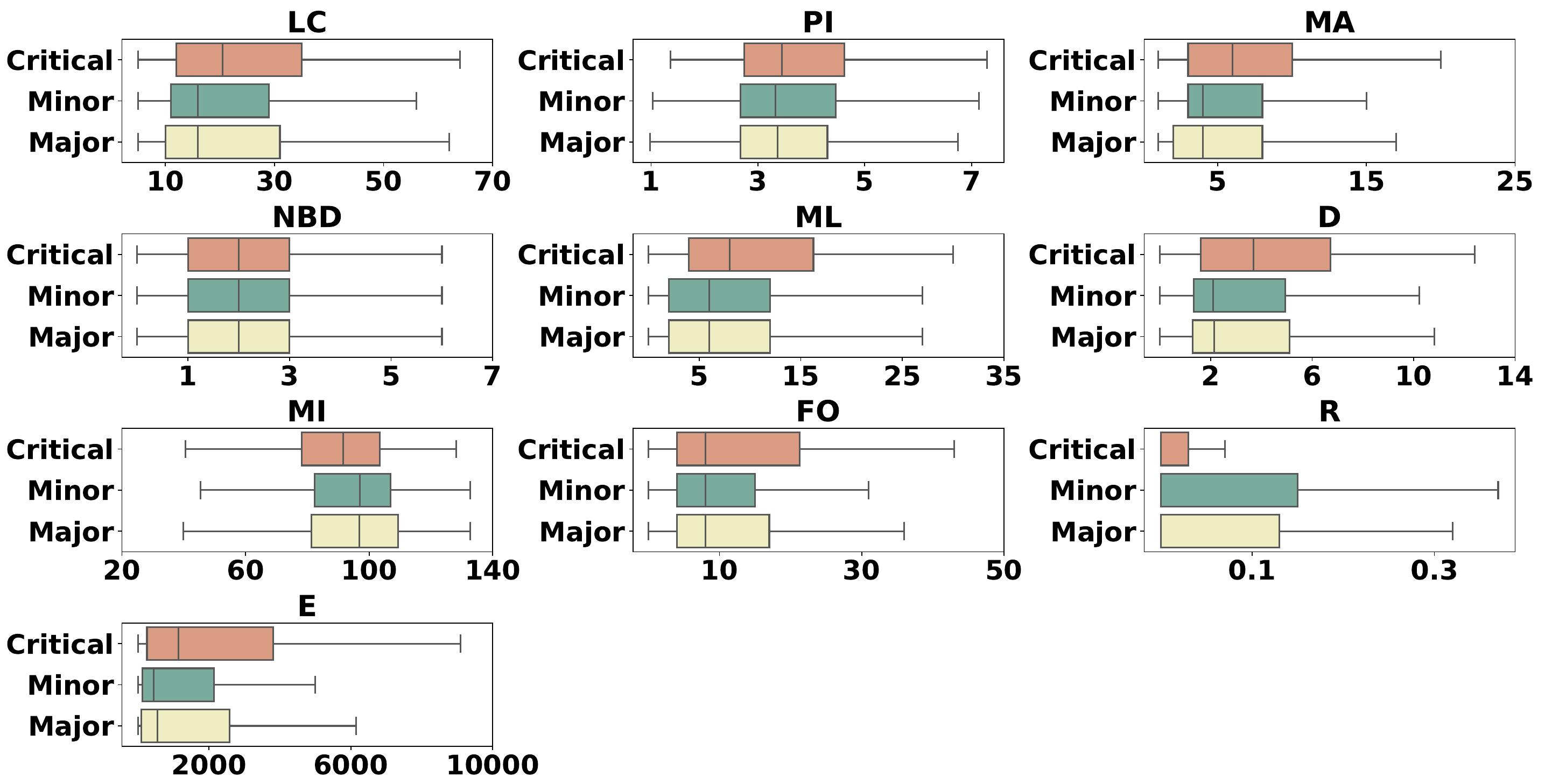}
   \caption{Bugs.jar}
   \label{fig:severity_busjar}
\end{subfigure}

\caption[]{Comparing source code metrics between different severity groups for Defects4J and Bugs.jar datasets.}
%\textcolor{red}{In figure b, the order should have been Critical, Major, and Minor. It's not serious at this stage, but little annoying}
\label{fig:severity_metric_boxplot}
\end{figure*}

Fig \ref{fig:severity_metric_boxplot} compares the
distributions of different code metrics after grouping the bugs into different bug severity categories. Clearly, the usefulness of code metrics in distinguishing bug severity is not convincing. To provide better insights, we also show the Wilcoxon Rank-Sum test and Cliff's Delta size values for both datasets in Table \ref{tab:p_val_severity_all}. In the table, the first two major rows (for Defect4J, and Bugs.jar) show results for the aggregated analysis, and the last two rows show results for the two individual projects: \verb|accumulo| and \verb|jackrabbit-oak|.

In contrast to the previous section, in many cases, the results are not significantly different between different bug severity categories. 
Considering the aggregated analysis in both datasets, only Halstead difficulty, and effort show desired behavior: methods with higher severity bugs have higher difficulty and effort distribution. The Maintainability index (MI) performs very poorly: only in two cases, it exhibits statistically different distributions, but with negligible effect size. Readability shows the desired behavior for the Bugs.jar dataset when the comparison was done between critical and major, and critical and minor. The statistically different distributions among these groups suggest that a method with lower readability has a higher chance of having more severe bugs. Although source lines of code (LC) were very helpful in differentiating between buggy and non-buggy methods (Figure~\ref{fig:jxpath_14_b},  Table~\ref{tab:wicoxon_buggy_nonbuggy_per_project} and Table~\ref{tab:cliff_buggy_nonbuggy_per_project}), its performance is significantly worse when it comes to distinguishing different levels of bug severity. This is interesting because numerous previous studies (e.g.,~\cite{Emam:2001,gil2017correlation,TSE:2013}) have shown that size (source lines of code) is the most important code metric, and none of the other code metrics provide any new information if their correlation with size is normalized. Our result challenges that claim because we find the superior performance of other code metrics (e.g., Halstead difficulty, and Halstead effort) than size when distinguishing methods with different bug severity.

In many cases, there is no statistical difference between these code metrics when compared against bug severity. It means that there are some cases where the code is complex which makes it error-prone, but severe bugs are not necessary in the complex code. We may intuitively say that there may be bugs in a simple method containing an SQL query for reading data from a database. In such a method, the code complexity metrics suggest a simple method, but if the developers do not handle input validation, the code will be vulnerable to SQL Injection. On the other hand, we may have a very complex and big method containing nested loops with different switch cases and exception-handling statements. The code complexity metrics would suggest it is a complex method, but this method may try to handle the UI/GUI part of the system only, which is not vulnerable to any critical or high-severity bugs. We will explore the real examples from our datasets regarding our intuition in RQ3.\\

%\hspace{1cm}

\noindent\fbox{%
    \parbox{\columnwidth}{%
    \textbf{Summary of RQ1.2 Results:}
None of the selected code metrics show promising results in distinguishing bug severity. While Difficulty and Effort metrics have the best performance in the aggregated datasets, the Maintainable Index and Proxy indentation metrics have the weakest performance. Although the Line of Code metric has an excellent performance in distinguishing buggyness of the methods, it exhibits extremely poor performance in finding bug severity, in both the aggregated datasets and the individual projects.
    }%
}

\begin{table*}[tbp]
\caption{
\label{tab:p_val_severity_all} Cliff's Delta Size for bugs with different severity values. A red cell indicates that the null hypothesis is rejected (i.e., the distributions are statistically significantly different according to the Wilcoxon test), whereas a blue cell means they are not statistically different. Effect sizes are N∼Negligible, and S∼Small. For example, the first cell shows that in the Defect4J dataset, the distribution of source lines of code (LC) between methods with high and low severe bugs is not statistically different (blue) and the difference has a negligible effect size.}
\begin{center}
      \begin{tabular}{c|c|c|c|c|c|c|c|c|c|c|c}
        \toprule
        \textbf{Project} & \textbf{Distributions} & \textbf{LC} & \textbf{PI}&\textbf{MA} & \textbf{NBD} & \textbf{ML} & \textbf{D} & \textbf{MI} & \textbf{FO} & \textbf{R} & \textbf{E}
        \\
        \midrule
       Defects4J (all) 
       & High-Low & \cellcolor[HTML]{68CBD0}N& \cellcolor[HTML]{68CBD0}N& \cellcolor[HTML]{68CBD0}N& \cellcolor[HTML]{68CBD0}N& \cellcolor[HTML]{68CBD0}N& \cellcolor[HTML]{E67E7B}S& \cellcolor[HTML]{68CBD0}N& \cellcolor[HTML]{E67E7B}S& \cellcolor[HTML]{00D2CB}N& \cellcolor[HTML]{E67E7B}N\\
       
       \midrule
       
       & Critical-Major  & \cellcolor[HTML]{E67E7B}N & \cellcolor[HTML]{68CBD0}N & \cellcolor[HTML]{E67E7B}S & \cellcolor[HTML]{E67E7B}N & \cellcolor[HTML]{E67E7B}S & \cellcolor[HTML]{E67E7B}S & \cellcolor[HTML]{E67E7B}N & \cellcolor[HTML]{68CBD0}N & \cellcolor[HTML]{E67E7B}N & \cellcolor[HTML]{E67E7B}S \\
       
       & Critical-Minor  & \cellcolor[HTML]{E67E7B}{\color[HTML]{000000}N} & \cellcolor[HTML]{68CBD0}N & \cellcolor[HTML]{E67E7B}N & \cellcolor[HTML]{68CBD0}N & \cellcolor[HTML]{E67E7B}N & \cellcolor[HTML]{E67E7B}S & \cellcolor[HTML]{68CBD0}N & \cellcolor[HTML]{E67E7B}N & \cellcolor[HTML]{E67E7B}N & \cellcolor[HTML]{E67E7B}S \\
       
       \multirow{-3}{*}{Bugs.jar (all)} & Major-Minor& \cellcolor[HTML]{68CBD0}N& \cellcolor[HTML]{68CBD0}N & \cellcolor[HTML]{68CBD0}N& \cellcolor[HTML]{68CBD0}N & \cellcolor[HTML]{68CBD0}N& \cellcolor[HTML]{68CBD0}N & \cellcolor[HTML]{68CBD0}N& \cellcolor[HTML]{68CBD0}N & \cellcolor[HTML]{68CBD0}N& \cellcolor[HTML]{68CBD0}N \\ 
       
       \midrule
        & Critical-Major  & \cellcolor[HTML]{68CBD0}N  & \cellcolor[HTML]{68CBD0}N& \cellcolor[HTML]{E67E7B}S& \cellcolor[HTML]{E67E7B}S& \cellcolor[HTML]{68CBD0}S& \cellcolor[HTML]{68CBD0}N& \cellcolor[HTML]{68CBD0}N& \cellcolor[HTML]{68CBD0}N& \cellcolor[HTML]{68CBD0}N& \cellcolor[HTML]{68CBD0}N \\
       
       & Critical-Minor  
       & \cellcolor[HTML]{68CBD0}{\color[HTML]{000000} S} & \cellcolor[HTML]{E67E7B}S & \cellcolor[HTML]{E67E7B}S & \cellcolor[HTML]{E67E7B}S & \cellcolor[HTML]{E67E7B}S & \cellcolor[HTML]{E67E7B}S & \cellcolor[HTML]{68CBD0}S & \cellcolor[HTML]{E67E7B}S & \cellcolor[HTML]{68CBD0}N & \cellcolor[HTML]{E67E7B}S    \\
       
       \multirow{-3}{*}{accumulo} & Major-Minor     & \cellcolor[HTML]{68CBD0}N & \cellcolor[HTML]{E67E7B}S & \cellcolor[HTML]{68CBD0}N & \cellcolor[HTML]{68CBD0}N & \cellcolor[HTML]{68CBD0}N & \cellcolor[HTML]{68CBD0}N & \cellcolor[HTML]{68CBD0}N & \cellcolor[HTML]{E67E7B}S & \cellcolor[HTML]{68CBD0}N & \cellcolor[HTML]{68CBD0}N \\ 

    %   ----------------------
        \midrule
        & Critical-Major  & \cellcolor[HTML]{68CBD0}N  & \cellcolor[HTML]{68CBD0}N & \cellcolor[HTML]{68CBD0}N & \cellcolor[HTML]{68CBD0}N & \cellcolor[HTML]{E67E7B}N & \cellcolor[HTML]{68CBD0}N & \cellcolor[HTML]{68CBD0}N & \cellcolor[HTML]{E67E7B}S & \cellcolor[HTML]{68CBD0}N & \cellcolor[HTML]{68CBD0}N \\
           
           & Critical-Minor  
           & \cellcolor[HTML]{68CBD0}{\color[HTML]{000000} N} & \cellcolor[HTML]{E67E7B}S & \cellcolor[HTML]{68CBD0}N & \cellcolor[HTML]{E67E7B}S & \cellcolor[HTML]{68CBD0}S & \cellcolor[HTML]{E67E7B}S & \cellcolor[HTML]{68CBD0}N & \cellcolor[HTML]{68CBD0}N & \cellcolor[HTML]{E67E7B}S & \cellcolor[HTML]{68CBD0}N \\
           
           \multirow{-3}{*}{jackrabbit-oak} & Major-Minor     & \cellcolor[HTML]{E67E7B}S & \cellcolor[HTML]{E67E7B}S & \cellcolor[HTML]{E67E7B}S & \cellcolor[HTML]{E67E7B}S & \cellcolor[HTML]{E67E7B}S & \cellcolor[HTML]{E67E7B}S & \cellcolor[HTML]{E67E7B}N & \cellcolor[HTML]{E67E7B}S & \cellcolor[HTML]{E67E7B}N & \cellcolor[HTML]{E67E7B}S     \\ 
      \bottomrule
      \end{tabular}
    \end{center}
\end{table*}

\subsection{Static Analysis Tools \& Bug Severity (RQ2)}
The second goal of our study is to find out if static analysis tools can detect the bugs and their severity labels.

\subsubsection{Motivation}
Since source code metrics do not show promising results in predicting the bug severity, we explore other approaches. To do so, we study the static analysis tools that are widely used in practice for different purposes (e.g., finding programming errors, coding standard violations, syntax violations, and security vulnerabilities) by technical teams. This approach is favorable because of its higher speed than the human code review and its capability to work offline with low required resources. Since these tools are already integrated into the deployment process of many companies, if we could use them for predicting the severity we can leverage them for prioritizing the bugs without requiring any other tools/methods. Therefore, in this RQ, first, we need to find out if these tools can detect real-world complex bugs, and then we study their performance in predicting the detected bug's severity.

\subsubsection{Approach}
\label{static-tools}
We answer one RQ with its two sub-RQs to find the ability of the static analysis tools to detect bugs and their severity.\\\\
\textbf{RQ2}: What is the capability of static analysis tools in finding bugs and their severity?
\begin{itemize}
\itemsep-.2em
    \item \textbf{RQ2-1}: How effective static analysis tools are in detecting buggy methods?
    \item \textbf{RQ2-2}: Can static analysis tools estimate detected bugs' severity?
\end{itemize}

\textbf{Design of RQ2:} To answer this RQ, we selected two static analysis tools named SpotBugs, \cite{spotbugs} and Infer \cite{infer}. We selected SpotBugs because it not only can detect bugs but also report a Rank value which indicates the severity. We used the popular Infer tool only to detect buggy methods regardless of their severity values. Since there is no need for a statistical test in this RQ, there is no restriction on the number of samples in each group. Therefore, we use the USL values (mentioned in Section \ref{sec:Dataset}) directly without applying any merging. 

SpotBugs has different configurations that affect its performance in finding bugs. It has some specific detector modules that focus on specific bug types, but we used the standard detectors, which is also the default option. Also, this tool has a configuration option named \verb|effort| which adjusts the taken effort in finding bugs. We used the \verb|effort=max|, which is the highest level of effort, so in this way, we have provided as much as the computation cost this tool requires to work on our datasets.

There are different ways to provide projects' source code to these tools, such as providing only the buggy class (es), the package containing the buggy class (es), or the whole project. By providing only the buggy class, the tool may not be able to find the bugs that are across different classes since it only analyses that specific class, but when we provide the whole package or the whole project, it analyses any file in these modules to find the bugs accurately which also needs more resources and time. We provided the whole project to be analyzed with tools to make the result more robust. However, this takes 3 and 52 hours for the Defect4J and Bugs.jar dataset, respectively, on a regular computer with 16GB RAM and 16 CPU cores.

Although we successfully applied these tools on all of the Defects4J projects, for several projects of the Bugs.jar dataset we faced build errors due to dependency issues. We were successful in building \verb|flink|, \verb|commons-math|, and \verb|accumulo| projects which consist of 248 buggy methods.

SpotBugs reports the \verb|Confidence| value for each bug instance, which indicates the level of confidence the tool has in reporting this warning. This attribute may have different values such as 1 to match high-confidence warnings, 2 to match normal-confidence warnings, or 3 to match low-confidence warnings. We considered all of the reported warnings regardless of their confidence values.

Also, this tool reports the \verb|Rank| property which indicates the bug rank.This tool categorizes bugs into 4 different levels; scariest, scary, troubling, and concern. The reported Rank property represents an integer value ranging from 1 to 20. Within this range, 1 to 4 corresponds to the scariest, 5 to 9 to scary, 10 to 14 to troubling, and 15 to 20 to bugs of concern~\cite{spotbugs}. The lower the number, the more important it is. This property has a similar semantic to the bug severity value, so we used this value to see the tool's capability in finding the bugs' severity.

\begin{table}[h!]
    \caption{Translation between the SpotBugs reported rank (SRR) value and the unified severity label (USL) value.
\label{tab:spotbugs_severity} }
    \begin{center}
      \begin{tabular}{c|c}
        \toprule
        \textbf{USL} & \textbf{SRR}\\
        \midrule
        Critical, Blocker & 1 to 4\\
        Major, High & 5 to 9\\
        Medium & 10 to 14\\
        Low, Trivial, Minor & 15 to 20\\
        \bottomrule
      \end{tabular}
    \end{center}
  \end{table}

The translation between the range of this value and our USLs is provided in Table \ref{tab:spotbugs_severity}. As the table shows, we consider 1 to 4 as Critical and Blocker USL, 5 to 9 as Major and High USL, 10 to 14 as Medium USL, and 15 to 20 as Minor, Low, and Trivial USL. For example, when the SpotBugs reports any number in the range of [1,4], and the actual severity value is either Critical or Blocker we say that it detects the severity correctly.

The Infer tool reports the \verb|IssueType| attribute with these categories: 1) Error 2) Warning 3) Info, Advice, Like. These values can be interpreted as the bug importance, but we found that all of our bugs are mapped to the Error category, based on their severity values. Thus it is not possible to leverage the \verb|IssueType| property to predict the bug severity, and we cannot use this tool to predict the severity. Therefore, we leverage this tool only to see if it can detect buggy methods, regardless of their severity values.

Both of these tools report the roots of detected bugs according to their defined patterns, by locating the start and end lines of the bug. Since we are considering bugs at the method-level granularity, we consider a method to be buggy whenever these tools report a bug in that method regardless of the reported start and end lines.

\subsubsection{Results}

\label{sec:Answer to RQ2}

\textbf{RQ2-1 Answer:} After applying SpotBugs and Infer tools we count the number of buggy methods that are reported as buggy (TP), the number of non-buggy methods which are not reported as buggy (TN), the number of buggy methods that are not detected by tools (FN), and the number of non-buggy methods which are reported as buggy methods (FP). If any of these tools report more than one bug for each method we only count that method once to prevent the duplication problem.

\begin{table}[h]
\caption{
\label{tab:static_tools_buggy} Static analysis tools performance in finding buggy methods. The Infer tool requires the project under test to be compiled, which did not work for the Bugs.jar dataset due to different build systems and heavy dependencies.}

      \begin{center}
      \begin{tabular}{c|c|c|c|c|c}
        \toprule
        \textbf{Dataset} & \textbf{Tool} & \textbf{ACC} & \textbf{Precision} & \textbf{Recall} &\textbf{F1}
        \\
        \midrule
        \multirow{2}{*}{D4J}  & SpotBugs & 93\% & 7.1\% & 7.2\% & 7.1\%\\        
         & Infer & 95\% & 9\% & 1.9\% & 3.1\%\\
        \midrule
        Bugs.jar & SpotBugs & 95\% & 7.6\% & 4.3\% & 5.4\% \\
        \bottomrule
      \end{tabular}
    \end{center}
  \end{table}

We calculated the Accuracy (ACC), Precision, Recall, and F1 values which are shown in Table \ref{tab:static_tools_buggy}. These values show that these tools have poor performance in detecting buggy methods of both datasets. While the accuracy value is so high, other metric values are relatively low which is because of the imbalanced dataset we have and accuracy is not the best metric for measuring performance for imbalanced datasets. The precision value range is 7\%-9\% which means that they may report many non-buggy methods as buggy methods, so the technical team may spend extra time on finding bugs in these methods, and this increases the cost of the maintenance phase without any useful result. However, the recall value (predicting buggy methods as non-buggy methods) is a more important metric in our case which has a smaller range of 2\%-7\%. This small value indicates that these tools miss many bugs which may lead to harmful consequences. The high accuracy with low precision and recall means that tools are making many Type I errors (false positives) and Type II errors (false negatives). The calculated F1 score shows that SpotBugs has a better performance in finding bugs in the Defect4J dataset than the Bugs.jar dataset, and the Infer tool has a worse performance than SpotBugs for the Defects4J dataset.

Our results are consistent with the previous study \cite{habib2018many} in 2018 where the authors studied the capability of SpotBugs, Infer, and ErrorProne tools in finding buggy methods of an older version of the Defects4J dataset. They concluded that 95.5\% of buggy methods are not detected by these tools. Surprisingly, although these tools have been updated several times since four years ago, their performance in finding bugs is not improved much.

Since these tools match the provided code with their predefined generic bug patterns, we may intuitively say that in cases where the bugs are complex or related to the software specification, they are missed by these tools. We will perform a qualitative study of some randomly sampled bugs to find the reasons behind this poor performance, in RQ3.\\

\noindent\fbox{%
    \parbox{\columnwidth}{%
    \textbf{Summary of RQ2.1 Results:}
Both SpotBugs and Infer tools have significantly low performance in finding bugs even with many existing developed patterns. The precision range for both datasets is 7\%-9\% and the Recall range is 2\%-7\%, which implies the extremely poor performance of these tools.
    }%
}\\\\

\textbf{RQ2.2 Answer:} Since only the SpotBugs tool can detect the severity of bugs, in this section we provide results only for this tool on both Defects4J and Bugs.jar datasets.

\begin{figure}
\centering
\begin{subfigure}[b]{0.85\columnwidth}
   \includegraphics[width=1\linewidth]{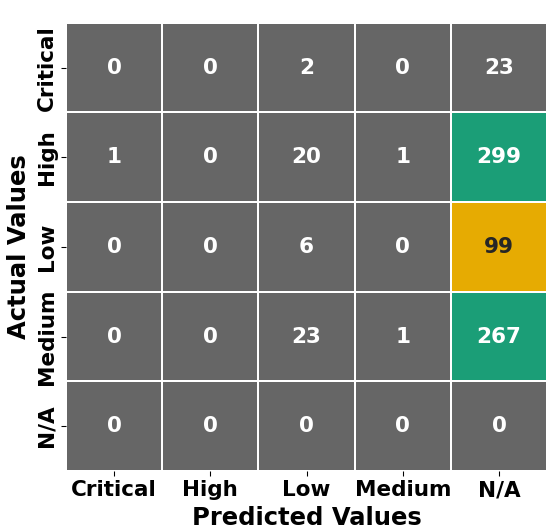}
   \caption{Defects4J}
\end{subfigure}

\begin{subfigure}[b]{0.85\columnwidth}
   \includegraphics[width=1\linewidth]{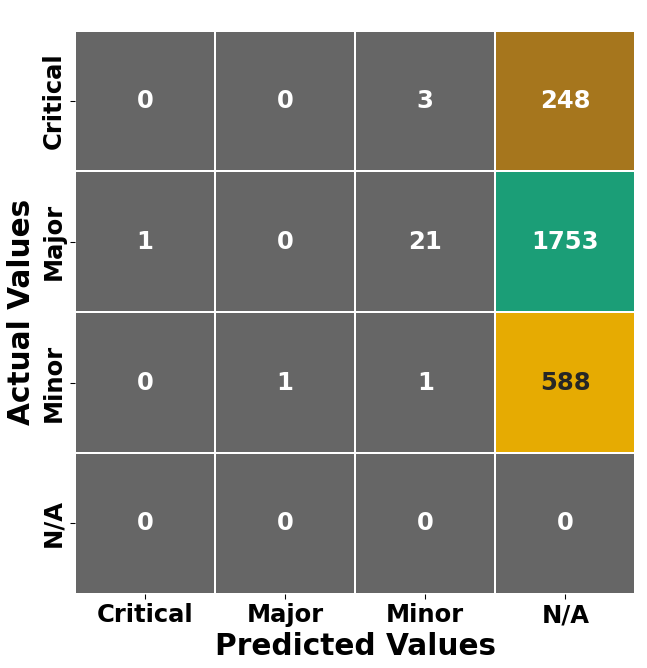}
   \caption{Bugs.jar}
\end{subfigure}

\caption[]{Confusion matrices of bug severity prediction by SpotBugs for Defects4J and Bugs.jar datasets. In this figure, the 'N/A' column shows the number of bugs that were not detected by the tool. Therefore, in the first row of (a) Defects4J, there are 23 critical bugs in this dataset that were not detected.}
\label{fig:severity_confusion_matrix}
\end{figure}

We provided the confusion matrices for the SpotBugs tool in Fig \ref{fig:severity_confusion_matrix}. These figures show that this tool labels most of the bugs with wrong severity labels. For example, it detects two critical bugs as low-severity bugs and 20 high-severity bugs as low-severity bugs, in the Defects4J dataset. These figures show that SpotBugs did not detect the severity of any Critical bugs in the Defect4J and Bugs.jar datasets.

Table \ref{tab:static_tools_severity} shows different descriptive statistics of this tool's performance. Since we have multi-class labels, we used \verb|macro average| \cite{macroaverage} for calculating these values. In this method, the final result is computed by taking the arithmetic mean of all the per-class descriptive scores.

In this table we observe small accuracy values for both datasets which show their weakness in finding the correct bug severity. Although the accuracy metric is not a good metric for imbalanced datasets, we report this value only for the sake of completeness and we don't use this metric for any comparison or conclusion. For the mentioned reason we report other metrics such as precision, recall, and F1 which are suitable for imbalanced datasets. The F1 score has been used in this paper to value the performance of these tools and accuracy is reported for completeness. The F1 score shows that although SpotBugs exhibits better performance for the Defect4J dataset, its performance is still very poor. We studied bug characteristics manually in RQ3 to find the reasons for this weak performance.

\begin{table}[h!]
\caption{
\label{tab:static_tools_severity} SpotBugs performance in finding bug severity of Defects4J and Bugs.jar datasets.
}
      \begin{center}
      \begin{tabular}{c|c|c|c|c}
        \toprule
        \textbf{Dataset} &  \textbf{Accuracy} & \textbf{Precision} & \textbf{Recall} &
        \textbf{F1}
        \\
        \midrule
        D4J & 0.94\% & 12.35\% & 1.25\% & 1.67\%\\
        Bugs.jar & 0.03\% & 1.00\% & 0.04\% &0.08\% \\
        \bottomrule
      \end{tabular}
    \end{center}
  \end{table}

\noindent\fbox{%
    \parbox{\columnwidth}{%
    \textbf{Summary of RQ2.2 Results:}
SpotBugs has significantly better performance in finding the bug severity of the Defects4J dataset than the Bugs.jar dataset. However, this tool wrongly assigns lower severity values to bugs in many cases. The overall performance shows that there is still much work to be done for this tool to accurately estimate bug severity.    }%
}\\

\subsection{Manual Analysis of Bug Characteristics (RQ3)}
Our final goal, in RQ3, is to dig more into the results of RQ1 and RQ2 and study the bug characteristics in different important scenarios such as (a) when there is a contradiction between code metrics and severity and (b) when the static analysis tools miss/mislabel severe bugs.%, and (c) when code metrics and static analysis tools can be complementary.

\subsubsection{Motivation}
To understand why source code metrics cannot indicate the bug severity, we need to study the cases where there is a contradiction between the bug severity and source code metrics. Also, it seems important to find out if there is any relationship between the bug's type and its severity. The answer to this question can help researchers to consider bug types as important features when providing new approaches for bug prediction to have better accuracy, and also developers may pay more attention to specific bug types during the manual severity assignment process.

To find out why static analysis tools miss many severe bugs or predict lower severity than the actual severity, in most cases, we need to study if this problem exists because of the nature of static analysis tools or if there is potential for improvement. The answer to this question will help researchers to improve the developed approaches to enhance the bug severity performance and also developers can get insights into how confidently they should use these tools in their projects.

%Finally, since real bugs are very complex due to the complexity of the current software and various new technologies, it would be useful to consider multiple approaches instead of relying on a single approach to predict the severity. Therefore, we also study if code metrics and static analysis tools are complementary in terms of bug severity prediction. This may lead to a technique that combines these approaches to predict the severity more accurately.

\subsubsection{Approach}
We answer one RQ with two sub-RQs to study the bug characteristics. \\\\
\textbf{RQ3}: What are the characteristics of bugs with different severity values?

\begin{itemize}
\itemsep-.2em
    \item \textbf{RQ2-1}: How do code metrics perform in distinguishing bug severity for different types of bugs?
    \item \textbf{RQ2-2}: Why do static analysis tools miss or mislabel severe bugs?
\end{itemize}
%\textbf{RQ3-3}: Do the results of code metrics and static analysis tools have overlap?\\

\label{sec:RQ3-1}
\textbf{Design of RQ3-1:} As we show in the RQ1 results section, there is a correlation between bug existence and code complexity metrics, but there is no significant correlation between bug severity and code complexity metrics. Therefore, we manually investigated a random sample of bugs to find factors that lead to contradictions between bug severity values and code metrics. The goal of this sub-RQ is to study the characteristics of buggy methods with high severity (e.g., Blocker, Critical, Major) which have low code metrics measurement, and to study buggy methods with low severity (e.g., Low, Minor, Trivial) which have high code metrics measurement.

To facilitate the manual analysis, we split all bugs into two high-level categories of severity, which are: severe (including Blocker, Critical, Major, and High severity USLs from RQ1) and non-severe (including Medium, Low, Minor, and Trivial USLs from RQ1).

To find the two lists of the most and the least complex methods, we calculate a representative sum complexity value from different code metrics. Since our code metric values do not have the same range, we normalize them by using the Robust Scaler algorithm \cite{robustscaler}, which is robust against outliers. It follows a similar algorithm to the MinMax scale, but it uses the interquartile range rather than the min-max. The formula for Robust Scaler is shown in Eq\ref{Eq4}:

\begin{equation}\label{Eq4}
Robust Scaler=\frac{x_i - Q_1(x)}{Q_3(x)-Q_1(x)}
\end{equation}

Where $Q1$ is the 1st quartile, and $Q3$ is the third quartile.\\

To find the Sum Complexity (SC) value---which indicates the total complexity of the method---we take the sum of different code metrics (we multiply Readability and Maintainable Index by -1 to reverse their value). We provide the same weight to all the code metrics and the possible impact of this decision is discussed in the \hyperref[sec:Threats to Validity]{Threats to Validity} section. Table \ref{tab:sum_complexity} shows two sample bugs with the calculated code metrics, the normalized code metric, and also the Sum Complexity values.

\begin{table*}[tbp]
\centering
\caption{
\label{tab:sum_complexity} Two sample bugs from Severe and Non-Severe categories with their code metric values, normalized values, and the Sum Complexity values. The value in each row and the value of Sum Complexity are rounded to two precision.
}
    
\begin{center}
\begin{tabular}{c|cccc|cccc}
        \toprule
          & \multicolumn{2}{c}{\textbf{Severe}} & \multicolumn{2}{c}{\textbf{Non-Severe}} \\
        \multirow{1}{*}{\textbf{Metrics}} & Raw & Normalized & Raw& Normalized \\
          \midrule
          \textbf{LC} & 13.00 & -0.35 & 532.00 & 21.27 \\
          \textbf{PI} & 2.89 & -0.25 & 8.61 & 2.75 \\
          \textbf{MA} & 3.00 & -0.37 & 194.00 & 23.50 \\
          \textbf{NBD} & 1.00 & -0.33 & 12.00 & 3.33 \\
          \textbf{ML} & 2.00 & -0.53 & 219.00 & 13.93 \\
          \textbf{D} & 1.86 & -0.39 & 18.84 & 2.20 \\
          \textbf{FO} & 4.00 & -0.35 & 40.00 & 22.14 \\
          \textbf{E} & 220.14 & -0.27 & 214131.00 & 45.50 \\
          \textbf{MI} & -103.93 & -0.53 &23.86 &4.15\\
          \textbf{R} & -0.22 & -3.66 & 0 &0\\
          \midrule
          \textbf{Sum Complexity} & 143.74 & -7.07 & 215459.26 & 138.80 \\
          \bottomrule
          
\end{tabular}
\end{center}
\end{table*}

After calculating the SC value, to find the contradiction, we sorted our samples based on the SC (normalized version) value in ascending and descending orders separately and kept the first 80 samples in both cases for both datasets which resulted in 360 samples ($80 * 2  \textnormal{ (Severe | Non-Severe)} * 2 \textnormal{ (Defects4J | Bugs.jar)}$).

%\begin{displaymath}
%60 * 2  \textnormal{ (Severe | Non-Severe)} * 2 \textnormal{ (Defects4J | Bugs.jar)} = 240
%\end{displaymath}

Then, for the manual analysis, we randomly select 80 bugs out of these 360 bugs distributed evenly between Severe and Non-Severe bugs (our two categories in this RQ) and also between Defects4J and Bugs.jar (our datasets). Then, we provide the buggy code and the corresponding fixed code of these samples to the first and third authors.

We use a Card Sorting approach~\cite{miles1994qualitative} to label the bug type manually. In this approach, we asked the first and third authors to label each buggy method with one or more ``bug types'' of their choice, based on their own defined list of bug types (e.g., Network, UI/GUI, IndexOutofBound, Math calculation), by looking at the buggy code and the fixed code only. The total unique labels created by the two authors in this phase were nine low-level labels (mentioned in the \ref{sec:Answer to RQ3} section). 

Then, the first and third authors had a meeting to discuss their assigned labels and merged the nine lower-level 
 labels into the final set of four higher-level labels. 
During the merging process, while authors agreed on most of the samples' labels, several assigned labels had similar meanings but were expressed in different terms. We resolved those conflicts by assigning a more generic label that covers both original labels. For example, there is a sample where the first author assigned ``I/O bug'', as the label, but the third author assigned ``Network bug''. We merged these two labels in the I/O bug label which covers the Network bug too. 
% Also, there are a few cases where the first and third authors cannot make the decision because of the conflicts, so we resolved these conflicts by comparing both provided categories and the code.
Following this approach, we ended up with the following four labels: Integration, Edge/Boundary, Security, and Specification.\\

\textbf{Design of RQ3-2:} Based on RQ2 results, we found that the SpotBugs tool missed or mislabeled bug severity in many cases, which may lead to undesirable consequences. For example, mislabeling the high-severity bugs with low severity values will reduce the bug's importance. Thus the development team may postpone the fixing process. Therefore, in this sub-RQ, we want to find out the reasons behind why these tools are not able to find the bugs or their severity values.

To do so, we first manually analyze 30 randomly chosen severe bugs (Blocker, Critical, Major, and High severity categories) in both datasets that are missed by the static analysis tools. In this step, we mainly want to know if the lack of bug patterns in static analysis tools prevents them from detecting these bugs, or if these bugs' nature is such that static analysis tools cannot detect them regardless of what pattern/rule they implement. 

We then study a set of 30 randomly selected samples of severe bugs (Blocker, Critical, Major, and High severity categories) that are detected by SpotBugs tools, but the reported severity is lower than the actual severity values, e.g., when the actual severity value is Critical, but the tools report the Low severity value.\\

\subsubsection{Results}

\label{sec:Answer to RQ3}

\textbf{RQ3-1 Answer:} After applying the Card Sorting approach which is explained in Section \ref{sec:RQ3-1} we found nine low-level bug labels such as UI (User Interface) bugs, multi-threading bugs, and math calculation bugs. 

\begin{table}[h!]
\caption{
\label{tab:bug_type_low_categories} Distribution of the randomly sampled bugs in the nine low-level bug labels.
}   

    \begin{center}
      \begin{tabular}{c|c}
        \toprule
        \textbf{Low-level Label} & \textbf{Number}
        \\
        \midrule
 Wrong/missing objects and parameters & 29 \\
        Input/state checking  & 21 \\
        Completely wrong implementation & 8  \\
        Exception handling  & 7 \\
        Null checking  & 5  \\
        Math/String calculation & 4  \\
        I/O & 3  \\
        Multi-threading  & 2 \\
        UI  & 1  \\
        \bottomrule
      \end{tabular}
    \end{center}
    
  \end{table}

Table \ref{tab:bug_type_low_categories} shows bug types and their distributions. \textbf{Wrong / missing objects and parameters type} refers to code where the developer passes a wrong parameter to a method or uses the wrong object to call.  \textbf{Input/state checking} bugs happen when the code does not validate the function inputs or it does not consider the specific case of an object state for having different logic. \textbf{Completely wrong implementation} bug type shows cases where the fixed code is changed completely to implement new logic compared to the buggy code. \textbf{Exception handling} bugs occur when the method does not handle exceptions when a variable/object has a special value. \textbf{Null checking} bugs happen when the code accesses the null object or does not handle special specifications when the method argument is null. \textbf{Math/String calculation} bugs occur when the mathematical calculations or string manipulations are performed wrongly. 
\textbf{I/O} bugs occur in a method where the code accesses any I/O such as a file or network. \textbf{Multi-threading} bugs are due to not handling threading issues such as deadlock. Finally, \textbf{UI} bugs are related to improper implementation of the user interfaces including the command line interface and graphical.

\begin{table*}[tbp]
\caption{
\label{tab:bug_type_categories} Distribution of the four higher-level bug labels where there are contradictions between severity values and Sum Complexity (SC) values.
}
    \begin{center}
      \begin{tabular}{c|c|c}
        \toprule
        \textbf{High-level Label} & \textbf{Severe Bug \& Low SC} & \textbf{Non-severe Bug \& High SC}
        \\
        \midrule
        Specification & 26 & 28  \\
        Edge/Boundary & 4 & 9  \\
        Security & 7 & 3  \\
        Integration & 3 & 0  \\
        \bottomrule
      \end{tabular}
    \end{center}
  \end{table*}

Table \ref{tab:bug_type_categories} shows the distributions of the four higher-level bug labels by grouping the contradictions ---severe bugs but low Sum Complexity (SC), and non-severe bugs but high Sum Complexity (SC). The number of Security faults is higher when the bugs are labeled as severe, but the code metric values are low which indicates that security bugs may appear in simple methods but may lead to severe bugs. Furthermore, Edge/Boundary faults are mostly available in samples with high Sum Complexity (SC) values where the code is complex by having many different branches and statements where developers tend to miss edge/boundary cases, but these bugs are not severe necessarily.

In the following, we will provide definitions for each of these four bug-type categories and give some samples per category. Just a reminder that these categories have been defined based on our dataset and are the result of merging the labels from the Card Sorting phase.\\

\textbf{Specification faults:}
This category, which is our largest category, exhibits bugs where developers have not implemented the full specification (e.g., the logic related to a partition in the input space has been ignored, i.e., misrepresented). This bug type is called Specification faults since they may not occur if the developers follow the specification of the software completely \cite{hemmati2015effective}. Patches for these bugs contain changing conditional statements, math calculations, string manipulation, or a complete rewrite of the code.

For example, there is a bug with a Low severity value in \verb|Closure-119| project where the buggy method is too complex (containing 120 lines of code having many nested \verb|switch-case| and conditional statements), but the patch is very simple which handles a small part of input space. Part of this method and its patch are shown in Fig \ref{fig:closure_119}. Long methods such as our mentioned example are usually bug-prone, and they are detected as buggy by code metrics correctly, but the bug could have a Low severity value since these long and complicated methods may handle a non-sensitive task such as handling the GUI part.

\begin{figure}
\centering
% \fcolorbox{black}{white}{
% \includegraphics[width=0.97\linewidth]{figs/closure119.png}}
 \begin{minted}{diff}
    case Token.ASSIGN:
        if (parent.getFirstChild() == n) {
            isSet = true;
            type = getValueType(n.getNext());
        }
        break;
    case Token.GETPROP:
        return;
    case Token.FUNCTION:
        Node gramps = parent.getParent();
        if (gramps == null || 
            NodeUtil.isFunctionExpression(parent)) {
            return;
        }
        isSet = true;
        type = Name.Type.FUNCTION;
        break;
+   case Token.CATCH:
    case Token.INC:
    case Token.DEC:
        isSet = true;
        type = Name.Type.OTHER;
        break;
    default:
\end{minted}
\caption{Sample bug with the Low severity value in the Closure-119 project of the Defects4J dataset.}
\label{fig:closure_119} 
\end{figure}
Also, Fig \ref{fig:lang_30} shows part of a buggy method from the \verb|Lang30| project which has a Low severity value. In this sample, the patch is not as simple as in the previous example, and part of the code is rewritten, but the severity is still low.

\begin{figure}
\centering
% \fcolorbox{black}{white}{
% \includegraphics[width=0.97\linewidth]{figs/lang30_2.png}}
\begin{minted}{diff}
 char ch = cs.charAt(i);
 for (int j = 0; j < searchLength; j++) {
    if (searchChars[j] == ch) {
-        if (i < csLast && j < searchLast && 
-           ch >= Character.MIN_HIGH_SURROGATE && 
-           ch <= Character.MAX_HIGH_SURROGATE) {
- // missing low surrogate, fine, like String.indexOf(String)
-            if (searchChars[j + 1] == cs.charAt(i + 1)) {
+        if (Character.isHighSurrogate(ch)) {
+           if (j == searchLast) {
+ // missing low surrogate, fine, like String.indexOf(String)
+               return true;
+            }
+            if (i < csLast && 
+                searchChars[j + 1] == cs.charAt(i + 1)) {
                    return true;
             }
            } else {
                // ch is in the Basic Multilingual Plane
                return true;
                }
            }
        }
    }
    return false;
 }
\end{minted}
\caption[]{Sample bug with the Low severity value in the Lang-30 project of the Defects4J dataset.}
\label{fig:lang_30}
\end{figure}

The number of Specification faults in both severe and non-severe categories is almost equal as shown in Table \ref{tab:bug_type_categories}, so further information is strongly required to detect the bug severity in this case.\\

\textbf{Edge/Boundary cases:}
This kind of bug happens when the developers do not handle edge/boundary cases properly. Multiple samples in this category contain severe bugs, but with simple code according to our code metrics, containing only a few statements. For example, Fig \ref{fig:math_36} shows the buggy method in the \verb|Math-36| project, where the developer did not handle the edge case (\verb|Double.isNaN(result)|), but the patch shows that the studied method should have performed an extra calculation when the \verb|result| variable is \verb|NaN|. This bug may lead to a software crash, unauthorized access, or undesired behavior. 

\begin{figure}
\centering
% \fcolorbox{black}{white}{
% \includegraphics[width=0.97\linewidth]{figs/math_36_2.png}}
 \begin{minted}{diff}
 public double doubleValue() {
     double result = numerator.doubleValue() / 
     denominator.doubleValue();
+    if (Double.isNaN(result)) {
+    // Numerator and/or denominator must be out of range:
+    // Calculate how far to shift them to put them in range.
+        int shift = Math.max(numerator.bitLength(), 
+        denominator.bitLength()) - Double.MAX_EXPONENT;
+        result = numerator.shiftRight(shift).doubleValue() / 
+            denominator.shiftRight(shift).doubleValue();
+    }
     return result;
 }
\end{minted}
   \caption{Sample bug with High severity value in the Math-36 project of the Defects4J dataset.}
   \label{fig:math_36} 
\end{figure}

Another severe bug exists in the \verb|accumulo| project where the buggy method renames a requested table. In this method, the code does not check if the requested table exists already or whether the new requested name is already taken by another table, so this bug probably leads to a software crash or overwriting the existing table.

We found that most of the Edge/Boundary bugs are not severe in our datasets, but they appear in complex methods. Measuring the bug consequences in the whole system would be required to find the bug impact and eventually the severity. Edge/Boundary case testing could be a possible solution to determine the bug impact based on the number of failed tests.\\

\textbf{Security faults:}
Security faults can be exploited to gain unauthorized access/privileges by intruders. Since it is not easy to find if there is a security fault, solely from the source code, we found that some bugs are highly potential to be exploited by intruders by considering keywords and the method context.

For example, there is a bug with Major severity in the \texttt{jackrabbit-oak-3324\_5f863af6} project which is shown in Fig \ref{fig:jackrabbit-security}. This simple (based on the code metric values) method checks whether the requested path argument has access permission to be used. Handling permissions incorrectly may potentially lead to unauthorized access, and this usually has harmful consequences such as data loss or breaking the system completely.

\begin{figure}
\centering
% \fcolorbox{black}{white}{
% \includegraphics[width=0.97\linewidth]{figs/jackrabbit-security.png}}
 \begin{minted}{diff}
 @Override
 public boolean isGranted(@Nonnull String path, long permissions) {
-    Iterator<PermissionEntry> it = getEntryIterator(
-    new EntryPredicate(path, 
-    Permissions.respectParentPermissions(permissions)));
-    return hasPermissions(it, permissions, path);
+    EntryPredicate predicate = new EntryPredicate(
+        path, Permissions.respectParentPermissions(permissions));
+    return hasPermissions(getEntryIterator(predicate), 
+    predicate, permissions, path);
 }
\end{minted}
\caption{Sample bug with the Major severity value in the jackrabbit-oak-3324\_5f863af6 project of Bugs.jar dataset.}
\label{fig:jackrabbit-security}

\end{figure}

After investigating all the samples in this category, we found that Security faults are severe (Blocker, Critical, Major, High) in most cases. However, these bugs are available in simple methods (based on our code metrics) that handle a sensitive task which may lead to harmful consequences and result in severe bugs.

Although we found most of our Security faults happen in simple methods, more specific methodologies such as security testing are required to find these bugs' severity values. If we can find the security fault type (e.g., Injection, Data Exposure, Broken Access Control, etc) the bug severity can be found easily since security fault types typically have a specific ranking based on their probable consequences in a few defined standards.\\

\begin{figure}
\centering
% \fcolorbox{black}{white}{
%  \includegraphics[width=0.97\linewidth]{figs/test3.png}}
 \begin{minted}{diff}
 protected Object functionFloor(EvalContext context) {
     assertArgCount(1);
     double v = InfoSetUtil.
     doubleValue(getArg1().computeValue(context));
+    if (Double.isNaN(v) || Double.isInfinite(v)) {
+        return new Double(v);
+    }
     return new Double(Math.floor(v));
 }
\end{minted}
\caption{Sample bug with High severity value in the JxPath-14 project of the Defects4J dataset.}
\label{fig:jxpath_14}
\end{figure}

\textbf{Integration faults:}
This bug label exhibits cases where the bug is related to another component, module, or class that our buggy method is using, so it would be impossible to access the buggy method without considering the called module(s). For example, Fig \ref{fig:jxpath_14} shows a buggy code and its corresponding fixed code in the \verb|JxPath-14| project. The buggy method seems to be very simple containing only three statements, so based on the code metric values this is not a complex code, but the actual severity value is high. The bug is related to the \verb|floor()| method which is from another module (\verb|math|). It seems that this module may not handle \verb|NaN| and \verb|Infinite| values or it may handle them differently than the studied project's (\verb|JxPath|) specification. From the patch, we see that the developer needs to return a different value when the \verb|v| parameter is \verb|NaN| or \verb|Infinite|, so the developer had to handle this case in the code before calling the \verb|Math.floor()| function. 

As another example, the buggy method in the \verb|Cli-2| project uses the \verb|NumberUtils.createNumber()| method to convert the variable with String type to the Number type, but the patch shows that the developer has rewritten the logic of converting String to Number inside the code without using the \verb|NumberUtils| anymore. In this case, the called component may contain a bug or it may be developed with different specifications which does not satisfy the caller component's developer's requirements. Therefore, for detecting the severity value of this kind of bug accurately it is required to consider both caller and called module code together.

Since we found only three Integration faults, it is less reliable to conclude the severity of bugs in this category. However, our observations show that estimating the severity of faults in this category by only considering the caller method seems impossible. Thus an analysis in a coarser granularity (e.g, class/module/package) is suggested.\newline

\noindent\fbox{%
    \parbox{\columnwidth}{%
    \textbf{Summary of RQ3.1 Results:}
Most of our studied bugs were placed in the Specification categories, evenly, regardless of their severity values. Our studied bugs in the Edge/Boundary category are complex (high SC value) but mainly have a Low severity value. Security bugs have High severity, but they are mostly not complex (low SC value). Although we need more samples to confirm, our dataset indicates that bugs related to security are always severe.}%
}\\\\

\textbf{RQ3-2 Answer:} In this section, we provide the results of our manual investigation of severe bugs (Blocker, Critical, Major, High) that are not detected by static analysis tools and go through some examples of severe bugs that are miss-labeled by SpotBugs.

\begin{figure}
\centering
% \fcolorbox{black}{white}{
% \includegraphics[width=0.97\linewidth]{figs/cli2.png}}
 \begin{minted}{diff}
 protected void burstToken(String token, 
 boolean stopAtNonOption){
    int tokenLength = token.length();
    for (int i = 1; i < tokenLength; i++){
        String ch = String.valueOf(token.charAt(i));
        boolean hasOption = options.hasOption(ch);
        if (hasOption){
            tokens.add("-" + ch);
            currentOption = options.getOption(ch);
            if (currentOption.hasArg() && 
                (token.length() != (i + 1))){
                tokens.add(token.substring(i + 1));
                break;
            }
        }
        else if (stopAtNonOption){
            process(token.substring(i));
        }
        else{
-           tokens.add("-" + ch);
+           tokens.add(token);
+           break;
        }
    }
 }
\end{minted}
\caption[]{Critical bug in the Cli-2 project of the Defects4J dataset.}
\label{fig:cli_2}
\end{figure}

Fig \ref{fig:cli_2} shows a Critical bug in \verb|Cli-2| project, which is not detected by SpotBugs and Infer because it cannot be fitted into any rule. The provided patch seems to be simple, but the bug is too complex since it is related to a specific case where the developer should use \verb|token| variable and also the \verb|break| statement in the last \verb|else| branch to handle a special case required by the specification.

\begin{figure}
\centering
% \fcolorbox{black}{white}{
% \includegraphics[width=0.97\linewidth]{figs/math11.png}}
 \begin{minted}{diff}
 public double density(final double[] vals) throws 
    DimensionMismatchException {
    final int dim = getDimension();
    if (vals.length != dim) {
        throw new DimensionMismatchException(vals.length, 
        dim);
    }
-   return FastMath.pow(2 * FastMath.PI, -dim / 2) *
+   return FastMath.pow(2 * FastMath.PI, -0.5 * dim) *
            FastMath.pow(covarianceMatrixDeterminant, -0.5) *
            getExponentTerm(vals);
}
\end{minted}
\caption[]{Critical bug in the Math-11 project of the Defects4J dataset.}
\label{fig:math_11}
\end{figure}

Another example is a Critical bug in the \verb|Math-11| project which is shown in Fig \ref{fig:math_11}. The patch is very simple, and the buggy statement looks almost identical to the fixed statement such that it is even very hard for a human to find the issue without knowing the special values of \verb|dim| variable that leads to the bug.

\begin{figure}
\centering
 \begin{minted}{diff}
 if (size >= 2 && elementPairs.get(0) instanceof Separator) {
     Separator sep = (Separator) elementPairs.get(0);
+    if (sep.iAfterParser==null && sep.iAfterPrinter==null) {
         PeriodFormatter f = toFormatter(
         elementPairs.subList(2,size), notPrinter, notParser);
         sep = sep.finish(f.getPrinter(), f.getParser());
         return new PeriodFormatter(sep, sep);
+     }
 }
\end{minted}
\caption[]{High-severity bug in the Times-27 project of the Defects4J dataset.}
\label{fig:times_27}
\end{figure}

Fig \ref{fig:times_27} shows the High severity bug in \texttt{Times-27} project where the \texttt{PeriodFormatter} class is not working well and the fix is as simple as having the \texttt{null} checking before the logic. While the fixed code is only one line and simple, it is hard for static analysis tools to detect this since we may have different logic based on the nullability of objects and there is no general error pattern because of this.

\begin{figure}
\centering
 \begin{minted}{diff}
+ if (entryEncoding != null) {
+   return new JarArchiveOutputStream(out, entryEncoding);
+   } else {
        return new JarArchiveOutputStream(out);
+   }

  if (CPIO.equalsIgnoreCase(archiverName)) {
+      if (entryEncoding != null) {
+        return new ArjArchiveInputStream(in, entryEncoding);
+      } else {
           return new ArjArchiveInputStream(in);
+      }
 }
\end{minted}
\caption[]{High-severity bug in the Compress-29 project of the Defects4J dataset.}
\label{fig:comrpress_29}
\end{figure}

The \texttt{Compress-29} bug which has a high severity is shown in Fig \ref{fig:comrpress_29} where the polymorphism is applied with the method overloading approach. Although this is the standard approach to change the behavior of code in compile-time, the decision to use a particular overloaded function completely depends on the other lines of the code. There is no issue from the static analysis tools in this case since no patterns can fit into this situation.

\begin{figure}
\centering
 \begin{minted}{diff}
  public double getSumSquaredErrors() {
-    return sumYY - sumXY * sumXY / sumXX;
+    return Math.max(0d, sumYY - sumXY * sumXY / sumXX);
  }
\end{minted}
\caption[]{Major bug in the Math-105 project of the Defects4J dataset.}
\label{fig:math_105}
\end{figure}

We found other severe bugs that cannot be generalized in the bug patterns such as a bug in the \verb|Math-105| project shown in fig \ref{fig:math_105} where the \verb|getSumSquaredErrors()| in the buggy code returns the maximum number between the error value and zero wrongly. The patch shows that it should return the calculated error value only (there is no need for the max function). 

\begin{figure}
\centering
 \begin{minted}{diff}
 if (cs1 instanceof String && cs2 instanceof String) {
     return cs1.equals(cs2);
 }
- return CharSequenceUtils.regionMatches(cs1, false, 0, 
-        cs2, 0, Math.max(cs1.length(), cs2.length()));
+ return cs1.length() == cs2.length() &&
+        CharSequenceUtils.regionMatches(cs1, false, 0, 
         cs2, 0, cs1.length());
}
\end{minted}
\caption[]{High-severity bug in the Codec-18 project of the Defects4J dataset.}
\label{fig:codec_18}
\end{figure}

Also, \verb|Codec-18| project (shown in fig \ref{fig:codec_18}) contains a bug where the \verb|equals| method checks if two provided \verb|CharSequence| objects are equal. The patch shows that the developer implemented the equality condition wrongly by not considering the maximum length of these strings which are patched in the fixed version.

\begin{figure}
\centering
 \begin{minted}{diff}
+ else if (current == JobStatus.FAILING) {
+        if (transitionState(current, JobStatus.CANCELLING)) {
+                return;
+          }
+   }
+ else if (current == JobStatus.RESTARTING) {
+    synchronized (progressLock) {
+        if (transitionState(current, JobStatus.CANCELED)) {
+                postRunCleanup();
+                progressLock.notifyAll();
+                LOG.info("Canceled during restart.");
+                return;
+            }
+        }
+ }
\end{minted}
\caption[]{Major bug in the FLINK-3011 project of the Bugs.jar dataset.}
\label{fig:flink-3011}
\end{figure}

Another example is the \texttt{FLINK-3011} major bug in the bugs.jar dataset (shown in fig \ref{fig:flink-3011})  where the state of a running job is not handled comprehensively, so it cannot cancel a failing/restarting job. Since state handling is one of the most complex parts of every application, especially in multi-threaded applications our static analysis tools were not able to detect this bug based on their pre-defined patterns.

In all of the mentioned examples, we found that these bugs are so unique that they cannot be mapped to any of the predefined patterns. Detecting these severe bugs by only leveraging the source code (even at the coarse-level granularity such as class/package/module) is almost impossible. Therefore, more sophisticated patterns need to be implemented by static analysis tools. One may also try using dynamic approaches such as testing to detect these types of bugs.

Furthermore, we found some samples where there is a specific rule for the bug in the tools, but the bug is not detected by any of the tools. For example, the buggy method in the \verb|Cli-9| project copies the values of a \verb|mapping| variable (defined at the class level) to the provided argument of the method. The provided patch shows that there is a specific condition when the \verb|mapping| variable is \verb|null|, so the buggy method should return the provided argument instead of copying data. Although the SpotBugs tool has some predefined rules regarding the \verb|null| handling of variables in the method it cannot report this method as buggy.

One possibility for this low effectiveness of static analysis tools is that since these tools are heavily used in production as a part of the development environment or building process, there is a high chance that all of these projects were scanned by these tools after each commit/release, so all bugs that can be detected by these tools have been fixed already.

By investigating the results of the SpotBugs tool to see why this tool mislabeled the severity of many bugs we found that 70 bugs are detected to have lower severity values than their actual severity value. After analyzing these samples, we found two scenarios. The first one would be samples that SpotBugs reported another bug in the buggy method than the actual bug. In the second case, Spotbugs reports the bug correctly, but it assigns the lower severity labels.

\begin{figure}
\centering
 \begin{minted}{diff}
 public int compareTo(Fraction object) {
-    double nOd = doubleValue();
-    double dOn = object.doubleValue();
+    long nOd = ((long) numerator) * object.denominator;
+    long dOn = ((long) denominator) * object.numerator;
     return (nOd < dOn) ? -1 : ((nOd > dOn) ? +1 : 0);
 }
\end{minted}
\caption[]{High-severity bug in the Math-91 project of the Defects4j dataset.}
\label{fig:math-91}
\end{figure}

For example, the reported bug in the \verb|Math-91| (fig \ref{fig:math-91}) project has High severity, but SpotBugs reports Low severity. The actual reported bug is due to the equality problem of two \verb|Fraction| objects because of the limited precision \footnote{\url{https://issues.apache.org/jira/browse/MATH-252}}, but the SpotBugs reports the \verb|CO_COMPARETO_INCORRECT_FLOATING| as the issue type which is in the \verb|BAD_PRACTICE| category with the following description:

``This method compares double or float values using a pattern like this: $val1 > val2~?~1 : val1 < val2 ~?~ -1 : 0$"

Although this statement exists in the buggy method, the actual reported bug is not happening because of this statement.

\begin{figure}
\centering
 \begin{minted}{diff}
-    currEntry = new TarArchiveEntry(headerBuf);
+    try {
+         currEntry = new TarArchiveEntry(headerBuf);
+    } catch (IllegalArgumentException e) {
+         IOException ioe = new IOException(
+                     "Error detected parsing the header");
+         ioe.initCause(e);
+         throw ioe;
+    }
\end{minted}
\caption[]{High-severity bug in the Compress-12 project of the Defects4j dataset.}
\label{fig:compress_12}
\end{figure}

Another example is the bug in the \verb|Compress-12| project where the actual reported bug is related to throwing the \verb|IllegalArgumentException| instead of \verb|IOException| on corrupted files \footnote{\url{https://issues.apache.org/jira/browse/COMPRESS-178}}. This can lead to serious issues such as crash or wrong exception handling in different layers where this exception is being handled. The path for this bug is shown in fig \ref{fig:compress_12}. However, SpotBugs reported this method containing a Low-severity bug with the \verb|I18N| issue type in the \verb|DM_DEFAULT_ENCODING| category with the following description:

``Found a call to a method which will perform a byte to String (or String to byte) conversion, and will assume that the default platform encoding is suitable."

In the mentioned examples we found that SpotBugs reported a different bug with Low severity instead of the actual bug with High severity.

\begin{figure}
\centering
 \begin{minted}{diff}
-    babcs.set(bs);
-    matcher.reset(babcs);
-    return matcher.matches();
+    try {
+      matcher.reset(
+        new String(bs.getBackingArray(), encoding));
+      return matcher.matches();
+    } catch (UnsupportedEncodingException e) {
+      e.printStackTrace();
+    }
\end{minted}
\caption[]{High-severity bug in the accumulo-209-397f86f6 project of the Bugs.jar dataset.}
\label{fig:accumulo-209}
\end{figure}

\begin{figure}
\centering
 \begin{minted}{diff}
  int m = problem.getMeasurements().length;
- int p = problem.getAllParameters().length;
+ int p = problem.getUnboundParameters().length;
  if (m <= p) {
    throw new EstimationException("no degrees of freedom ({0}
    measurements, {1} parameters)",
            new Object[] { new Integer(m), new Integer(p)});
}
-  double[] errors = new double[
-                    problem.getAllParameters().length];
+  double[] errors = new double[
+                    problem.getUnboundParameters().length];
\end{minted}
\caption[]{Critical bug in the Math-100 project of the Defects4j dataset.}
\label{fig:math-100}
\end{figure}

{
The \texttt{Math-100} project with the Critical bug crashes with \texttt{ArrayOutOfBounds} exception in the \texttt{getCovariances()} and \texttt{guessParametersErrors()} functions \footnote{\url{https://issues.apache.org/jira/browse/MATH-200}} but the SpotBugs tool reports bugs in this function with \texttt{Bx-DM\_NUMBER\_CTOR} pattern with the following description:

``Method invokes inefficient Number constructor; use static valueOf instead``

Although the reported bug from SpotBugs tools may apply to the buggy method, SpotBugs is finding another possible bug which is not the actual bug available in this project, so the reported severity from SpotBugs is lower than the actual severity. The partial actual fix for this bug is shown in fig \ref{fig:math-100}.

\begin{figure}
\centering
 \begin{minted}{diff}
  if (dec == null && exp == null
-     && isDigits(numeric.substring(1))
-     && (numeric.charAt(0) == '-' || 
-     Character.isDigit(numeric.charAt(0)))) {
+     && (numeric.charAt(0) == '-' && 
+     isDigits(numeric.substring(1)) || isDigits(numeric))) {
        try {
          return createLong(numeric);
        }
\end{minted}
\caption[]{High-severity bug in the Lang-58 project of the Defects4j dataset.}
\label{fig:lang-58}
\end{figure}

Another High-severity bug in \texttt{Lang-58} project of Defects4J dataset exists where the reported bug is because of throwing a \texttt{NumberFormatException} when passing special values to the \texttt{createNumber} method and it cannot parse the input correctly \footnote{\url{https://issues.apache.org/jira/browse/LANG-300}}. The SpotBugs tool reports another bug for this function that points to a different possible bug that is not the same as the actual bug. SpotBugs report \texttt{SF-SF\_SWITCH\_FALLTHROUGH} with the following description:

``Switch statement found where one case falls through to the next case``

Although the reported issue by SpotBugs exists in the studied function, this is not the reason for the actual reported bug (the fix is shown in fig \ref{fig:lang-58}) and SpotBugs is reporting low severity since this bug pattern is defined to have a low severity generally by SpotBugs.

Another example is \verb|accumulo-209_397f86f6| project with Major bug with the multi-byte character encoding issue\footnote{\url{https://github.com/bugs-dot-jar/accumulo/tree/bugs-dot-jar_ACCUMULO-209_397f86f6}} (part of the patch is shown in fig \ref{fig:accumulo-209}), and the SpotBugs reported a \verb|DM_DEFAULT_ENCODING| as the issue type in \verb|I18N| category with the following description:

``Found a call to a method which will perform a byte to String (or String to byte) conversion, and will assume that the default platform encoding is suitable."

It seems that SpotBugs has detected the bug correctly, but it assigns a low severity to it. Based on our best knowledge, SpotBugs assigns the severity of the reported bugs based on their implemented patterns, so it always reports Low severity for this kind of bug. However, in practice, one specific issue may lead to a severe or non-severe bug in different cases such as the number of affected users or the project timeline.\\

\noindent\fbox{%
    \parbox{\columnwidth}{%
    \textbf{Summary of RQ3.2 Results:}
Based on our manual investigation, SpotBugs missed most of the studied bugs regardless of their severity value for two reasons. First, SpotBugs finds other possible bugs than the actual reported bugs. Second, there is no matching rule for the actual bugs, and SpotBugs reports another bug based on its available pre-defined bug pattern. In specific cases there are matching (almost the same) rules, however, the SpotBugs cannot detect their severity accurately. In this case, the reported severity is lower than the actual severity value since SpotBugs reports the severity value based on its general pre-defined bug patterns, but in practice, each specific bug may have different severity values based on different conditions such as the number of affected users or the project timeline.}}%
\\\\

%% file: discussion.tex
\section{Discussion on Practical Implications}
\label{sec:discussion}

We now summarize our findings and discuss how these findings can be useful in practice and bug prediction research.

In \textbf{RQ1}, we evaluated the effectiveness of code metrics in detecting bugs and their severity at the method-level source code granularity. Our results suggest that code metrics are good indicators of bug-prone methods. Our results are interesting because the true effectiveness of code metrics has been historically debated~\cite{chowdhury2022revisiting}. While some studies claimed code metrics are useful to understand software maintenance~\cite{Johnson:2019, landman2014empirical,chowdhury2022revisiting}, such as finding or predicting bug-prone source code, other studies found code metrics useless~\cite{Shepperd:1988,gil2017correlation}. In recent studies, it was found that code metrics are useful at the method-level granularity, although they are not useful at the file/class-level granularity~\cite{landman2014empirical,chowdhury2022revisiting}. This observation, however, does not match with the findings of Pascarella et al.~\cite{pascarella2020performance}; the authors observed that code metrics were not useful in predicting bugs at the method level.

We also observed that although all the code metrics are helpful for bug detection, only a few of them help understand bug severity. Notably, when bug severity is considered, some code metrics---such as readability, Halstead effort, and difficulty---outperform the famous lines of source code (LC). This a very unique observation, as according to multiple research, LC is the metric that rules them all---LC always outperforms other code metrics in understanding maintenance effort~\cite{chowdhury2022revisiting, gil2017correlation, chowdhury:2022,Emam:2001,TSE:2013}.

We conclude that practitioners should be aware of code metrics-induced code smells because we observed a clear difference in code metrics distribution between bug-prone and bug-free source code methods. Researchers should focus on producing more code metrics that are more geared toward bug severity estimation. We see potential future research where code metrics can be used in combination with large language models (LLMs) 
such as CodeBERT~\cite{feng2020codebert,ehsan-llm} or GPT \cite{openai2023gpt4}. Through word embeddings, LLMs can potentially find and suggest important tokens that often produce severe bugs.

In RQ2, we observed that existing static bug detection tools perform extremely poorly in detecting real-world bugs. These tools suffer from both false positives and false negatives, confirming earlier findings~\cite{habib2018many,thung2012extent}. In a few cases, when these tools successfully captured real-world bugs, their estimation of bug severity was inaccurate---a unique finding of our study. We suggest that practitioners should be careful in relying on the existing static bug detection tools. In RQ3, we provided a qualitative analysis explaining why these tools fail in detecting bugs and estimating their severity. For example, we found that a less complex source code method that contains no bug patterns can contain severe bugs for containing security-related APIs.

We believe that our observations will be instrumental for future research on producing accurate bug severity prediction models. For example, based on our observation, we suggest future studies to combine static analysis and code metrics results to get the benefits of both. This is useful since the metrics and the tools seem to focus on different aspects of severity. By comparing the calculated SC (mentioned in \ref{sec:RQ3-1}) and the severity predicted by the SpotBugs tool we found some samples where using both of these results will be more useful. For example, in the \verb|Math-87| project, the buggy method \verb|computeShiftIncrement()| has High severity value. The SpotBugs tool reports a Low severity value, but the SC value shows that this method is a complex method regarding the code metrics we used. In this case, the code metric values exhibit the severity better. Conversely, there is a buggy method named \verb|parse()| in the \verb|Closure-68| project with a Medium severity value. The SC value for this buggy method is very large which shows the complexity of the method, so we may assume this is a severe bug, but the SpotBugs reports a Low severity value. In the mentioned example, the SpotBugs prediction is closer to the actual severity value than the code metric values.

We want to emphasize that although many of our findings are negative, it does not mean we do not need this nature of research. It is in fact as important to publish negative results for the community so that future researchers know what does not work and they do not spend more time on the same tools and techniques. We also have to point out that none of the findings are ``obvious'' before being empirically and systematically examined. Therefore, we think that identifying what did not work by itself is an important contribution to the research community.

%% file: limitations.tex
\section{Threats to Validity}
\label{sec:Threats to Validity}

Similar to any empirical study there are some threats that may impact our results, so in this section, we provide possible threats and our actions to mitigate these threats.

\subsection{External Validity}
Our selected projects may not be consistent with the closed-source projects or may not cover various software domains or other programming languages. We did our best to select two large and popular datasets containing real-world bugs of a diverse set of open-source projects that have been used in many studies \cite{martinez2017automatic,shamshiri2015automatically,pearson2017evaluating,saha2017elixir}. The next threat would be the selection of static analysis tools, and for mitigating this problem we used the most popular tools which are used in practice and state-of-the-art research. We exclude Google Error Prone from the list of our tools since we found that it is not possible to handle older Java versions (which exist in our dataset) with the latest versions of this tool. Also, the selected code metrics may not reveal the potential of all available code metrics in finding bug severity and this may impact our result generalizability, but to mitigate this issue we tried to leverage as many as available method-level code metrics that are used in previous research for defect-related problems (e.g, defect prediction).

\subsection{Internal Validity}
Another threat would be our strategy in selecting the buggy methods since we selected all methods that are changed/removed during the bug fixing patch as a buggy method which is a common approach in the state-of-the-art approaches (referenced in \ref{sec:Dataset}). The reason behind this is that bug fixing usually requires changing multiple methods and specifying one single method to be the source of a bug requires other information such as a bug report description. To mitigate this threat we consider each of those methods as an independent sample, so code metrics and static analysis tools can consider them independently. The next threat is the method of identifying reported bugs by static analyzers' tools. In this study, we consider the tools' report at the method-level granularity without considering the exact reported lines by these tools. We did this because there is no information regarding the exact bug location (line) in our datasets, and we want to keep the same method-level granularity for RQ1 and RQ2. To mitigate this threat whenever any of our tools report a bug inside the method, we count it as a correct detection to not underestimate the tools' performance.

\subsection{Construct Validity}
\label{construct}
For collecting ground truths about bug severity, we relied on practitioner-assigned labels, which can be inaccurate. Tian et al.~\cite{tian2016unreliability} observed that a duplicate bug report can have multiple severity labels. Practitioner-assigned severity labels, however, are perhaps the only way to understand bug severity and are widely adopted in bug severity research (e.g.,~\cite{tian2012information, tian2013drone, ramay2019deep, lamkanfi2010predicting}). In our study, however, this threat was mitigated as we have merged similar severity for most of our analysis. That is, although different practitioners may assign different severity labels to the same bug report, the labels are still expected to be similar (e.g., critical and major, instead of critical and minor). 
When calculating the sum complexity for answering RQ3-1, we have provided the same weight to all the different code metrics. These can be inaccurate because some code metrics can be more important than others. 

\subsection{Conclusion Validity}
Our results may not generalize to other tools and datasets. Our results of static analysis tools may be affected by using projects that have already applied the studied static analyzer tools warnings during the software development life cycle. Since it is not easy to understand if it happened, our results should be considered as an assessment of those bugs that are not already fixed by tools. Our selection of code metrics may not be sufficient to expose the bug severity, but to mitigate this issue we considered the 10 popular metrics which have established performance in finding bugs in our study.

%% file: related_work.tex
\section{Related Work}
\label{sec:Related Work}

There has been much research in the bug-prediction field by focusing on different aspects using diverse granularity levels and various techniques \cite{hall2011systematic,hosseini2017systematic,wahono2015systematic,li2020systematic}. Previous studies can be categorized from different aspects such as bug reproducibility~\cite{sahoo2010empirical}, working on the characteristics of source code \cite{khomh2012exploratory,palomba2018diffuseness,bacchelli2010popular,bavota2012does,pai2007empirical,spadini2018relation,rahman2019source}, or bug description \cite{sun2019bug,wang2010detect,khatiwada2018just}. From the source code granularity aspect, there has been research in file level \cite{zimmermann2007predicting,giger2011comparing}, package level \cite{zimmermann2007predicting}, class level \cite{shivaji2009reducing}, and method level \cite{giger2012method}.
For bug severity prediction tasks, almost all of the existing works have leveraged bug reports using natural language techniques or different classic and deep neural network models. In the rest of this section, we briefly discuss the related work on generic bug prediction, bug severity prediction, and static analysis tools.

\subsection{Bug Prediction}

Bug prediction studies have used different code metrics such as LOC ~\cite{pascarella2020performance,antinyan2014identifying,shin2010evaluating,chowdhury:2022}, McCabe ~\cite{antinyan2014identifying}, Halstead metrics ~\cite{antinyan2017evaluating}, C\&K metrics ~\cite{subramanyam2003empirical,jureczko2010using}. Code metrics have been applied on different granularity levels such as package/class level~\cite{Okutan:2014,Koru:2005,zimmermann2007predicting}, method level~\cite{pascarella2020performance,giger2012method,ferenc:2020,chowdhury:2022,grund2021codeshovel,Mo:2022}, and line level~\cite{wattanakriengkrai2020predicting}. One major problem with the high granularity levels (package/class) is that they are practically less helpful for the developers~\cite{giger2012method,pascarella2020performance} since it requires significant effort to locate bugs at the package/class components. Unfortunately, line-level granularity can suffer from too many false positives, because multiple lines can be similar just by chance~\cite{Daniela:2014,Francisco:2017}. Consequently, method level granularity has been the new focus of the community~\cite{pascarella2020performance,giger2012method,ferenc:2020,chowdhury:2022,grund2021codeshovel,Mo:2022}, especially for developing bug prediction models, and several studies show positive and encouraging results~\cite{giger2012method,ferenc:2020,Mo:2022}.

\subsection{Bug Severity Prediction}
Due to time and budget constraints, practitioners often try to solve the severe bugs first. This, however, requires manually labeling the severity of the reported bugs which is time-consuming. Consequently, previous research has focused on predicting the severity of bug reports using different machine learning and natural language processing techniques.

Tian et al. \cite{tian2012information} leveraged bug reports and their severity labels in the past to recommend fine-grained severity labels for newly reported bugs by measuring the similarity between bug reports and using nearest neighbors classification. The authors improved the f-measure of the state-of-the-art approach significantly. In a separate work~\cite{tian2013drone}, the authors found a positive impact on prediction accuracy with multi-factor analysis---they have produced features from different factors such as text, related reports, and temporal information. 

Ramay et al \cite{ramay2019deep} proposed an approach by using a deep learning model, natural language techniques, and emotion analysis by using bug reports for predicting bug severity. They mentioned that this method improved the f-measure of state-of-the-art approaches by 7.90\% on average. Lamkanfi et al~\cite{lamkanfi2010predicting} proposed bug severity prediction using text mining algorithms. Their analysis with Mozilla, Eclipse, and GNOME software projects suggests that with enough training data accurate bug report severity models can be built.
% Their evaluation of 30,000 bug reports extracted from Eclipse, Mozilla, and Netbeans projects shows the effectiveness of their approach in predicting severity. 

Tan et al \cite{tan2020bug} leveraged question-and-answer pairs from Stack Overflow and combined them with related bug reports to make an enhanced version of the bug reports. They predict the severity of bug reports using different machine learning models which led to improvements by approximately 23\% of the average f-measure. In a recent study, Agrawal et al.~\cite{agrawal2021developing} investigated the usefulness of word embedding in predicting the severity of bug reports, and observed mixed results. Interested readers are encouraged to read the survey paper by Gomes et al.~\cite{gomes2019bug} to understand the research gaps and opportunities in bug report severity prediction.

The objective of our study is significantly different than those previous studies: our focus was not to understand the severity of bug reports but to understand and predict source code methods with different bug severity.
The authors in \cite{zhou2006empirical}  employed statistical (logistic regression) and machine learning (naive Bayes, random forest, and NNge) methods to investigate the fault-proneness prediction usefulness of OO design metrics with regard to ungraded, high, and low severity faults.
Although relevant, this study is much more limited than ours from multiple aspects. They only investigate one C++ project, with just over 2,000 faulty and non-faulty methods and around 500 bugs, and seven OO metrics. Whereas ours looks into over 77,000 methods including over 3,000 faulty methods from 19 Java projects. Therefore, even the metric analysis part of our experiment is much larger scale, let aside the static analysis contribution. 

\subsection{Static Analysis Tools}
The application of static analysis tools is common in software projects, as suggested by the findings of Lavazza et al.~\cite{lavazza2020empirical}. The authors found that approximately 50\% of the issues reported by the SpotBugs tool have disappeared in the next revision. Various research has been done to evaluate the true effectiveness of various static analysis tools for bug detection tasks. Some researchers leveraged these tools as an oracle for their own provided techniques. For example, Tomassi~\cite{tomassi2018bugs} used SpotBugs and ErrorProne to detect bugs in the BugSwarm dataset, and they found that these tools were not effective in finding the bugs since their results showed only one
successful bug detection by SpotBugs. Ayewah et al. \cite{ayewah2007using} leveraged FindBugs to find bugs in Google's internal code base, and they found that integrating this tool into Google’s Mondrian code review system would help developers see the potential bug existence in the code. Habib and Pradel~\cite{habib2018many} studied SpotBugs, Infer, and Google Error Prone tools to find their capabilities in detecting Java real bugs. They concluded that these tools are mostly complementary to each other, and they miss the large majority of the bugs. Similar results---less effectiveness of bug detection tools---were also observed by Thung et al.~\cite{thung2012extent}. 
Dura et al.~\cite{dura2021javadl} introduced the JavaDL which is a Datalog-based declarative specification language for bug pattern detection in Java code, and compared it against the SpotBugs and ErrorProne tools. The authors found that JavaDL has comparable performance to these tools. Habib and Pradel \cite{habib2019neural} proposed a method to consider bug detection as a classification problem by using the neural networks and Google Error Prone as an oracle.

To the best of our knowledge, none of the previous studies have investigated the relationship between various popular code metrics and bug severity in method-level granularity, nor the effectiveness of static analysis tools for bug severity detection.

%% file: conclusion.tex
\section{Conclusion and Future Work}
\label{sec:Conclusion}
In this paper, we studied 10 source code metrics and two static analysis tools to understand and distinguish buggy code and to find their capability in estimating the bugs' severity levels. Our quantitative and qualitative studies on the Defects4J and Bugs.jar datasets---containing 3,358 buggy methods from 19 Java open-source projects---showed that code metrics are good indicators of buggyness, but neither code metrics nor static analysis tools are significant estimators of bug severity. We found that there is no relationship between the code complexity and bug severity, and static analysis tools miss many bugs due to the lack of specialized patterns.
Manual inspection of severe bugs reveals that some bug types (e.g., integration, and security bugs) are often more severe than others (e.g., edge/boundary, and specification bugs).

Potential future directions of this research are studying the power of dynamic analysis and testing in estimating bug severity. In addition, one can use the findings of this study regarding static analysis tools' limitations and try to enrich their rule sets to better identify severe bugs. It is also promising to combine static analysis and code metrics for better severity prediction. Lastly, the advances in large language models can be leveraged to estimate the severity of a bug based on the pre-trained models on very large datasets of code bases and further fine-tune them on smaller datasets containing severity keywords in the fixes.

%% file: paper.bbl
\begin{thebibliography}{131}
\expandafter\ifx\csname natexlab\endcsname\relax\def\natexlab#1{#1}\fi
\providecommand{\url}[1]{\texttt{#1}}
\providecommand{\href}[2]{#2}
\providecommand{\path}[1]{#1}
\providecommand{\DOIprefix}{doi:}
\providecommand{\ArXivprefix}{arXiv:}
\providecommand{\URLprefix}{URL: }
\providecommand{\Pubmedprefix}{pmid:}
\providecommand{\doi}[1]{\href{http://dx.doi.org/#1}{\path{#1}}}
\providecommand{\Pubmed}[1]{\href{pmid:#1}{\path{#1}}}
\providecommand{\bibinfo}[2]{#2}
\ifx\xfnm\relax \def\xfnm[#1]{\unskip,\space#1}\fi
%Type = Article
\bibitem[{Agrawal and Goyal(2021)}]{agrawal2021developing}
\bibinfo{author}{Agrawal, R.}, \bibinfo{author}{Goyal, R.}, \bibinfo{year}{2021}.
\newblock \bibinfo{title}{Developing bug severity prediction models using word2vec}.
\newblock \bibinfo{journal}{International Journal of Cognitive Computing in Engineering} \bibinfo{volume}{2}, \bibinfo{pages}{104--115}.
%Type = Article
\bibitem[{Alenezi et~al.(2019)Alenezi, Akour and Al~Sghaier}]{alenezi2019impact}
\bibinfo{author}{Alenezi, M.}, \bibinfo{author}{Akour, M.}, \bibinfo{author}{Al~Sghaier, H.}, \bibinfo{year}{2019}.
\newblock \bibinfo{title}{The impact of co-evolution of code production and test suites through software releases in open source software systems}.
\newblock \bibinfo{journal}{International Journal of Innovative Technology and Exploring Engineering (IJITEE)} \bibinfo{volume}{9}, \bibinfo{pages}{2737--2739}.
%Type = Inproceedings
\bibitem[{AlOmar et~al.(2019)AlOmar, Mkaouer, Ouni and Kessentini}]{alomar2019impact}
\bibinfo{author}{AlOmar, E.A.}, \bibinfo{author}{Mkaouer, M.W.}, \bibinfo{author}{Ouni, A.}, \bibinfo{author}{Kessentini, M.}, \bibinfo{year}{2019}.
\newblock \bibinfo{title}{On the impact of refactoring on the relationship between quality attributes and design metrics}, in: \bibinfo{booktitle}{2019 ACM/IEEE International Symposium on Empirical Software Engineering and Measurement (ESEM)}, \bibinfo{organization}{IEEE}. pp. \bibinfo{pages}{1--11}.
%Type = Inproceedings
\bibitem[{Aniche et~al.(2013)Aniche, Oliva and Gerosa}]{aniche2013asserts}
\bibinfo{author}{Aniche, M.F.}, \bibinfo{author}{Oliva, G.A.}, \bibinfo{author}{Gerosa, M.A.}, \bibinfo{year}{2013}.
\newblock \bibinfo{title}{What do the asserts in a unit test tell us about code quality? a study on open source and industrial projects}, in: \bibinfo{booktitle}{2013 17th European Conference on Software Maintenance and Reengineering}, \bibinfo{organization}{IEEE}. pp. \bibinfo{pages}{111--120}.
%Type = Inproceedings
\bibitem[{Antinyan et~al.(2014)Antinyan, Staron, Meding, {\"O}sterstr{\"o}m, Wikstrom, Wranker, Henriksson and Hansson}]{antinyan2014identifying}
\bibinfo{author}{Antinyan, V.}, \bibinfo{author}{Staron, M.}, \bibinfo{author}{Meding, W.}, \bibinfo{author}{{\"O}sterstr{\"o}m, P.}, \bibinfo{author}{Wikstrom, E.}, \bibinfo{author}{Wranker, J.}, \bibinfo{author}{Henriksson, A.}, \bibinfo{author}{Hansson, J.}, \bibinfo{year}{2014}.
\newblock \bibinfo{title}{Identifying risky areas of software code in agile/lean software development: An industrial experience report}, in: \bibinfo{booktitle}{2014 Software Evolution Week-IEEE Conference on Software Maintenance, Reengineering, and Reverse Engineering (CSMR-WCRE)}, \bibinfo{organization}{IEEE}. pp. \bibinfo{pages}{154--163}.
%Type = Article
\bibitem[{Antinyan et~al.(2017)Antinyan, Staron and Sandberg}]{antinyan2017evaluating}
\bibinfo{author}{Antinyan, V.}, \bibinfo{author}{Staron, M.}, \bibinfo{author}{Sandberg, A.}, \bibinfo{year}{2017}.
\newblock \bibinfo{title}{Evaluating code complexity triggers, use of complexity measures and the influence of code complexity on maintenance time}.
\newblock \bibinfo{journal}{Empirical Software Engineering} \bibinfo{volume}{22}, \bibinfo{pages}{3057--3087}.
%Type = Misc
\bibitem[{ATLASSIAN(2023)}]{Jira2023}
\bibinfo{author}{ATLASSIAN}, \bibinfo{year}{2023}.
\newblock \bibinfo{title}{Jira software}.
\newblock \bibinfo{howpublished}{\url{https://www.atlassian.com/software/jira}}.
%Type = Inproceedings
\bibitem[{Ayewah et~al.(2007)Ayewah, Pugh, Morgenthaler, Penix and Zhou}]{ayewah2007using}
\bibinfo{author}{Ayewah, N.}, \bibinfo{author}{Pugh, W.}, \bibinfo{author}{Morgenthaler, J.D.}, \bibinfo{author}{Penix, J.}, \bibinfo{author}{Zhou, Y.}, \bibinfo{year}{2007}.
\newblock \bibinfo{title}{Using findbugs on production software}, in: \bibinfo{booktitle}{Companion to the 22nd ACM SIGPLAN conference on Object-oriented programming systems and applications companion}, pp. \bibinfo{pages}{805--806}.
%Type = Inproceedings
\bibitem[{Bacchelli and Bird(2013)}]{bacchelli2013expectations}
\bibinfo{author}{Bacchelli, A.}, \bibinfo{author}{Bird, C.}, \bibinfo{year}{2013}.
\newblock \bibinfo{title}{Expectations, outcomes, and challenges of modern code review}, in: \bibinfo{booktitle}{2013 35th International Conference on Software Engineering (ICSE)}, \bibinfo{organization}{IEEE}. pp. \bibinfo{pages}{712--721}.
%Type = Inproceedings
\bibitem[{Bacchelli et~al.(2010)Bacchelli, D’Ambros and Lanza}]{bacchelli2010popular}
\bibinfo{author}{Bacchelli, A.}, \bibinfo{author}{D’Ambros, M.}, \bibinfo{author}{Lanza, M.}, \bibinfo{year}{2010}.
\newblock \bibinfo{title}{Are popular classes more defect prone?}, in: \bibinfo{booktitle}{International Conference on Fundamental Approaches to Software Engineering}, \bibinfo{organization}{Springer}. pp. \bibinfo{pages}{59--73}.
%Type = Inproceedings
\bibitem[{Bavota et~al.(2012)Bavota, De~Carluccio, De~Lucia, Di~Penta, Oliveto and Strollo}]{bavota2012does}
\bibinfo{author}{Bavota, G.}, \bibinfo{author}{De~Carluccio, B.}, \bibinfo{author}{De~Lucia, A.}, \bibinfo{author}{Di~Penta, M.}, \bibinfo{author}{Oliveto, R.}, \bibinfo{author}{Strollo, O.}, \bibinfo{year}{2012}.
\newblock \bibinfo{title}{When does a refactoring induce bugs? an empirical study}, in: \bibinfo{booktitle}{2012 IEEE 12th International Working Conference on Source Code Analysis and Manipulation}, \bibinfo{organization}{IEEE}. pp. \bibinfo{pages}{104--113}.
%Type = Inproceedings
\bibitem[{Bennett and Rajlich(2000)}]{bennett2000software}
\bibinfo{author}{Bennett, K.H.}, \bibinfo{author}{Rajlich, V.T.}, \bibinfo{year}{2000}.
\newblock \bibinfo{title}{Software maintenance and evolution: a roadmap}, in: \bibinfo{booktitle}{Proceedings of the Conference on the Future of Software Engineering}, pp. \bibinfo{pages}{73--87}.
%Type = Inproceedings
\bibitem[{Bhat and Nagappan(2006)}]{bhat2006evaluating}
\bibinfo{author}{Bhat, T.}, \bibinfo{author}{Nagappan, N.}, \bibinfo{year}{2006}.
\newblock \bibinfo{title}{Evaluating the efficacy of test-driven development: industrial case studies}, in: \bibinfo{booktitle}{Proceedings of the 2006 ACM/IEEE international symposium on Empirical software engineering}, pp. \bibinfo{pages}{356--363}.
%Type = Article
\bibitem[{B{\"o}rstler and Paech(2016)}]{borstler2016role}
\bibinfo{author}{B{\"o}rstler, J.}, \bibinfo{author}{Paech, B.}, \bibinfo{year}{2016}.
\newblock \bibinfo{title}{The role of method chains and comments in software readability and comprehension—an experiment}.
\newblock \bibinfo{journal}{IEEE Transactions on Software Engineering} \bibinfo{volume}{42}, \bibinfo{pages}{886--898}.
%Type = Article
\bibitem[{Buse and Weimer(2009)}]{buse2009learning}
\bibinfo{author}{Buse, R.P.}, \bibinfo{author}{Weimer, W.R.}, \bibinfo{year}{2009}.
\newblock \bibinfo{title}{Learning a metric for code readability}.
\newblock \bibinfo{journal}{IEEE Transactions on software engineering} \bibinfo{volume}{36}, \bibinfo{pages}{546--558}.
%Type = Misc
\bibitem[{Celerity()}]{ibm}
\bibinfo{author}{Celerity}, .
\newblock \bibinfo{title}{The true cost of a software bug: Part one}.
\newblock \bibinfo{howpublished}{\url{https://www.celerity.com/insights/the-true-cost-of-a-software-bug}}.
\newblock \bibinfo{note}{[Online; last accessed 01-Sep-2022]}.
%Type = Article
\bibitem[{Chowdhury et~al.(2022a)Chowdhury, Holmes, Zaidman and Kazman}]{chowdhury2022revisiting}
\bibinfo{author}{Chowdhury, S.}, \bibinfo{author}{Holmes, R.}, \bibinfo{author}{Zaidman, A.}, \bibinfo{author}{Kazman, R.}, \bibinfo{year}{2022}a.
\newblock \bibinfo{title}{Revisiting the debate: Are code metrics useful for measuring maintenance effort?}
\newblock \bibinfo{journal}{Empirical Software Engineering} \bibinfo{volume}{27}, \bibinfo{pages}{158}.
%Type = Article
\bibitem[{Chowdhury et~al.(2024)Chowdhury, Uddin, Hemmati and Holmes}]{chowdhury-method-2024}
\bibinfo{author}{Chowdhury, S.}, \bibinfo{author}{Uddin, G.}, \bibinfo{author}{Hemmati, H.}, \bibinfo{author}{Holmes, R.}, \bibinfo{year}{2024}.
\newblock \bibinfo{title}{Method-level bug prediction: Problems and promises}.
\newblock \bibinfo{journal}{ACM Trans. Softw. Eng. Methodol.} \URLprefix \url{https://doi.org/10.1145/3640331}. \bibinfo{note}{just Accepted}.
%Type = Inproceedings
\bibitem[{Chowdhury et~al.(2022b)Chowdhury, Uddin and Holmes}]{chowdhury:2022}
\bibinfo{author}{Chowdhury, S.}, \bibinfo{author}{Uddin, G.}, \bibinfo{author}{Holmes, R.}, \bibinfo{year}{2022}b.
\newblock \bibinfo{title}{An empirical study on maintainable method size in java}, in: \bibinfo{booktitle}{19$^{th}$ International Conference on Mining Software Repositories}.
%Type = Article
\bibitem[{Curtis et~al.(1979)Curtis, Sheppard, Milliman, Borst and Love}]{curtis1979measuring}
\bibinfo{author}{Curtis, B.}, \bibinfo{author}{Sheppard, S.B.}, \bibinfo{author}{Milliman, P.}, \bibinfo{author}{Borst, M.}, \bibinfo{author}{Love, T.}, \bibinfo{year}{1979}.
\newblock \bibinfo{title}{Measuring the psychological complexity of software maintenance tasks with the halstead and mccabe metrics}.
\newblock \bibinfo{journal}{IEEE Transactions on software engineering} , \bibinfo{pages}{96--104}.
%Type = Article
\bibitem[{Dura et~al.(2021)Dura, Reichenbach and S{\"o}derberg}]{dura2021javadl}
\bibinfo{author}{Dura, A.}, \bibinfo{author}{Reichenbach, C.}, \bibinfo{author}{S{\"o}derberg, E.}, \bibinfo{year}{2021}.
\newblock \bibinfo{title}{Javadl: automatically incrementalizing java bug pattern detection}.
\newblock \bibinfo{journal}{Proceedings of the ACM on Programming Languages} \bibinfo{volume}{5}, \bibinfo{pages}{1--31}.
%Type = Article
\bibitem[{{El Emam} et~al.(2001){El Emam}, {Benlarbi}, {Goel} and {Rai}}]{Emam:2001}
\bibinfo{author}{{El Emam}, K.}, \bibinfo{author}{{Benlarbi}, S.}, \bibinfo{author}{{Goel}, N.}, \bibinfo{author}{{Rai}, S.N.}, \bibinfo{year}{2001}.
\newblock \bibinfo{title}{The confounding effect of class size on the validity of object-oriented metrics}.
\newblock \bibinfo{journal}{IEEE Transactions on Software Engineering} \bibinfo{volume}{27}, \bibinfo{pages}{630--650}.
%Type = Misc
\bibitem[{ErrorProne(2023)}]{error_prone}
\bibinfo{author}{ErrorProne}, \bibinfo{year}{2023}.
\newblock \bibinfo{title}{Errorprone}.
\newblock \URLprefix \url{https://errorprone.info/}. \bibinfo{note}{[Online; accessed 20-April-2022}.
%Type = Article
\bibitem[{Feng et~al.(2020)Feng, Guo, Tang, Duan, Feng, Gong, Shou, Qin, Liu, Jiang et~al.}]{feng2020codebert}
\bibinfo{author}{Feng, Z.}, \bibinfo{author}{Guo, D.}, \bibinfo{author}{Tang, D.}, \bibinfo{author}{Duan, N.}, \bibinfo{author}{Feng, X.}, \bibinfo{author}{Gong, M.}, \bibinfo{author}{Shou, L.}, \bibinfo{author}{Qin, B.}, \bibinfo{author}{Liu, T.}, \bibinfo{author}{Jiang, D.}, et~al., \bibinfo{year}{2020}.
\newblock \bibinfo{title}{Codebert: A pre-trained model for programming and natural languages}.
\newblock \bibinfo{journal}{arXiv preprint arXiv:2002.08155} .
%Type = Misc
\bibitem[{Ferenc et~al.(2020a)Ferenc, B{\'a}n, Gr{\'o}sz and Gyim{\'o}thy}]{ferenc2020deep}
\bibinfo{author}{Ferenc, R.}, \bibinfo{author}{B{\'a}n, D.}, \bibinfo{author}{Gr{\'o}sz, T.}, \bibinfo{author}{Gyim{\'o}thy, T.}, \bibinfo{year}{2020}a.
\newblock \bibinfo{title}{Deep learning in static, metric-based bug prediction}.
%Type = Article
\bibitem[{Ferenc et~al.(2020b)Ferenc, Gyimesi, Gyimesi, Tóth and Gyimóthy}]{ferenc:2020}
\bibinfo{author}{Ferenc, R.}, \bibinfo{author}{Gyimesi, P.}, \bibinfo{author}{Gyimesi, G.}, \bibinfo{author}{Tóth, Z.}, \bibinfo{author}{Gyimóthy, T.}, \bibinfo{year}{2020}b.
\newblock \bibinfo{title}{An automatically created novel bug dataset and its validation in bug prediction}.
\newblock \bibinfo{journal}{The Journal of systems and software} \bibinfo{volume}{169}, \bibinfo{pages}{110691}.
%Type = Misc
\bibitem[{FindBugs(2015)}]{findbugs}
\bibinfo{author}{FindBugs}, \bibinfo{year}{2015}.
\newblock \bibinfo{title}{Findbugs}.
\newblock \URLprefix \url{http://findbugs.sourceforge.net/}. \bibinfo{note}{[Online; accessed 20-April-2022]}.
%Type = Inproceedings
\bibitem[{Giger et~al.(2012)Giger, D'Ambros, Pinzger and Gall}]{giger2012method}
\bibinfo{author}{Giger, E.}, \bibinfo{author}{D'Ambros, M.}, \bibinfo{author}{Pinzger, M.}, \bibinfo{author}{Gall, H.C.}, \bibinfo{year}{2012}.
\newblock \bibinfo{title}{Method-level bug prediction}, in: \bibinfo{booktitle}{Proceedings of the 2012 ACM-IEEE International Symposium on Empirical Software Engineering and Measurement}, \bibinfo{organization}{IEEE}. pp. \bibinfo{pages}{171--180}.
%Type = Inproceedings
\bibitem[{Giger et~al.(2011)Giger, Pinzger and Gall}]{giger2011comparing}
\bibinfo{author}{Giger, E.}, \bibinfo{author}{Pinzger, M.}, \bibinfo{author}{Gall, H.C.}, \bibinfo{year}{2011}.
\newblock \bibinfo{title}{Comparing fine-grained source code changes and code churn for bug prediction}, in: \bibinfo{booktitle}{Proceedings of the 8th Working Conference on Mining Software Repositories}, pp. \bibinfo{pages}{83--92}.
%Type = Article
\bibitem[{Gil and Lalouche(2017)}]{gil2017correlation}
\bibinfo{author}{Gil, Y.}, \bibinfo{author}{Lalouche, G.}, \bibinfo{year}{2017}.
\newblock \bibinfo{title}{On the correlation between size and metric validity}.
\newblock \bibinfo{journal}{Empirical Software Engineering} \bibinfo{volume}{22}, \bibinfo{pages}{2585--2611}.
%Type = Article
\bibitem[{Gomes et~al.(2019)Gomes, da~Silva~Torres and C{\^o}rtes}]{gomes2019bug}
\bibinfo{author}{Gomes, L.A.F.}, \bibinfo{author}{da~Silva~Torres, R.}, \bibinfo{author}{C{\^o}rtes, M.L.}, \bibinfo{year}{2019}.
\newblock \bibinfo{title}{Bug report severity level prediction in open source software: A survey and research opportunities}.
\newblock \bibinfo{journal}{Information and software technology} \bibinfo{volume}{115}, \bibinfo{pages}{58--78}.
%Type = Inproceedings
\bibitem[{Grund et~al.(2021)Grund, Chowdhury, Bradley, Hall and Holmes}]{grund2021codeshovel}
\bibinfo{author}{Grund, F.}, \bibinfo{author}{Chowdhury, S.}, \bibinfo{author}{Bradley, N.C.}, \bibinfo{author}{Hall, B.}, \bibinfo{author}{Holmes, R.}, \bibinfo{year}{2021}.
\newblock \bibinfo{title}{Codeshovel: Constructing method-level source code histories}, in: \bibinfo{booktitle}{2021 IEEE/ACM 43rd International Conference on Software Engineering (ICSE)}, \bibinfo{organization}{IEEE}. pp. \bibinfo{pages}{1510--1522}.
%Type = Inproceedings
\bibitem[{Habib and Pradel(2018)}]{habib2018many}
\bibinfo{author}{Habib, A.}, \bibinfo{author}{Pradel, M.}, \bibinfo{year}{2018}.
\newblock \bibinfo{title}{How many of all bugs do we find? a study of static bug detectors}, in: \bibinfo{booktitle}{2018 33rd IEEE/ACM International Conference on Automated Software Engineering (ASE)}, \bibinfo{organization}{IEEE}. pp. \bibinfo{pages}{317--328}.
%Type = Article
\bibitem[{Habib and Pradel(2019)}]{habib2019neural}
\bibinfo{author}{Habib, A.}, \bibinfo{author}{Pradel, M.}, \bibinfo{year}{2019}.
\newblock \bibinfo{title}{Neural bug finding: A study of opportunities and challenges}.
\newblock \bibinfo{journal}{arXiv preprint arXiv:1906.00307} .
%Type = Article
\bibitem[{Hall et~al.(2011)Hall, Beecham, Bowes, Gray and Counsell}]{hall2011systematic}
\bibinfo{author}{Hall, T.}, \bibinfo{author}{Beecham, S.}, \bibinfo{author}{Bowes, D.}, \bibinfo{author}{Gray, D.}, \bibinfo{author}{Counsell, S.}, \bibinfo{year}{2011}.
\newblock \bibinfo{title}{A systematic literature review on fault prediction performance in software engineering}.
\newblock \bibinfo{journal}{IEEE Transactions on Software Engineering} \bibinfo{volume}{38}, \bibinfo{pages}{1276--1304}.
%Type = Book
\bibitem[{Halstead(1977)}]{halstead1977elements}
\bibinfo{author}{Halstead, M.H.}, \bibinfo{year}{1977}.
\newblock \bibinfo{title}{Elements of Software Science (Operating and programming systems series)}.
\newblock \bibinfo{publisher}{Elsevier Science Inc.}
%Type = Inproceedings
\bibitem[{Hata et~al.(2012)Hata, Mizuno and Kikuno}]{hata2012bug}
\bibinfo{author}{Hata, H.}, \bibinfo{author}{Mizuno, O.}, \bibinfo{author}{Kikuno, T.}, \bibinfo{year}{2012}.
\newblock \bibinfo{title}{Bug prediction based on fine-grained module histories}, in: \bibinfo{booktitle}{2012 34th international conference on software engineering (ICSE)}, \bibinfo{organization}{IEEE}. pp. \bibinfo{pages}{200--210}.
%Type = Inproceedings
\bibitem[{Hemmati(2015)}]{hemmati2015effective}
\bibinfo{author}{Hemmati, H.}, \bibinfo{year}{2015}.
\newblock \bibinfo{title}{How effective are code coverage criteria?}, in: \bibinfo{booktitle}{2015 IEEE International Conference on Software Quality, Reliability and Security}, \bibinfo{organization}{IEEE}. pp. \bibinfo{pages}{151--156}.
%Type = Inproceedings
\bibitem[{{Herzig} and {Zeller}(2013)}]{Herzig:2013}
\bibinfo{author}{{Herzig}, K.}, \bibinfo{author}{{Zeller}, A.}, \bibinfo{year}{2013}.
\newblock \bibinfo{title}{The impact of tangled code changes}, in: \bibinfo{booktitle}{2013 10th Working Conference on Mining Software Repositories}, pp. \bibinfo{pages}{121--130}.
%Type = Inproceedings
\bibitem[{Hess and Kromrey(2004)}]{hess2004robust}
\bibinfo{author}{Hess, M.R.}, \bibinfo{author}{Kromrey, J.D.}, \bibinfo{year}{2004}.
\newblock \bibinfo{title}{Robust confidence intervals for effect sizes: A comparative study of cohen’sd and cliff’s delta under non-normality and heterogeneous variances}, in: \bibinfo{booktitle}{annual meeting of the American Educational Research Association}, \bibinfo{organization}{Citeseer}.
%Type = Inproceedings
\bibitem[{Hindle et~al.(2008)Hindle, Godfrey and Holt}]{hindle2008reading}
\bibinfo{author}{Hindle, A.}, \bibinfo{author}{Godfrey, M.W.}, \bibinfo{author}{Holt, R.C.}, \bibinfo{year}{2008}.
\newblock \bibinfo{title}{Reading beside the lines: Indentation as a proxy for complexity metric}, in: \bibinfo{booktitle}{2008 16th IEEE International Conference on Program Comprehension}, \bibinfo{organization}{IEEE}. pp. \bibinfo{pages}{133--142}.
%Type = Article
\bibitem[{Hosseini et~al.(2017)Hosseini, Turhan and Gunarathna}]{hosseini2017systematic}
\bibinfo{author}{Hosseini, S.}, \bibinfo{author}{Turhan, B.}, \bibinfo{author}{Gunarathna, D.}, \bibinfo{year}{2017}.
\newblock \bibinfo{title}{A systematic literature review and meta-analysis on cross project defect prediction}.
\newblock \bibinfo{journal}{IEEE Transactions on Software Engineering} \bibinfo{volume}{45}, \bibinfo{pages}{111--147}.
%Type = Misc
\bibitem[{Infer(2023)}]{infer}
\bibinfo{author}{Infer}, \bibinfo{year}{2023}.
\newblock \bibinfo{title}{Infer}.
\newblock \URLprefix \url{https://fbinfer.com/}. \bibinfo{note}{[Online; accessed 20-April-2022]}.
%Type = Inproceedings
\bibitem[{{Johnson} et~al.(2019){Johnson}, {Lubo}, {Yedla}, {Aponte} and {Sharif}}]{Johnson:2019}
\bibinfo{author}{{Johnson}, J.}, \bibinfo{author}{{Lubo}, S.}, \bibinfo{author}{{Yedla}, N.}, \bibinfo{author}{{Aponte}, J.}, \bibinfo{author}{{Sharif}, B.}, \bibinfo{year}{2019}.
\newblock \bibinfo{title}{An empirical study assessing source code readability in comprehension}, in: \bibinfo{booktitle}{2019 IEEE International Conference on Software Maintenance and Evolution}, pp. \bibinfo{pages}{513--523}.
%Type = Article
\bibitem[{Jureczko and Spinellis(2010)}]{jureczko2010using}
\bibinfo{author}{Jureczko, M.}, \bibinfo{author}{Spinellis, D.}, \bibinfo{year}{2010}.
\newblock \bibinfo{title}{Using object-oriented design metrics to predict software defects}.
\newblock \bibinfo{journal}{Models and Methods of System Dependability. Oficyna Wydawnicza Politechniki Wroc{\l}awskiej} , \bibinfo{pages}{69--81}.
%Type = Inproceedings
\bibitem[{Just et~al.(2014)Just, Jalali and Ernst}]{just2014defects4j}
\bibinfo{author}{Just, R.}, \bibinfo{author}{Jalali, D.}, \bibinfo{author}{Ernst, M.D.}, \bibinfo{year}{2014}.
\newblock \bibinfo{title}{Defects4j: A database of existing faults to enable controlled testing studies for java programs}, in: \bibinfo{booktitle}{Proceedings of the 2014 International Symposium on Software Testing and Analysis}, pp. \bibinfo{pages}{437--440}.
%Type = Article
\bibitem[{Kafura and Reddy(1987)}]{kafura1987use}
\bibinfo{author}{Kafura, D.}, \bibinfo{author}{Reddy, G.R.}, \bibinfo{year}{1987}.
\newblock \bibinfo{title}{The use of software complexity metrics in software maintenance}.
\newblock \bibinfo{journal}{IEEE Transactions on Software Engineering} , \bibinfo{pages}{335--343}.
%Type = Article
\bibitem[{Kanwal and Maqbool(2012)}]{kanwal2012bug}
\bibinfo{author}{Kanwal, J.}, \bibinfo{author}{Maqbool, O.}, \bibinfo{year}{2012}.
\newblock \bibinfo{title}{Bug prioritization to facilitate bug report triage}.
\newblock \bibinfo{journal}{Journal of Computer Science and Technology} \bibinfo{volume}{27}, \bibinfo{pages}{397--412}.
%Type = Inproceedings
\bibitem[{Kasto and Whalley(2013)}]{kasto2013measuring}
\bibinfo{author}{Kasto, N.}, \bibinfo{author}{Whalley, J.}, \bibinfo{year}{2013}.
\newblock \bibinfo{title}{Measuring the difficulty of code comprehension tasks using software metrics}, in: \bibinfo{booktitle}{Proceedings of the Fifteenth Australasian Computing Education Conference-Volume 136}, pp. \bibinfo{pages}{59--65}.
%Type = Article
\bibitem[{Khatiwada et~al.(2018)Khatiwada, Tushev and Mahmoud}]{khatiwada2018just}
\bibinfo{author}{Khatiwada, S.}, \bibinfo{author}{Tushev, M.}, \bibinfo{author}{Mahmoud, A.}, \bibinfo{year}{2018}.
\newblock \bibinfo{title}{Just enough semantics: An information theoretic approach for ir-based software bug localization}.
\newblock \bibinfo{journal}{Information and Software Technology} \bibinfo{volume}{93}, \bibinfo{pages}{45--57}.
%Type = Article
\bibitem[{Khomh et~al.(2012)Khomh, Penta, Gu{\'e}h{\'e}neuc and Antoniol}]{khomh2012exploratory}
\bibinfo{author}{Khomh, F.}, \bibinfo{author}{Penta, M.D.}, \bibinfo{author}{Gu{\'e}h{\'e}neuc, Y.G.}, \bibinfo{author}{Antoniol, G.}, \bibinfo{year}{2012}.
\newblock \bibinfo{title}{An exploratory study of the impact of antipatterns on class change-and fault-proneness}.
\newblock \bibinfo{journal}{Empirical Software Engineering} \bibinfo{volume}{17}, \bibinfo{pages}{243--275}.
%Type = Article
\bibitem[{Kondo et~al.(2020)Kondo, German, Mizuno and Choi}]{kondo2020impact}
\bibinfo{author}{Kondo, M.}, \bibinfo{author}{German, D.M.}, \bibinfo{author}{Mizuno, O.}, \bibinfo{author}{Choi, E.H.}, \bibinfo{year}{2020}.
\newblock \bibinfo{title}{The impact of context metrics on just-in-time defect prediction}.
\newblock \bibinfo{journal}{Empirical Software Engineering} \bibinfo{volume}{25}, \bibinfo{pages}{890--939}.
%Type = Inproceedings
\bibitem[{Kononenko et~al.(2016)Kononenko, Baysal and Godfrey}]{kononenko2016code}
\bibinfo{author}{Kononenko, O.}, \bibinfo{author}{Baysal, O.}, \bibinfo{author}{Godfrey, M.W.}, \bibinfo{year}{2016}.
\newblock \bibinfo{title}{Code review quality: How developers see it}, in: \bibinfo{booktitle}{Proceedings of the 38th international conference on software engineering}, pp. \bibinfo{pages}{1028--1038}.
%Type = Article
\bibitem[{Koru and Liu(2005)}]{Koru:2005}
\bibinfo{author}{Koru, A.}, \bibinfo{author}{Liu, H.}, \bibinfo{year}{2005}.
\newblock \bibinfo{title}{Building effective defect-prediction models in practice}.
\newblock \bibinfo{journal}{IEEE Software} \bibinfo{volume}{22}, \bibinfo{pages}{23--29}.
%Type = Inproceedings
\bibitem[{Lamkanfi et~al.(2010)Lamkanfi, Demeyer, Giger and Goethals}]{lamkanfi2010predicting}
\bibinfo{author}{Lamkanfi, A.}, \bibinfo{author}{Demeyer, S.}, \bibinfo{author}{Giger, E.}, \bibinfo{author}{Goethals, B.}, \bibinfo{year}{2010}.
\newblock \bibinfo{title}{Predicting the severity of a reported bug}, in: \bibinfo{booktitle}{2010 7th IEEE Working Conference on Mining Software Repositories (MSR 2010)}, \bibinfo{organization}{IEEE}. pp. \bibinfo{pages}{1--10}.
%Type = Inproceedings
\bibitem[{Landman et~al.(2014)Landman, Serebrenik and Vinju}]{landman2014empirical}
\bibinfo{author}{Landman, D.}, \bibinfo{author}{Serebrenik, A.}, \bibinfo{author}{Vinju, J.}, \bibinfo{year}{2014}.
\newblock \bibinfo{title}{Empirical analysis of the relationship between cc and sloc in a large corpus of java methods}, in: \bibinfo{booktitle}{2014 IEEE International Conference on Software Maintenance and Evolution}, \bibinfo{organization}{IEEE}. pp. \bibinfo{pages}{221--230}.
%Type = Inproceedings
\bibitem[{Lavazza et~al.(2020)Lavazza, Tosi and Morasca}]{lavazza2020empirical}
\bibinfo{author}{Lavazza, L.}, \bibinfo{author}{Tosi, D.}, \bibinfo{author}{Morasca, S.}, \bibinfo{year}{2020}.
\newblock \bibinfo{title}{An empirical study on the persistence of spotbugs issues in open-source software evolution}, in: \bibinfo{booktitle}{International Conference on the Quality of Information and Communications Technology}, pp. \bibinfo{pages}{144--151}.
%Type = Article
\bibitem[{Le~Goues et~al.(2011)Le~Goues, Nguyen, Forrest and Weimer}]{le2011genprog}
\bibinfo{author}{Le~Goues, C.}, \bibinfo{author}{Nguyen, T.}, \bibinfo{author}{Forrest, S.}, \bibinfo{author}{Weimer, W.}, \bibinfo{year}{2011}.
\newblock \bibinfo{title}{Genprog: A generic method for automatic software repair}.
\newblock \bibinfo{journal}{Ieee transactions on software engineering} \bibinfo{volume}{38}, \bibinfo{pages}{54--72}.
%Type = Article
\bibitem[{Li et~al.(2020)Li, Shepperd and Guo}]{li2020systematic}
\bibinfo{author}{Li, N.}, \bibinfo{author}{Shepperd, M.}, \bibinfo{author}{Guo, Y.}, \bibinfo{year}{2020}.
\newblock \bibinfo{title}{A systematic review of unsupervised learning techniques for software defect prediction}.
\newblock \bibinfo{journal}{Information and Software Technology} \bibinfo{volume}{122}, \bibinfo{pages}{106287}.
%Type = Inproceedings
\bibitem[{Long and Rinard(2016)}]{long2016automatic}
\bibinfo{author}{Long, F.}, \bibinfo{author}{Rinard, M.}, \bibinfo{year}{2016}.
\newblock \bibinfo{title}{Automatic patch generation by learning correct code}, in: \bibinfo{booktitle}{Proceedings of the 43rd Annual ACM SIGPLAN-SIGACT Symposium on Principles of Programming Languages}, pp. \bibinfo{pages}{298--312}.
%Type = Article
\bibitem[{M{\"a}ntyl{\"a} and Lassenius(2008)}]{mantyla2008types}
\bibinfo{author}{M{\"a}ntyl{\"a}, M.V.}, \bibinfo{author}{Lassenius, C.}, \bibinfo{year}{2008}.
\newblock \bibinfo{title}{What types of defects are really discovered in code reviews?}
\newblock \bibinfo{journal}{IEEE Transactions on Software Engineering} \bibinfo{volume}{35}, \bibinfo{pages}{430--448}.
%Type = Article
\bibitem[{Martin(2007)}]{martin2007professionalism}
\bibinfo{author}{Martin, R.C.}, \bibinfo{year}{2007}.
\newblock \bibinfo{title}{Professionalism and test-driven development}.
\newblock \bibinfo{journal}{Ieee Software} \bibinfo{volume}{24}, \bibinfo{pages}{32--36}.
%Type = Article
\bibitem[{Martinez et~al.(2017)Martinez, Durieux, Sommerard, Xuan and Monperrus}]{martinez2017automatic}
\bibinfo{author}{Martinez, M.}, \bibinfo{author}{Durieux, T.}, \bibinfo{author}{Sommerard, R.}, \bibinfo{author}{Xuan, J.}, \bibinfo{author}{Monperrus, M.}, \bibinfo{year}{2017}.
\newblock \bibinfo{title}{Automatic repair of real bugs in java: A large-scale experiment on the defects4j dataset}.
\newblock \bibinfo{journal}{Empirical Software Engineering} \bibinfo{volume}{22}, \bibinfo{pages}{1936--1964}.
%Type = Misc
\bibitem[{Mashhadi(2023)}]{Ehsan-Empirical2023}
\bibinfo{author}{Mashhadi, E.}, \bibinfo{year}{2023}.
\newblock \bibinfo{title}{Bug severity empirical study}.
\newblock \bibinfo{howpublished}{\url{https://github.com/EhsanMashhadi/BugSeverityEmpiricalStudy}}.
%Type = Inproceedings
\bibitem[{Mashhadi et~al.(2023)Mashhadi, Ahmadvand and Hemmati}]{ehsan-llm}
\bibinfo{author}{Mashhadi, E.}, \bibinfo{author}{Ahmadvand, H.}, \bibinfo{author}{Hemmati, H.}, \bibinfo{year}{2023}.
\newblock \bibinfo{title}{Method-level bug severity prediction using source code metrics and llms}, in: \bibinfo{booktitle}{2023 IEEE 34th International Symposium on Software Reliability Engineering (ISSRE)}, pp. \bibinfo{pages}{635--646}.
\newblock \DOIprefix\doi{10.1109/ISSRE59848.2023.00055}.
%Type = Inproceedings
\bibitem[{Mashhadi and Hemmati(2021)}]{mashhadi2021applying}
\bibinfo{author}{Mashhadi, E.}, \bibinfo{author}{Hemmati, H.}, \bibinfo{year}{2021}.
\newblock \bibinfo{title}{Applying codebert for automated program repair of java simple bugs}, in: \bibinfo{booktitle}{2021 IEEE/ACM 18th International Conference on Mining Software Repositories (MSR)}, \bibinfo{organization}{IEEE}. pp. \bibinfo{pages}{505--509}.
%Type = Inproceedings
\bibitem[{{Matter} et~al.(2009){Matter}, {Kuhn} and {Nierstrasz}}]{Matter:2009}
\bibinfo{author}{{Matter}, D.}, \bibinfo{author}{{Kuhn}, A.}, \bibinfo{author}{{Nierstrasz}, O.}, \bibinfo{year}{2009}.
\newblock \bibinfo{title}{Assigning bug reports using a vocabulary-based expertise model of developers}, in: \bibinfo{booktitle}{2009 6th IEEE International Working Conference on Mining Software Repositories}, pp. \bibinfo{pages}{131--140}.
%Type = Article
\bibitem[{McCabe(1976)}]{mccabe1976complexity}
\bibinfo{author}{McCabe, T.J.}, \bibinfo{year}{1976}.
\newblock \bibinfo{title}{A complexity measure}.
\newblock \bibinfo{journal}{IEEE Transactions on software Engineering} , \bibinfo{pages}{308--320}.
%Type = Inproceedings
\bibitem[{McClure(1978)}]{mcclure1978model}
\bibinfo{author}{McClure, C.L.}, \bibinfo{year}{1978}.
\newblock \bibinfo{title}{A model for program complexity analysis}, in: \bibinfo{booktitle}{Proceedings of the 3rd international conference on Software engineering}, pp. \bibinfo{pages}{149--157}.
%Type = Misc
\bibitem[{Microsoft(2023)}]{visual_studio}
\bibinfo{author}{Microsoft}, \bibinfo{year}{2023}.
\newblock \bibinfo{title}{Visual studio3}.
\newblock \URLprefix \url{https://docs.microsoft.com/en-us/visualstudio/code-quality/code-metrics-maintainability-index-range-and-meaning?view=vs-2022\&viewFallbackFrom=vs-2022\%3A}. \bibinfo{note}{[Online; accessed 20-April-2022]}.
%Type = Book
\bibitem[{Miles and Huberman(1994)}]{miles1994qualitative}
\bibinfo{author}{Miles, M.B.}, \bibinfo{author}{Huberman, A.M.}, \bibinfo{year}{1994}.
\newblock \bibinfo{title}{Qualitative data analysis: An expanded sourcebook}.
\newblock \bibinfo{publisher}{sage}.
%Type = Inproceedings
\bibitem[{Mo et~al.(2016)Mo, Cai, Kazman, Xiao and Feng}]{mo2016decoupling}
\bibinfo{author}{Mo, R.}, \bibinfo{author}{Cai, Y.}, \bibinfo{author}{Kazman, R.}, \bibinfo{author}{Xiao, L.}, \bibinfo{author}{Feng, Q.}, \bibinfo{year}{2016}.
\newblock \bibinfo{title}{Decoupling level: a new metric for architectural maintenance complexity}, in: \bibinfo{booktitle}{2016 IEEE/ACM 38th International Conference on Software Engineering (ICSE)}, \bibinfo{organization}{IEEE}. pp. \bibinfo{pages}{499--510}.
%Type = Article
\bibitem[{Mo et~al.(2022)Mo, Wei, Feng and Li}]{Mo:2022}
\bibinfo{author}{Mo, R.}, \bibinfo{author}{Wei, S.}, \bibinfo{author}{Feng, Q.}, \bibinfo{author}{Li, Z.}, \bibinfo{year}{2022}.
\newblock \bibinfo{title}{An exploratory study of bug prediction at the method level}.
\newblock \bibinfo{journal}{Information and software technology} \bibinfo{volume}{144}, \bibinfo{pages}{106794}.
%Type = Article
\bibitem[{Nawrocki and Wojciechowski(2001)}]{nawrocki2001experimental}
\bibinfo{author}{Nawrocki, J.}, \bibinfo{author}{Wojciechowski, A.}, \bibinfo{year}{2001}.
\newblock \bibinfo{title}{Experimental evaluation of pair programming}.
\newblock \bibinfo{journal}{European Software Control and Metrics (Escom)} , \bibinfo{pages}{99--101}.
%Type = Article
\bibitem[{Neysiani et~al.(2020)Neysiani, Babamir and Aritsugi}]{neysiani2020efficient}
\bibinfo{author}{Neysiani, B.S.}, \bibinfo{author}{Babamir, S.M.}, \bibinfo{author}{Aritsugi, M.}, \bibinfo{year}{2020}.
\newblock \bibinfo{title}{Efficient feature extraction model for validation performance improvement of duplicate bug report detection in software bug triage systems}.
\newblock \bibinfo{journal}{Information and Software Technology} \bibinfo{volume}{126}, \bibinfo{pages}{106344}.
%Type = Article
\bibitem[{Okutan and Yıldız(2014)}]{Okutan:2014}
\bibinfo{author}{Okutan, A.}, \bibinfo{author}{Yıldız, O.T.}, \bibinfo{year}{2014}.
\newblock \bibinfo{title}{Software defect prediction using bayesian networks}.
\newblock \bibinfo{journal}{Empirical software engineering : an international journal} \bibinfo{volume}{19}, \bibinfo{pages}{154--181}.
%Type = Inproceedings
\bibitem[{Oman and Hagemeister(1992)}]{oman1992metrics}
\bibinfo{author}{Oman, P.}, \bibinfo{author}{Hagemeister, J.}, \bibinfo{year}{1992}.
\newblock \bibinfo{title}{Metrics for assessing a software system's maintainability}, in: \bibinfo{booktitle}{Proceedings Conference on Software Maintenance 1992}, \bibinfo{organization}{IEEE Computer Society}. pp. \bibinfo{pages}{337--338}.
%Type = Misc
\bibitem[{OpenAI(2023)}]{openai2023gpt4}
\bibinfo{author}{OpenAI}, \bibinfo{year}{2023}.
\newblock \bibinfo{title}{Gpt-4 technical report}.
\newblock \href{http://arxiv.org/abs/2303.08774}{\tt arXiv:2303.08774}.
%Type = Article
\bibitem[{Pai and Dugan(2007)}]{pai2007empirical}
\bibinfo{author}{Pai, G.J.}, \bibinfo{author}{Dugan, J.B.}, \bibinfo{year}{2007}.
\newblock \bibinfo{title}{Empirical analysis of software fault content and fault proneness using bayesian methods}.
\newblock \bibinfo{journal}{IEEE Transactions on software Engineering} \bibinfo{volume}{33}, \bibinfo{pages}{675--686}.
%Type = Article
\bibitem[{Palomba et~al.(2018)Palomba, Bavota, Penta, Fasano, Oliveto and Lucia}]{palomba2018diffuseness}
\bibinfo{author}{Palomba, F.}, \bibinfo{author}{Bavota, G.}, \bibinfo{author}{Penta, M.D.}, \bibinfo{author}{Fasano, F.}, \bibinfo{author}{Oliveto, R.}, \bibinfo{author}{Lucia, A.D.}, \bibinfo{year}{2018}.
\newblock \bibinfo{title}{On the diffuseness and the impact on maintainability of code smells: a large scale empirical investigation}.
\newblock \bibinfo{journal}{Empirical Software Engineering} \bibinfo{volume}{23}, \bibinfo{pages}{1188--1221}.
%Type = Inproceedings
\bibitem[{Pantiuchina et~al.(2018)Pantiuchina, Lanza and Bavota}]{pantiuchina2018improving}
\bibinfo{author}{Pantiuchina, J.}, \bibinfo{author}{Lanza, M.}, \bibinfo{author}{Bavota, G.}, \bibinfo{year}{2018}.
\newblock \bibinfo{title}{Improving code: The (mis) perception of quality metrics}, in: \bibinfo{booktitle}{2018 IEEE International Conference on Software Maintenance and Evolution (ICSME)}, \bibinfo{organization}{IEEE}. pp. \bibinfo{pages}{80--91}.
%Type = Article
\bibitem[{Pascarella et~al.(2020)Pascarella, Palomba and Bacchelli}]{pascarella2020performance}
\bibinfo{author}{Pascarella, L.}, \bibinfo{author}{Palomba, F.}, \bibinfo{author}{Bacchelli, A.}, \bibinfo{year}{2020}.
\newblock \bibinfo{title}{On the performance of method-level bug prediction: A negative result}.
\newblock \bibinfo{journal}{Journal of Systems and Software} \bibinfo{volume}{161}, \bibinfo{pages}{110493}.
%Type = Inproceedings
\bibitem[{Pearson et~al.(2017)Pearson, Campos, Just, Fraser, Abreu, Ernst, Pang and Keller}]{pearson2017evaluating}
\bibinfo{author}{Pearson, S.}, \bibinfo{author}{Campos, J.}, \bibinfo{author}{Just, R.}, \bibinfo{author}{Fraser, G.}, \bibinfo{author}{Abreu, R.}, \bibinfo{author}{Ernst, M.D.}, \bibinfo{author}{Pang, D.}, \bibinfo{author}{Keller, B.}, \bibinfo{year}{2017}.
\newblock \bibinfo{title}{Evaluating and improving fault localization}, in: \bibinfo{booktitle}{2017 IEEE/ACM 39th International Conference on Software Engineering (ICSE)}, \bibinfo{organization}{IEEE}. pp. \bibinfo{pages}{609--620}.
%Type = Inproceedings
\bibitem[{Pecorelli et~al.(2019)Pecorelli, Palomba, Di~Nucci and De~Lucia}]{pecorelli2019comparing}
\bibinfo{author}{Pecorelli, F.}, \bibinfo{author}{Palomba, F.}, \bibinfo{author}{Di~Nucci, D.}, \bibinfo{author}{De~Lucia, A.}, \bibinfo{year}{2019}.
\newblock \bibinfo{title}{Comparing heuristic and machine learning approaches for metric-based code smell detection}, in: \bibinfo{booktitle}{2019 IEEE/ACM 27th International Conference on Program Comprehension (ICPC)}, \bibinfo{organization}{IEEE}. pp. \bibinfo{pages}{93--104}.
%Type = Inproceedings
\bibitem[{Polo et~al.(2001)Polo, Piattini and Ruiz}]{polo2001using}
\bibinfo{author}{Polo, M.}, \bibinfo{author}{Piattini, M.}, \bibinfo{author}{Ruiz, F.}, \bibinfo{year}{2001}.
\newblock \bibinfo{title}{Using code metrics to predict maintenance of legacy programs: A case study}, in: \bibinfo{booktitle}{Proceedings IEEE International Conference on Software Maintenance. ICSM 2001}, \bibinfo{organization}{IEEE}. pp. \bibinfo{pages}{202--208}.
%Type = Inproceedings
\bibitem[{Posnett et~al.(2011)Posnett, Hindle and Devanbu}]{posnett2011simpler}
\bibinfo{author}{Posnett, D.}, \bibinfo{author}{Hindle, A.}, \bibinfo{author}{Devanbu, P.}, \bibinfo{year}{2011}.
\newblock \bibinfo{title}{A simpler model of software readability}, in: \bibinfo{booktitle}{Proceedings of the 8th working conference on mining software repositories}, pp. \bibinfo{pages}{73--82}.
%Type = Article
\bibitem[{Rafique and Mi{\v{s}}i{\'c}(2012)}]{rafique2012effects}
\bibinfo{author}{Rafique, Y.}, \bibinfo{author}{Mi{\v{s}}i{\'c}, V.B.}, \bibinfo{year}{2012}.
\newblock \bibinfo{title}{The effects of test-driven development on external quality and productivity: A meta-analysis}.
\newblock \bibinfo{journal}{IEEE Transactions on Software Engineering} \bibinfo{volume}{39}, \bibinfo{pages}{835--856}.
%Type = Article
\bibitem[{Rahman and Williams(2019)}]{rahman2019source}
\bibinfo{author}{Rahman, A.}, \bibinfo{author}{Williams, L.}, \bibinfo{year}{2019}.
\newblock \bibinfo{title}{Source code properties of defective infrastructure as code scripts}.
\newblock \bibinfo{journal}{Information and Software Technology} \bibinfo{volume}{112}, \bibinfo{pages}{148--163}.
%Type = Inproceedings
\bibitem[{Ralph and Tempero(2018)}]{ralph2018construct}
\bibinfo{author}{Ralph, P.}, \bibinfo{author}{Tempero, E.}, \bibinfo{year}{2018}.
\newblock \bibinfo{title}{Construct validity in software engineering research and software metrics}, in: \bibinfo{booktitle}{Proceedings of the 22nd International Conference on Evaluation and Assessment in Software Engineering 2018}, pp. \bibinfo{pages}{13--23}.
%Type = Article
\bibitem[{Ramay et~al.(2019)Ramay, Umer, Yin, Zhu and Illahi}]{ramay2019deep}
\bibinfo{author}{Ramay, W.Y.}, \bibinfo{author}{Umer, Q.}, \bibinfo{author}{Yin, X.C.}, \bibinfo{author}{Zhu, C.}, \bibinfo{author}{Illahi, I.}, \bibinfo{year}{2019}.
\newblock \bibinfo{title}{Deep neural network-based severity prediction of bug reports}.
\newblock \bibinfo{journal}{IEEE Access} \bibinfo{volume}{7}, \bibinfo{pages}{46846--46857}.
%Type = Inproceedings
\bibitem[{Saha et~al.(2014)Saha, Khurshid and Perry}]{saha2014empirical}
\bibinfo{author}{Saha, R.K.}, \bibinfo{author}{Khurshid, S.}, \bibinfo{author}{Perry, D.E.}, \bibinfo{year}{2014}.
\newblock \bibinfo{title}{An empirical study of long lived bugs}, in: \bibinfo{booktitle}{2014 Software Evolution Week-IEEE Conference on Software Maintenance, Reengineering, and Reverse Engineering (CSMR-WCRE)}, \bibinfo{organization}{IEEE}. pp. \bibinfo{pages}{144--153}.
%Type = Inproceedings
\bibitem[{Saha et~al.(2018)Saha, Lyu, Lam, Yoshida and Prasad}]{saha2018bugs}
\bibinfo{author}{Saha, R.K.}, \bibinfo{author}{Lyu, Y.}, \bibinfo{author}{Lam, W.}, \bibinfo{author}{Yoshida, H.}, \bibinfo{author}{Prasad, M.R.}, \bibinfo{year}{2018}.
\newblock \bibinfo{title}{Bugs. jar: a large-scale, diverse dataset of real-world java bugs}, in: \bibinfo{booktitle}{Proceedings of the 15th international conference on mining software repositories}, pp. \bibinfo{pages}{10--13}.
%Type = Inproceedings
\bibitem[{Saha et~al.(2017)Saha, Lyu, Yoshida and Prasad}]{saha2017elixir}
\bibinfo{author}{Saha, R.K.}, \bibinfo{author}{Lyu, Y.}, \bibinfo{author}{Yoshida, H.}, \bibinfo{author}{Prasad, M.R.}, \bibinfo{year}{2017}.
\newblock \bibinfo{title}{Elixir: Effective object-oriented program repair}, in: \bibinfo{booktitle}{2017 32nd IEEE/ACM International Conference on Automated Software Engineering (ASE)}, \bibinfo{organization}{IEEE}. pp. \bibinfo{pages}{648--659}.
%Type = Inproceedings
\bibitem[{Sahoo et~al.(2010)Sahoo, Criswell and Adve}]{sahoo2010empirical}
\bibinfo{author}{Sahoo, S.K.}, \bibinfo{author}{Criswell, J.}, \bibinfo{author}{Adve, V.}, \bibinfo{year}{2010}.
\newblock \bibinfo{title}{An empirical study of reported bugs in server software with implications for automated bug diagnosis}, in: \bibinfo{booktitle}{Proceedings of the 32nd ACM/IEEE International Conference on Software Engineering-Volume 1}, pp. \bibinfo{pages}{485--494}.
%Type = Misc
\bibitem[{Scikit-Learn(2007-2023a)}]{macroaverage}
\bibinfo{author}{Scikit-Learn}, \bibinfo{year}{2007-2023}a.
\newblock \bibinfo{title}{Macroaverage}.
\newblock \URLprefix \url{https://scikit-learn.org/stable/modules/generated/sklearn.metrics.f1_score.html}. \bibinfo{note}{[Online; accessed 20-April-2022]}.
%Type = Misc
\bibitem[{Scikit-Learn(2007-2023b)}]{robustscaler}
\bibinfo{author}{Scikit-Learn}, \bibinfo{year}{2007-2023}b.
\newblock \bibinfo{title}{Robustscaler}.
\newblock \URLprefix \url{https://scikit-learn.org/stable/modules/generated/sklearn.preprocessing.RobustScaler.html}. \bibinfo{note}{[Online; accessed 20-April-2022]}.
%Type = Inproceedings
\bibitem[{Servant and Jones(2017)}]{Francisco:2017}
\bibinfo{author}{Servant, F.}, \bibinfo{author}{Jones, J.A.}, \bibinfo{year}{2017}.
\newblock \bibinfo{title}{Fuzzy fine-grained code-history analysis}, in: \bibinfo{booktitle}{Proceedings of the International Conference on Software Engineering (ICSE)}, pp. \bibinfo{pages}{746--757}.
%Type = Inproceedings
\bibitem[{Shamshiri et~al.(2015)Shamshiri, Just, Rojas, Fraser, McMinn and Arcuri}]{shamshiri2015automatically}
\bibinfo{author}{Shamshiri, S.}, \bibinfo{author}{Just, R.}, \bibinfo{author}{Rojas, J.M.}, \bibinfo{author}{Fraser, G.}, \bibinfo{author}{McMinn, P.}, \bibinfo{author}{Arcuri, A.}, \bibinfo{year}{2015}.
\newblock \bibinfo{title}{Do automatically generated unit tests find real faults? an empirical study of effectiveness and challenges (t)}, in: \bibinfo{booktitle}{2015 30th IEEE/ACM International Conference on Automated Software Engineering (ASE)}, \bibinfo{organization}{IEEE}. pp. \bibinfo{pages}{201--211}.
%Type = Article
\bibitem[{Shaphiro and Wilk(1965)}]{shaphiro1965analysis}
\bibinfo{author}{Shaphiro, S.}, \bibinfo{author}{Wilk, M.}, \bibinfo{year}{1965}.
\newblock \bibinfo{title}{An analysis of variance test for normality}.
\newblock \bibinfo{journal}{Biometrika} \bibinfo{volume}{52}, \bibinfo{pages}{591--611}.
%Type = Article
\bibitem[{{Shepperd}(1988)}]{Shepperd:1988}
\bibinfo{author}{{Shepperd}, M.}, \bibinfo{year}{1988}.
\newblock \bibinfo{title}{A critique of cyclomatic complexity as a software metric}.
\newblock \bibinfo{journal}{Software Engineering Journal} \bibinfo{volume}{3}, \bibinfo{pages}{30--36}.
%Type = Inproceedings
\bibitem[{Shihab et~al.(2012)Shihab, Hassan, Adams and Jiang}]{shihab2012industrial}
\bibinfo{author}{Shihab, E.}, \bibinfo{author}{Hassan, A.E.}, \bibinfo{author}{Adams, B.}, \bibinfo{author}{Jiang, Z.M.}, \bibinfo{year}{2012}.
\newblock \bibinfo{title}{An industrial study on the risk of software changes}, in: \bibinfo{booktitle}{Proceedings of the ACM SIGSOFT 20th International Symposium on the Foundations of Software Engineering}, pp. \bibinfo{pages}{1--11}.
%Type = Article
\bibitem[{Shin et~al.(2010)Shin, Meneely, Williams and Osborne}]{shin2010evaluating}
\bibinfo{author}{Shin, Y.}, \bibinfo{author}{Meneely, A.}, \bibinfo{author}{Williams, L.}, \bibinfo{author}{Osborne, J.A.}, \bibinfo{year}{2010}.
\newblock \bibinfo{title}{Evaluating complexity, code churn, and developer activity metrics as indicators of software vulnerabilities}.
\newblock \bibinfo{journal}{IEEE transactions on software engineering} \bibinfo{volume}{37}, \bibinfo{pages}{772--787}.
%Type = Inproceedings
\bibitem[{Shivaji et~al.(2009)Shivaji, Whitehead, Akella and Kim}]{shivaji2009reducing}
\bibinfo{author}{Shivaji, S.}, \bibinfo{author}{Whitehead, E.J.}, \bibinfo{author}{Akella, R.}, \bibinfo{author}{Kim, S.}, \bibinfo{year}{2009}.
\newblock \bibinfo{title}{Reducing features to improve bug prediction}, in: \bibinfo{booktitle}{2009 IEEE/ACM International Conference on Automated Software Engineering}, \bibinfo{organization}{IEEE}. pp. \bibinfo{pages}{600--604}.
%Type = Article
\bibitem[{{Sjøberg} et~al.(2013){Sjøberg}, {Yamashita}, {Anda}, {Mockus} and {Dybå}}]{TSE:2013}
\bibinfo{author}{{Sjøberg}, D.I.K.}, \bibinfo{author}{{Yamashita}, A.}, \bibinfo{author}{{Anda}, B.C.D.}, \bibinfo{author}{{Mockus}, A.}, \bibinfo{author}{{Dybå}, T.}, \bibinfo{year}{2013}.
\newblock \bibinfo{title}{Quantifying the effect of code smells on maintenance effort}.
\newblock \bibinfo{journal}{IEEE Transactions on Software Engineering} \bibinfo{volume}{39}, \bibinfo{pages}{1144--1156}.
%Type = Inproceedings
\bibitem[{Spadini et~al.(2018)Spadini, Palomba, Zaidman, Bruntink and Bacchelli}]{spadini2018relation}
\bibinfo{author}{Spadini, D.}, \bibinfo{author}{Palomba, F.}, \bibinfo{author}{Zaidman, A.}, \bibinfo{author}{Bruntink, M.}, \bibinfo{author}{Bacchelli, A.}, \bibinfo{year}{2018}.
\newblock \bibinfo{title}{On the relation of test smells to software code quality}, in: \bibinfo{booktitle}{2018 IEEE international conference on software maintenance and evolution (ICSME)}, \bibinfo{organization}{IEEE}. pp. \bibinfo{pages}{1--12}.
%Type = Misc
\bibitem[{SpotBugs(2023)}]{spotbugs}
\bibinfo{author}{SpotBugs}, \bibinfo{year}{2023}.
\newblock \bibinfo{title}{Spotbugs}.
\newblock \URLprefix \url{https://spotbugs.github.io/}. \bibinfo{note}{[Online; accessed 20-April-2022]}.
%Type = Inproceedings
\bibitem[{Steidl et~al.(2014)Steidl, Hummel and Juergens}]{Daniela:2014}
\bibinfo{author}{Steidl, D.}, \bibinfo{author}{Hummel, B.}, \bibinfo{author}{Juergens, E.}, \bibinfo{year}{2014}.
\newblock \bibinfo{title}{Incremental origin analysis of source code files}, in: \bibinfo{booktitle}{Proceedings Working Conference on Mining Software Repositories (MSR)}, pp. \bibinfo{pages}{42 -- 51}.
%Type = Article
\bibitem[{Subramanyam and Krishnan(2003)}]{subramanyam2003empirical}
\bibinfo{author}{Subramanyam, R.}, \bibinfo{author}{Krishnan, M.S.}, \bibinfo{year}{2003}.
\newblock \bibinfo{title}{Empirical analysis of ck metrics for object-oriented design complexity: Implications for software defects}.
\newblock \bibinfo{journal}{IEEE Transactions on software engineering} \bibinfo{volume}{29}, \bibinfo{pages}{297--310}.
%Type = Article
\bibitem[{Sun et~al.(2015)Sun, Marakas and Aguirre-Urreta}]{sun2015effectiveness}
\bibinfo{author}{Sun, W.}, \bibinfo{author}{Marakas, G.}, \bibinfo{author}{Aguirre-Urreta, M.}, \bibinfo{year}{2015}.
\newblock \bibinfo{title}{The effectiveness of pair programming: Software professionals' perceptions}.
\newblock \bibinfo{journal}{IEEE Software} \bibinfo{volume}{33}, \bibinfo{pages}{72--79}.
%Type = Article
\bibitem[{Sun et~al.(2019)Sun, Zhou, Li, Ni and Lu}]{sun2019bug}
\bibinfo{author}{Sun, X.}, \bibinfo{author}{Zhou, W.}, \bibinfo{author}{Li, B.}, \bibinfo{author}{Ni, Z.}, \bibinfo{author}{Lu, J.}, \bibinfo{year}{2019}.
\newblock \bibinfo{title}{Bug localization for version issues with defect patterns}.
\newblock \bibinfo{journal}{IEEE Access} \bibinfo{volume}{7}, \bibinfo{pages}{18811--18820}.
%Type = Article
\bibitem[{Tan et~al.(2020)Tan, Xu, Wang, Zhang, Xu and Luo}]{tan2020bug}
\bibinfo{author}{Tan, Y.}, \bibinfo{author}{Xu, S.}, \bibinfo{author}{Wang, Z.}, \bibinfo{author}{Zhang, T.}, \bibinfo{author}{Xu, Z.}, \bibinfo{author}{Luo, X.}, \bibinfo{year}{2020}.
\newblock \bibinfo{title}{Bug severity prediction using question-and-answer pairs from stack overflow}.
\newblock \bibinfo{journal}{Journal of Systems and Software} \bibinfo{volume}{165}, \bibinfo{pages}{110567}.
%Type = Inproceedings
\bibitem[{Thung et~al.(2012)Thung, Lucia, Lo, Jiang, Rahman and Devanbu}]{thung2012extent}
\bibinfo{author}{Thung, F.}, \bibinfo{author}{Lucia}, \bibinfo{author}{Lo, D.}, \bibinfo{author}{Jiang, L.}, \bibinfo{author}{Rahman, F.}, \bibinfo{author}{Devanbu, P.T.}, \bibinfo{year}{2012}.
\newblock \bibinfo{title}{To what extent could we detect field defects? an empirical study of false negatives in static bug finding tools}, in: \bibinfo{booktitle}{Proceedings of the 27th IEEE/ACM International Conference on Automated Software Engineering}, pp. \bibinfo{pages}{50--59}.
%Type = Article
\bibitem[{Tian et~al.(2016)Tian, Ali, Lo and Hassan}]{tian2016unreliability}
\bibinfo{author}{Tian, Y.}, \bibinfo{author}{Ali, N.}, \bibinfo{author}{Lo, D.}, \bibinfo{author}{Hassan, A.E.}, \bibinfo{year}{2016}.
\newblock \bibinfo{title}{On the unreliability of bug severity data}.
\newblock \bibinfo{journal}{Empirical Software Engineering} \bibinfo{volume}{21}, \bibinfo{pages}{2298--2323}.
%Type = Inproceedings
\bibitem[{Tian et~al.(2012)Tian, Lo and Sun}]{tian2012information}
\bibinfo{author}{Tian, Y.}, \bibinfo{author}{Lo, D.}, \bibinfo{author}{Sun, C.}, \bibinfo{year}{2012}.
\newblock \bibinfo{title}{Information retrieval based nearest neighbor classification for fine-grained bug severity prediction}, in: \bibinfo{booktitle}{2012 19th Working Conference on Reverse Engineering}, \bibinfo{organization}{IEEE}. pp. \bibinfo{pages}{215--224}.
%Type = Inproceedings
\bibitem[{Tian et~al.(2013)Tian, Lo and Sun}]{tian2013drone}
\bibinfo{author}{Tian, Y.}, \bibinfo{author}{Lo, D.}, \bibinfo{author}{Sun, C.}, \bibinfo{year}{2013}.
\newblock \bibinfo{title}{Drone: Predicting priority of reported bugs by multi-factor analysis}, in: \bibinfo{booktitle}{2013 IEEE International Conference on Software Maintenance}, pp. \bibinfo{pages}{200--209}.
%Type = Inproceedings
\bibitem[{Tomassi(2018)}]{tomassi2018bugs}
\bibinfo{author}{Tomassi, D.A.}, \bibinfo{year}{2018}.
\newblock \bibinfo{title}{Bugs in the wild: examining the effectiveness of static analyzers at finding real-world bugs}, in: \bibinfo{booktitle}{Proceedings of the 2018 26th ACM Joint Meeting on European Software Engineering Conference and Symposium on the Foundations of Software Engineering}, pp. \bibinfo{pages}{980--982}.
%Type = Article
\bibitem[{Tosun et~al.(2010)Tosun, Bener, Turhan and Menzies}]{tosun2010practical}
\bibinfo{author}{Tosun, A.}, \bibinfo{author}{Bener, A.}, \bibinfo{author}{Turhan, B.}, \bibinfo{author}{Menzies, T.}, \bibinfo{year}{2010}.
\newblock \bibinfo{title}{Practical considerations in deploying statistical methods for defect prediction: A case study within the turkish telecommunications industry}.
\newblock \bibinfo{journal}{Information and Software Technology} \bibinfo{volume}{52}, \bibinfo{pages}{1242--1257}.
%Type = Inproceedings
\bibitem[{Tufano et~al.(2015)Tufano, Palomba, Bavota, Oliveto, Di~Penta, De~Lucia and Poshyvanyk}]{tufano2015and}
\bibinfo{author}{Tufano, M.}, \bibinfo{author}{Palomba, F.}, \bibinfo{author}{Bavota, G.}, \bibinfo{author}{Oliveto, R.}, \bibinfo{author}{Di~Penta, M.}, \bibinfo{author}{De~Lucia, A.}, \bibinfo{author}{Poshyvanyk, D.}, \bibinfo{year}{2015}.
\newblock \bibinfo{title}{When and why your code starts to smell bad}, in: \bibinfo{booktitle}{2015 IEEE/ACM 37th IEEE International Conference on Software Engineering}, \bibinfo{organization}{IEEE}. pp. \bibinfo{pages}{403--414}.
%Type = Article
\bibitem[{Uddin et~al.(2017)Uddin, Ghazali, Deris, Naseem and Shah}]{uddin2017survey}
\bibinfo{author}{Uddin, J.}, \bibinfo{author}{Ghazali, R.}, \bibinfo{author}{Deris, M.M.}, \bibinfo{author}{Naseem, R.}, \bibinfo{author}{Shah, H.}, \bibinfo{year}{2017}.
\newblock \bibinfo{title}{A survey on bug prioritization}.
\newblock \bibinfo{journal}{Artificial Intelligence Review} \bibinfo{volume}{47}, \bibinfo{pages}{145--180}.
%Type = Book
\bibitem[{Vucevic and Yaddow(2012)}]{vucevic2012testing}
\bibinfo{author}{Vucevic, D.}, \bibinfo{author}{Yaddow, W.}, \bibinfo{year}{2012}.
\newblock \bibinfo{title}{Testing the data warehouse practicum: Assuring data content, data structures and quality}.
\newblock \bibinfo{publisher}{Trafford Publishing}.
%Type = Article
\bibitem[{Wahono(2015)}]{wahono2015systematic}
\bibinfo{author}{Wahono, R.S.}, \bibinfo{year}{2015}.
\newblock \bibinfo{title}{A systematic literature review of software defect prediction}.
\newblock \bibinfo{journal}{Journal of software engineering} \bibinfo{volume}{1}, \bibinfo{pages}{1--16}.
%Type = Inproceedings
\bibitem[{Wang et~al.(2010)Wang, Lin, Zhang and Hu}]{wang2010detect}
\bibinfo{author}{Wang, D.}, \bibinfo{author}{Lin, M.}, \bibinfo{author}{Zhang, H.}, \bibinfo{author}{Hu, H.}, \bibinfo{year}{2010}.
\newblock \bibinfo{title}{Detect related bugs from source code using bug information}, in: \bibinfo{booktitle}{2010 IEEE 34th Annual Computer Software and Applications Conference}, \bibinfo{organization}{IEEE}. pp. \bibinfo{pages}{228--237}.
%Type = Article
\bibitem[{Wang et~al.(2019)Wang, Xia, Lo and Li}]{WANG:2019}
\bibinfo{author}{Wang, Q.}, \bibinfo{author}{Xia, X.}, \bibinfo{author}{Lo, D.}, \bibinfo{author}{Li, S.}, \bibinfo{year}{2019}.
\newblock \bibinfo{title}{Why is my code change abandoned?}
\newblock \bibinfo{journal}{Information and Software Technology} \bibinfo{volume}{110}, \bibinfo{pages}{108 -- 120}.
%Type = Book
\bibitem[{Watson et~al.(1996)Watson, Wallace and McCabe}]{watson1996structured}
\bibinfo{author}{Watson, A.H.}, \bibinfo{author}{Wallace, D.R.}, \bibinfo{author}{McCabe, T.J.}, \bibinfo{year}{1996}.
\newblock \bibinfo{title}{Structured testing: A testing methodology using the cyclomatic complexity metric}. volume \bibinfo{volume}{500}.
\newblock \bibinfo{publisher}{US Department of Commerce, Technology Administration, National Institute of~…}.
%Type = Article
\bibitem[{Wattanakriengkrai et~al.(2020)Wattanakriengkrai, Thongtanunam, Tantithamthavorn, Hata and Matsumoto}]{wattanakriengkrai2020predicting}
\bibinfo{author}{Wattanakriengkrai, S.}, \bibinfo{author}{Thongtanunam, P.}, \bibinfo{author}{Tantithamthavorn, C.}, \bibinfo{author}{Hata, H.}, \bibinfo{author}{Matsumoto, K.}, \bibinfo{year}{2020}.
\newblock \bibinfo{title}{Predicting defective lines using a model-agnostic technique}.
\newblock \bibinfo{journal}{arXiv preprint arXiv:2009.03612} .
%Type = Article
\bibitem[{Williams et~al.(2000)Williams, Kessler, Cunningham and Jeffries}]{williams2000strengthening}
\bibinfo{author}{Williams, L.}, \bibinfo{author}{Kessler, R.R.}, \bibinfo{author}{Cunningham, W.}, \bibinfo{author}{Jeffries, R.}, \bibinfo{year}{2000}.
\newblock \bibinfo{title}{Strengthening the case for pair programming}.
\newblock \bibinfo{journal}{IEEE software} \bibinfo{volume}{17}, \bibinfo{pages}{19--25}.
%Type = Article
\bibitem[{Wong et~al.(2016)Wong, Gao, Li, Abreu and Wotawa}]{wong2016survey}
\bibinfo{author}{Wong, W.E.}, \bibinfo{author}{Gao, R.}, \bibinfo{author}{Li, Y.}, \bibinfo{author}{Abreu, R.}, \bibinfo{author}{Wotawa, F.}, \bibinfo{year}{2016}.
\newblock \bibinfo{title}{A survey on software fault localization}.
\newblock \bibinfo{journal}{IEEE Transactions on Software Engineering} \bibinfo{volume}{42}, \bibinfo{pages}{707--740}.
%Type = Inproceedings
\bibitem[{Zaw et~al.(2020)Zaw, Hnin, Kyaw and Funabiki}]{zaw2020software}
\bibinfo{author}{Zaw, K.K.}, \bibinfo{author}{Hnin, H.W.}, \bibinfo{author}{Kyaw, K.Y.}, \bibinfo{author}{Funabiki, N.}, \bibinfo{year}{2020}.
\newblock \bibinfo{title}{Software quality metrics calculations for java programming learning assistant system}, in: \bibinfo{booktitle}{2020 IEEE Conference on Computer Applications (ICCA)}, \bibinfo{organization}{IEEE}. pp. \bibinfo{pages}{1--6}.
%Type = Article
\bibitem[{Zhou and Leung(2006)}]{zhou2006empirical}
\bibinfo{author}{Zhou, Y.}, \bibinfo{author}{Leung, H.}, \bibinfo{year}{2006}.
\newblock \bibinfo{title}{Empirical analysis of object-oriented design metrics for predicting high and low severity faults}.
\newblock \bibinfo{journal}{IEEE Transactions on software engineering} \bibinfo{volume}{32}, \bibinfo{pages}{771--789}.
%Type = Article
\bibitem[{Zhou et~al.(2010)Zhou, Xu and Leung}]{zhou2010ability}
\bibinfo{author}{Zhou, Y.}, \bibinfo{author}{Xu, B.}, \bibinfo{author}{Leung, H.}, \bibinfo{year}{2010}.
\newblock \bibinfo{title}{On the ability of complexity metrics to predict fault-prone classes in object-oriented systems}.
\newblock \bibinfo{journal}{Journal of Systems and Software} \bibinfo{volume}{83}, \bibinfo{pages}{660--674}.
%Type = Inproceedings
\bibitem[{Zimmermann et~al.(2007)Zimmermann, Premraj and Zeller}]{zimmermann2007predicting}
\bibinfo{author}{Zimmermann, T.}, \bibinfo{author}{Premraj, R.}, \bibinfo{author}{Zeller, A.}, \bibinfo{year}{2007}.
\newblock \bibinfo{title}{Predicting defects for eclipse}, in: \bibinfo{booktitle}{Third International Workshop on Predictor Models in Software Engineering (PROMISE'07: ICSE Workshops 2007)}, \bibinfo{organization}{IEEE}. pp. \bibinfo{pages}{9--9}.

\end{thebibliography}
